\documentclass[journal,twocolumn,12pt,twoside]{IEEEtran}
\usepackage{amsmath, amssymb,graphicx,bbold}
\usepackage{dutchcal}
\usepackage{xcolor, color}

\normalsize

\ifCLASSINFOpdf

\else

\fi

\newtheorem{Definition}{Definition}
\newtheorem{Example}{Example}
\newtheorem{Theorem}{Theorem}
\newtheorem{Lemma}{Lemma}
\newtheorem{Corollary}{Corollary}
\newtheorem{Remark}{Remark}
\newtheorem{Proposition}{Proposition}
\begin{document}
%
\title{On the Size of Error Ball in DNA Storage Channels}

\author{Aryan~Abbasian,~\IEEEmembership{}
        Mahtab~Mirmohseni,~\IEEEmembership{Senior~Member,~IEEE,}
        and~Masoumeh~Nasiri~Kenari,~\IEEEmembership{Senior~Member,~IEEE}
\thanks{Aryan Abbasian, Mahtab Mirmohseni and Masoumeh Nasiri Kenari are with the Department
of Electrical Engineering, Sharif University of Technology, Tehran,
Iran, e-mail: (aryan.abbasian@ee.sharif.edu; mirmohseni@sharif.edu ; mnasiri@sharif.edu)}
}

%
%

\markboth{IEEE Transactions on Molecular,Biological and Multi-Scale Communications}%
{}
%



\maketitle

\begin{abstract}
Recent experiments have demonstrated the feasibility of storing digital information in macromolecules such as DNA and protein. However, the DNA storage channel is prone to errors such as deletions, insertions, and substitutions. During the synthesis and reading phases of DNA strings, many noisy copies of the original string are generated. The problem of recovering the original string from these noisy copies is known as sequence reconstruction. A key concept in this problem is the error ball, which is the set of all possible sequences that can result from a limited number of errors applied to the original sequence. Levenshtein showed that the minimum number of noisy copies required for a given channel to recover the original sequence is equal to one plus the maximum size of the intersection of two error balls. Therefore, deriving the size of the error ball for any channel and any sequence is essential for solving the sequence reconstruction problem. In DNA storage systems, multiple types of errors such as deletion, insertion and substitution in a string could occur simultaneously.  In this work, we aim to derive the size of the error ball for channels with multiple types of errors and at most three edits. Specifically, we consider the channels with single-deletion double-substitution, single-deletion double-insertion and single-insertion single-substitution errors.

\end{abstract}


%
\IEEEpeerreviewmaketitle

\section{Introduction}
%
%
%
%
\IEEEPARstart{T}{he} need for long-term and dense data storage has motivated the search for new storage media. The current data storage devices, such as hard disk drives, are costly, have a limited lifespan, require electricity to operate, and generate a lot of heat. DNA molecules are a promising medium for archival data because of their ultra high density, durability, and low energy consumption [1]. DNA is a macromolecule composed of four bases: Adenine, Cytosine, Guanine, and Thymine. Therefore, a DNA string can be mathematically represented as a sequence of four letters. In DNA-based data storage, information can be stored in either natural (in vivo) or synthetic (in vitro) DNA strands. Synthetic DNA can offer higher densities, while DNA of living organisms, such as bacteria, is more cost-effective [2].
 
 In DNA-based data storage, first the original binary data is encoded into the quaternary sequences. Then, the DNA strands are synthesized and stored. In either in-vivo or in-vitro storages, these strands are expected to be copied many times. In the former, this is due to the biological processes like replication [3], and in the later due to the synthesis/sequencing processes [4]. These strands are prone to errors such as substitutions, insertions and deletions. The substitution occur when the correct symbol is replaced by an incorrect one. The deletion or insertion occur when a symbol is removed from or added to the sequence respectively. In the reading phase, having access to multiple noisy copies of a single strand, we must recover the original one. The problem of recovering a sequence of symbols from its noisy versions is called \emph{sequence reconstruction problem} [5].
 
 In 2001, Levenshtein in [5], [6] introduced a new communication paradigm where the transmitter transmits a sequence $\boldsymbol{\omega}$ over several noisy channels with the same error pattern which produce distinct outputs and the receiver must decode the transmitted sequence. He studied the minimum number of noisy channel outputs required to reconstruct any transmitted sequence. It is clear that this number must be larger than the size of the maximum intersection between the error balls of any two sequences. Levenshtein derived this value for various channels with different error patterns, see [5], [6]. For the more general case when the sequences belong to a set of codewords with a given minimum distance, Levenshtein used Hamming and Johnson graphs to derive the maximum size of the intersection of two error balls for substitutions and transpositions. Similar results were derived for deletions [10], [19] and insertions [9]. Levenshtein in [5], [6] also proposed reconstruction algorithms for channels with single type of error, such as substitutions, transpositions, deletions and insertions.   
 
   A prerequisite problem  to the sequence reconstruction problem is deriving the size of error ball for a particular channel for any given sequence. While the size of substitution ball and insertion sphere are only dependent on the length of the sequence and size of the alphabet, the size of deletion sphere is dependent on the structure of sequence itself. For example the size of single-deletion ball is equal to the number of runs in a sequence. In DNA-based data storage, multiple types of errors can occur simultaneously. 
  Authors in [11]-[13] and [8] derived the size of error ball for some channels with multiple types of error. In [12], the size of the single-deletion single-insertion error ball was derived. The technique of calculation was to first calculate the single-deletion sphere, then for each 1-subsequence calculate the single-insertion sphere and finally derive the size of the union of these spheres by subtracting the sequences which were counted more than once.Recently, the minimum, maximum and average size of single-deletion single-insertion ball was studied in [18] . In [11] the size of single-deletion single-substitution was derived for binary sequences and in [13] for non-binary sequences. The technique of calculation was similar to [12]. In [8] the size of single-deletion $t$-substitution and the single-insertion single-substitution error balls were derived for binary sequences. For calculation of error ball in  single-insertion single-substitution channel, first the single-insertion sphere was derived, then the single-substitution ball is calculated for each 1-supersequence and the size of the union of these balls is derived by subtraction of the common sequences.    

In this paper we continue to study the problem of calculating the size of error ball for channels with multiple types of error. The current synthesis technologies can produce strands of length 150 to 200. Also, recent experiments reported the probabilities of insertion, deletion and substitution to be $5.4\times 10^{-4}$ , $1.5\times 10^{-3}$ and $4.5\times 10^{-3}$ respectively [20]. Therefore, it is unlikely that DNA strands be corrupted by more than three edits. Thus, we concentrate on channels with three edits. 
 The first channel we consider is the single-deletion double-insertion channel. First we derive the single-deletion sphere and then for each 1-subsequence, we calculate the double-insertion sphere. We generalize the characterization of the intersection of two double-insertion balls into non-binary sequences, then derive the size of the intersection of three or more double-insertion balls. By using the inclusion-exclusion principle, we derive the size of error ball for this channel. The second channel we consider is single-deletion double-substitution. We derive the size of error ball for general non-binary sequences with the same technique we use for the previous channel. The third  channel we consider is single-insertion single-substitution channel. We generalize the technique used in [8] for non-binary sequences.  

The rest of this paper is organized as follows. In section II, we begin by preliminaries and reviewing previous results on the size of error ball. In sections III, IV, V we derive the size of error ball for single-deletion double-insertion channel, single-deletion double-substitution channel, single-insertion single-substitution channel . Section VI concludes the paper.
\section{Definitions, Preliminaries and Existing Results}
\subsection{Definitions}
The alphabet of size \emph{q} for integer $q\geq2$ is denoted by $\Sigma_{q}=\{0, 1, \ldots, q-1\}$, and $\Sigma^{n}_{q}$ is the set of all sequences of length \emph{n} in which every symbol belong to $\Sigma_{q}$. The bold letter refers to some sequence, such as \emph{\textbf{x}}, to distinguish it from symbols, and the \emph{i-}th symbol of a sequence $\emph{\textbf{x}}\in \Sigma^{n}_{q}$ is denoted by \emph{x}$_{i}$. The concatenation of two sequences $\emph{\textbf{x}}=\emph{x}_{1}\ldots\emph{x}_{n}$ and $\emph{\textbf{y}}=\emph{y}_{1}\ldots\emph{y}_{m}$ is denoted by $\emph{\textbf{xy}}=\emph{x}_{1}\ldots\emph{x}_{n}\emph{y}_{1}\ldots\emph{y}_{m}$. In particular, the $n$ times concatenation of $\emph{\textbf{x}}$ with itself is denoted by $\emph{\textbf{x}}^{n}$. Let $\mathcal{S}$ and $\mathcal{T}$ be two set of sequences. The set of sequences obtained by concatenating sequences in  $\mathcal{S}$ with $\mathcal{T}$ is denoted by $\mathcal{S}\circ\mathcal{T}$. When $\mathcal{T}=\{a\}$, we write $a\circ \mathcal{S}$ or $\mathcal{S}\circ a$.
We also denote $|\boldsymbol{\omega}|$ as the length of the sequence $\boldsymbol{\omega}$.   

The distance between two sequences may be defined in several ways. When the two sequences $\emph{\textbf{x}},\emph{\textbf{y}} \in \Sigma^{n}_{q}$ have the same length, the \emph{Hamming distance}, denoted by $d_{H}(\emph{\textbf{x}},\emph{\textbf{y}})$, is the minimum number of substitutions required to transform one sequence into another. Given a distance metric $d$ on a space $\emph{V}$, the ball with radius $t$ or simply \emph{t-ball} centered at $\boldsymbol{\omega}\in V$ is the set $\{\boldsymbol{\omega^{'}} : d(\boldsymbol{\omega}, \boldsymbol{\omega^{'}})\leq t\}$. The \emph{t-sphere} centered at $\boldsymbol{\omega}\in V$ is the set $\{\boldsymbol{\omega^{'}} : d(\boldsymbol{\omega}, \boldsymbol{\omega^{'}})=t\}$. The \emph{Hamming t-ball} centered at $\boldsymbol{\omega} \in \Sigma^{n}_{q}$, denoted by $\mathcal{H}_{t}(\boldsymbol{\omega})$, is the set $\{\textbf{\emph{z}} : d_{H}(\boldsymbol{\omega},\textbf{\emph{z}})\leq t\}$. Clearly, the size of the Hamming \emph{t}-ball is given by
\begin{align}
|\mathcal{H}_{t}(\boldsymbol{\omega})|=\sum_{i=0}^{t} \binom{n}{i}(q-1)^{i}.
\end{align}
 A sequence $\textbf{\emph{y}} \in \Sigma^{n-t}_{q}$ is a \emph{t-subsequence} of $\boldsymbol{\omega}$, if it can be derived after deletion of $t$ symbols from $\boldsymbol{\omega}$. Similarly, a sequence $\textbf{\emph{z}} \in \Sigma^{n+t}_{q}$ is a \emph{t-supersequence} of $\boldsymbol{\omega}$, if it can be formed by $t$ insertions into $\boldsymbol{\omega}$.
\begin{Definition}
The \emph{Deletion t-Sphere} of radius $t$ centered at $\boldsymbol{\omega} \in \Sigma^{n}_{q}$, denoted by $\mathcal{D}_{t}(\boldsymbol{\omega})$, is the set of all length $n-t$ subsequences of $\boldsymbol{\omega}$. Similarly, the \emph{Insertion t-Sphere} of radius $t$ centered at $\boldsymbol{\omega} \in \Sigma^{n}_{q}$, denoted by $\mathcal{I}_{t}(\boldsymbol{\omega})$, is the set of all length $n+t$ supersequences of $\boldsymbol{\omega}$.
\end{Definition}
Levenshtein showed that the size of the insertion \emph{t-}sphere only depends on the length of the sequence and the size of the alphabet [6]
\begin{equation}
|\mathcal{I}_{t}(\boldsymbol{\omega})|=\sum_{i=0}^{t}\binom{n+t}{i}(q-1)^{i}.
\end{equation}

\begin{Definition}
  A run in $\boldsymbol{\omega}$ is a maximal subsequence with consecutive similar symbols. The number of runs in the sequence $\boldsymbol{\omega}$ is denoted by $\rho(\boldsymbol{\omega})$. Any sequence like $\boldsymbol{\omega}$, can be represented as $\boldsymbol{\omega}=\sigma_{1}^{\ell_{1}}\sigma_{2}^{\ell_{2}}\ldots\sigma_{\rho(\boldsymbol{\omega})}^{\ell_{\rho(\boldsymbol{\omega})}} \in \Sigma^{n}_{q}$ where $\sigma_{i}\neq \sigma_{i+1}$.
  We define $\boldsymbol{\omega}^{(i)}$ be the sequence derived by deleting a symbol from the $i$-th run.
\end{Definition}
Unlike the substitution ball and insertion sphere, the size of the deletion sphere is dependent on the sequence itself, for example, $|\mathcal{D}_{1}(\boldsymbol{\omega})|=\rho(\boldsymbol{\omega})$.
 In DNA storage, multiple types of error can occur simultaneously. Therefore, we define the error ball, centered at $\boldsymbol{\omega} \in \Sigma^{n}_{q}$, denoted by $\mathcal{B}_{t_{1}, t_{2}, t_{3}}(\boldsymbol{\omega})$, as the set of sequences obtained by at most $t_{1}$ substitutions and exactly $t_{2}$ deletions and $t_{3}$ insertions. Note that, the order of occurrence of deletions, insertions and substitutions does not matter in deriving the size of error ball [8].
The \emph{Levenshtein distance} between two sequences $\textbf{\emph{x}}, \textbf{\emph{y}} \in \Sigma^{*}_{q}$, denoted by $d_{L}(\textbf{\emph{x}}, \textbf{\emph{y}})$, is the minimum number of insertions and deletions required to transform one sequence into another. The \emph{edit distance} between two sequences $\textbf{\emph{x}}, \textbf{\emph{y}} \in \Sigma^{*}_{q}$, denoted by $d_{E}(\textbf{\emph{x}}, \textbf{\emph{y}})$, is similarly defined as the minimum number of insertions, deletions and substitutions required to transform one sequence to another. 
\begin{Example}
  If $\emph{\textbf{x}}=0101$ and $\emph{\textbf{y}}=1010$, then $d_{L}(\emph{\textbf{x}}, \emph{\textbf{y}})=2$ while $d_{H}(\emph{\textbf{x}}, \emph{\textbf{y}})=4$.
\end{Example}
\begin{Definition}
  A sequence $\boldsymbol{\omega}=\omega_{1}\omega_{2}\ldots \omega_{n}$ is alternating if $\omega_{i}=\omega_{i+2}$ and $\omega_{1}\neq \omega_{2}$. An alternating subsequence of $\boldsymbol{\omega}$, $\boldsymbol{\omega}_{[i, j]}\triangleq \omega_{i}\omega_{i+1}\ldots \omega_{j}$ is a \emph{maximal alternating segment} if  $\boldsymbol{\omega}_{[i-1, j]}$ and $\boldsymbol{\omega}_{[i, j+1]}$ are not alternating. The number of maximal alternating segments of a sequence $\boldsymbol{\omega}$ is denoted by $\psi(\boldsymbol{\omega})$. For a sequence $\boldsymbol{\omega} \in \Sigma^{n}_{q}$, the \emph{maximal alternating segments profile } is the vector $(s_{1},s_{2},\ldots,s_{\psi(\boldsymbol{\omega})})$ where for $1\leq i \leq \psi(\boldsymbol{\omega})$, $s_{i}$ denotes the length of \emph{i}-th maximal alternating segment.
\end{Definition}
\begin{Example}
  If $\boldsymbol{\omega}=0021202$ then the maximal alternating segments are 0, 02, 212,202.
\end{Example}
 If a sequence $\boldsymbol{\omega} \in \Sigma^{n}_{q}$ has maximal alternating segments profile $(s_{1},s_{2},\ldots,s_{\psi(\boldsymbol{\omega})})$, then the size of single-deletion single-insertion error ball is [12]
\begin{equation}
  |\mathcal{B}_{0, 1, 1}(\boldsymbol{\omega})|=\rho(\boldsymbol{\omega})(n(q-1)-1)+2-\sum_{i=1}^{\psi(\boldsymbol{\omega})}\binom{s_{i}-1}{2},
\end{equation}

\subsection{Preliminaries and Existing Results}
Two pairs of sequences with the same Levenshtein distance, could have different intersection size of deletion balls. For example if $\emph{\textbf{x}}=0101$, $\emph{\textbf{y}}=1010$ and \emph{\textbf{x}}$^{'}=0111$, \emph{\textbf{y}}$^{'}=1110$ then $d_{L}($\emph{\textbf{x}},\emph{\textbf{y}}$)=d_{L}($\emph{\textbf{x}}$^{'}$,\emph{\textbf{y}}$^{'})=2$, but $|\mathcal{D}_{1}(\emph{\textbf{x}})\cap \mathcal{D}_{1}(\emph{\textbf{y}})|=2$ while $|\mathcal{D}_{1}($\emph{\textbf{x}}$^{'})\cap \mathcal{D}_{1}($\emph{\textbf{y}}$^{'})|=1$. In order to characterize the size of intersection of single error balls, the notion of Type-A-confusable sequences was first proposed in [14] was and extended to Type-B-confusable sequences in [15] for binary sequences. Our definition is similar to [16] for general sequences, and is based on length of different segments of the two sequences.
\begin{Definition}
  Two sequences $\emph{\textbf{x}},\emph{\textbf{y}}\in \Sigma^{n}_{q}$ are said to be Type-A-confusable if $\emph{\textbf{x}}=$\emph{\textbf{uvw}}, $\emph{\textbf{y}}=$\emph{\textbf{u}\textbf{v}$^{'}$\textbf{w}} and  \textbf{\emph{v$_{1}$}}$\neq$\textbf{\emph{v$_{1}^{'}$}},  \textbf{\emph{v$_{|\emph{\textbf{v}}|}$}}$\neq$\textbf{\emph{v$^{'}_{|\emph{\textbf{v}}|}$}} and both  \emph{\textbf{v}}$_{[2,|\emph{\textbf{v}}|]}=$ \emph{\textbf{v}}$^{'}_{[1,|\emph{\textbf{v}}|-1]}$ and  \emph{\textbf{v}}$_{[1,|\emph{\textbf{v}}|-1]}=$ \emph{\textbf{v}}$^{'}_{[2,|\emph{\textbf{v}}|]}$ are true.
\end{Definition}
\begin{Definition}
  Two sequences $\emph{\textbf{x}},\emph{\textbf{y}}\in \Sigma^{n}_{q}$ are said to be Type-B-confusable if $\emph{\textbf{x}}=$\emph{\textbf{uvw}}, $\emph{\textbf{y}}=$\emph{\textbf{u}\textbf{v}$^{'}$\textbf{w}} and  \textbf{\emph{v$_{1}$}}$\neq$\textbf{\emph{v$_{1}^{'}$}},  \textbf{\emph{v$_{|\emph{\textbf{v}}|}$}}$\neq$\textbf{\emph{v$^{'}_{|\emph{\textbf{v}}|}$}} and exactly one of \emph{\textbf{v}}$_{[2,|\emph{\textbf{v}}|]}=$ \emph{\textbf{v}}$^{'}_{[1,|\emph{\textbf{v}}|-1]}$ and  \emph{\textbf{v}}$_{[1,|\emph{\textbf{v}}|-1]}=$ \emph{\textbf{v}}$^{'}_{[2,|\emph{\textbf{v}}|]}$ is true.
\end{Definition}
\begin{Lemma}
  Let $\emph{\textbf{x}},\emph{\textbf{y}}\in \Sigma^{n}_{q}$ be two distinct sequences and $\alpha, \beta,\gamma, \delta \in \Sigma_{q}$ such that $\alpha\neq\beta$ and $\gamma\neq\delta$.
  
     (i) $\emph{\textbf{x}}, \emph{\textbf{y}}$ are Type-A-confusable if and only if either \emph{\textbf{x}}$=$\textbf{\emph{u}}$(\alpha\beta)^{m}\textbf{\emph{w}}$ and \emph{\textbf{y}}=\textbf{\emph{u}}$(\beta\alpha)^{m}\textbf{\emph{w}}$ or \emph{\textbf{x}}$=$\textbf{\emph{u}}$(\alpha\beta)^{m}\alpha\textbf{\emph{w}}$ and \emph{\textbf{y}}=\textbf{\emph{u}}$(\beta\alpha)^{m}\beta\textbf{\emph{w}}$.
     
      (ii) $\emph{\textbf{x}}, \emph{\textbf{y}}$ are Type-B-confusable if and only if either  \emph{\textbf{x}}=\textbf{\emph{u}}$\alpha\beta$\textbf{\emph{w}} and \emph{\textbf{y}}=\textbf{\emph{u}}$\beta\delta$\textbf{\emph{w}} where $\alpha, \beta\neq\delta$ or \emph{\textbf{x}}=\textbf{\emph{u}}$\alpha\beta$\textbf{\emph{z}}$\gamma$\textbf{\emph{w}} and \emph{\textbf{y}}=\textbf{\emph{u}}$\beta$\textbf{\emph{z}}$\gamma\delta$\textbf{\emph{w}} for some subsequences \textbf{\emph{u}}, \textbf{\emph{w}}, \textbf{\emph{z}}$\in \Sigma_{q}^{*}$.
  
\end{Lemma}
\begin{IEEEproof}
 The proof is simple and based on induction on the length of distinct segments of the two sequences.

    (i) If \emph{\textbf{x}}$=$\textbf{\emph{u}}$(\alpha\beta)^{m}\textbf{\emph{w}}$ and \emph{\textbf{y}}=\textbf{\emph{u}}$(\beta\alpha)^{m}\textbf{\emph{w}}$ or \emph{\textbf{x}}$=$\textbf{\emph{u}}$(\alpha\beta)^{m}\alpha\textbf{\emph{w}}$ and \emph{\textbf{y}}=\textbf{\emph{u}}$(\beta\alpha)^{m}\beta\textbf{\emph{w}}$ then clearly both conditions \emph{\textbf{v}}$_{[2,|\emph{\textbf{v}}|]}=$ \emph{\textbf{v}}$^{'}_{[1,|\emph{\textbf{v}}|-1]}$ and  \emph{\textbf{v}}$_{[1,|\emph{\textbf{v}}|-1]}=$ \emph{\textbf{v}}$^{'}_{[2,|\emph{\textbf{v}}|]}$ are satisfied.
      
   Conversely, suppose that $\emph{\textbf{x}}$ and $\emph{\textbf{y}}$ be Type-A-confusable. If  $|\emph{\textbf{v}}|=2$, then clearly, \emph{\textbf{x}}$=$\textbf{\emph{u}}$\alpha\beta$\textbf{\emph{w}} and \emph{\textbf{y}}$=$\textbf{\emph{u}}$\beta\alpha$\textbf{\emph{w}}. Now assume that  $|\emph{\textbf{v}}|=2m+1$ and the claim is true for all $2\leq |\emph{\textbf{v}}|\leq 2m$. Based on the definition, \emph{\textbf{v$_{2m+1}$}}$=$\emph{\textbf{v$^{'}_{2m}$}}$=\alpha$ and \emph{\textbf{v$^{'}_{2m+1}$}}$=$\emph{\textbf{v$_{2m}$}}$=\beta$. Therefore \emph{\textbf{v}}$=(\alpha\beta)^{m}\alpha$ and  \emph{\textbf{v$^{'}$}}$=(\beta\alpha)^{m}\beta$. The proof is similar when $|\emph{\textbf{v}}|=2m$ .
     
   (ii) If  \emph{\textbf{x}}=\textbf{\emph{u}}$\alpha\beta$\textbf{\emph{w}} and \emph{\textbf{y}}=\textbf{\emph{u}}$\beta\delta$\textbf{\emph{w}} then $ \textbf{\emph{v}}_{2}=\textbf{\emph{v}}^{'}_{1}$ but $ \textbf{\emph{v}}_{1}\neq\textbf{\emph{v}}^{'}_{2}$. If \emph{\textbf{x}}=\textbf{\emph{u}}$\alpha\beta$\textbf{\emph{z}}$\gamma$\textbf{\emph{w}} and \emph{\textbf{y}}=\textbf{\emph{u}}$\beta$\textbf{\emph{z}}$\gamma\delta$\textbf{\emph{w}} then \emph{\textbf{v}}$_{[2,|\emph{\textbf{v}}|]}=$ \emph{\textbf{v}}$^{'}_{[1,|\emph{\textbf{v}}|-1]}$ but $\emph{\textbf{v}}^{'}_{[2,|\emph{\textbf{v}}|]}\neq \emph{\textbf{v}}_{[1,|\emph{\textbf{v}}|-1]}$.
      
    Conversely, suppose that $\emph{\textbf{x}}$ and $\emph{\textbf{y}}$ be Type-B-confusable. If  $|\emph{\textbf{v}}|=2$, assume that \textbf{\emph{v}}$=\alpha\gamma$ and \textbf{\emph{v$^{'}$}}$=\beta\delta$. Without loss of generality, assume that  \textbf{\emph{v$_{2}$}}$=$\textbf{\emph{v$^{'}_{1}$}} or $\gamma=\beta$. Therefore \textbf{\emph{v}}$=\alpha\beta$ and \textbf{\emph{v$^{'}$}}$=\beta\delta$ where $\alpha\neq\delta$. Now assume that $|\emph{\textbf{v}}|\geq 3$ and \textbf{\emph{v$_{1}$}}$=\alpha$, \textbf{\emph{v}}$_{|\textbf{\emph{v}}|}=\gamma$, \textbf{\emph{v$^{'}_{1}$}}$=\beta$, \textbf{\emph{v$^{'}_{|\textbf{\emph{v}}|}$}}$=\delta$. Without loss of generality, assume that \emph{\textbf{v}}$_{[2,|\emph{\textbf{v}}|]}=$ \emph{\textbf{v}}$^{'}_{[1,|\emph{\textbf{v}}|-1]}$. Therefore \textbf{\emph{v$_{2}$}}$=\beta$ and \textbf{\emph{v$^{'}_{|\textbf{\emph{v}}|-1}$}}$=\gamma$. Now if we define \emph{\textbf{z}}$=$\emph{\textbf{v}}$_{[3,|\emph{\textbf{v}}|-1]}=$ \emph{\textbf{v}}$^{'}_{[2,|\emph{\textbf{v}}|-2]}$ the result follows.

\end{IEEEproof}
\begin{Theorem}
(\emph{Cai et al. [14] and Chrisnata et al. [15]}) Let $\textbf{\emph{x}},\textbf{\emph{y}}\in \Sigma^{n}_{q}$ be two distinct sequences.
\begin{itemize}
  \item If $d_{H}(\textbf{\emph{x}},\textbf{\emph{y}})=1$ then $|\mathcal{D}_{1}(\textbf{\emph{x}})\cap \mathcal{D}_{1}(\textbf{\emph{y}})|=1$ and $|\mathcal{I}_{1}(\textbf{\emph{x}})\cap \mathcal{I}_{1}(\textbf{\emph{y}})|=2$.
  \item If $d_{H}(\textbf{\emph{x}},\textbf{\emph{y}})\geq 2$ then $|\mathcal{B}(\textbf{\emph{x}})\cap \mathcal{B}(\textbf{\emph{y}})|=2$ if and only if $\textbf{\emph{x}}$ and $\textbf{\emph{y}}$ are Type-A-confusable,$|\mathcal{B}(\textbf{\emph{x}})\cap \mathcal{B}(\textbf{\emph{y}})|=1$ if and only if $\textbf{\emph{x}}$ and $\textbf{\emph{y}}$ are Type-B-confusable. Otherwise  $|\mathcal{B}(\textbf{\emph{x}})\cap \mathcal{B}(\textbf{\emph{y}})|=0$, where $\mathcal{B}\in\{\mathcal{D}_{1},\mathcal{I}_{1}\}$.
\end{itemize}
\end{Theorem}
\begin{IEEEproof} 
The result for sequences with Hamming distance one and Type-A-confusable sequences was proved by \emph{Cai et al.} in [14-Proposition 9]. In [15] \emph{Chrisnata et al.} proposed the concept of Type-B-confusability for binary sequences and proved the result for deletion spheres in Lemma 4. For general non-binary Type-B-confusable sequences, the proof is presented in Appendix A.
\end{IEEEproof}
\begin{Corollary}
Let $\boldsymbol{\omega}\in \Sigma^{n}_{q}$.
 Then $d_{H}(\boldsymbol{\omega}^{(i)}, \boldsymbol{\omega}^{(j)})=|j-i|$. Let $j-i\geq2$. Then $\boldsymbol{\omega}^{(i)}$ and $\boldsymbol{\omega}^{(j)}$ are Type-A-confusable if and only if $\ell_{i+1}=\ldots=\ell_{j-1}=1$ and $\sigma_{k}=\sigma_{k+2}, i\leq k \leq j-2$. Otherwise they are Type-B-confusable.
\end{Corollary}
\begin{IEEEproof} 
The first claim was proved for binary sequences in [8-Lemma 5] and the proof for non-binary sequences is exactly the same. We prove the second claim. Let $\boldsymbol{\omega}=\sigma_{1}^{\ell_{1}}\sigma_{2}^{\ell_{2}}\ldots\sigma_{\rho(\boldsymbol{\omega})}^{\ell_{\rho(\boldsymbol{\omega})}}$. Thus $\boldsymbol{\omega}^{(i)}=\emph{\textbf{uvw}}$ and $\boldsymbol{\omega}^{(j)}=$\emph{\textbf{u}\textbf{v}$^{'}$\textbf{w}} where $\emph{\textbf{u}}=\sigma_{1}^{\ell_{1}}\ldots\sigma_{i}^{\ell_{i}-1}$, $\emph{\textbf{v}}=\sigma_{i+1}^{\ell_{i+1}}\ldots\sigma_{j-1}^{\ell_{j-1}}\sigma_{j}$, \emph{\textbf{v}}$^{'}=\sigma_{i}\sigma_{i+1}^{\ell_{i+1}}\ldots\sigma_{j-1}^{\ell_{j-1}}$ and $\emph{\textbf{w}}=\sigma_{j}^{\ell_{j}-1}\ldots\sigma_{\rho(\boldsymbol{\omega})}^{\ell_{\rho(\boldsymbol{\omega})}}$. Based on Lemma 1, $\boldsymbol{\omega}^{(i)}$ and $\boldsymbol{\omega}^{(j)}$ are Type-A-confusable if and only if  $\ell_{i+1}=\ldots=\ell_{j-1}=1$ and $\sigma_{k}=\sigma_{k+2}, i\leq k \leq j-2$. If they are not Type-A-confusable then they must be Type-B-confusable since $\boldsymbol{\omega}\in \{\mathcal{I}_{1}(\boldsymbol{\omega}^{(i)})\cap\mathcal{I}_{1}(\boldsymbol{\omega}^{(j)})\}$.
\end{IEEEproof}
We use the inclusion-exclusion principle to derive the size of error ball for channels with single deletion and other types of error as well. Let $\boldsymbol{\omega}\in \Sigma^{n}_{q}$. In order to generate the ball $\mathcal{B}_{t_{1}, 1, t_{3}}(\boldsymbol{\omega})$ we can assume that first a deletion occurs in sequence $\boldsymbol{\omega}$, followed by at most $t_{1}$ substitutions and exactly $t_{3}$ insertions.
 \[ |\mathcal{B}_{t_{1}, 1, t_{3}}(\boldsymbol{\omega})|=|\bigcup_{i=1}^{\rho(\boldsymbol{\omega})} \mathcal{B}_{t_{1}, 0, t_{3}}(\boldsymbol{\omega}^{(i)})|\]
  \begin{equation}
  =\sum_{\emptyset\neq J\subseteq\{1,\ldots, \rho(\boldsymbol{\omega})\}}(-1)^{|J|+1}|\bigcap_{j\in J}\mathcal{B}_{t_{1}, 0, t_{3}}(\boldsymbol{\omega}^{(j)})|.
  \end{equation}
In order to derive the size of error ball for single-deletion multiple-insertion channel, we need to characterize the size of the intersection of two insertion error balls. In [5] and [6], Levenshtein derived the maximum size of intersection of two insertion spheres.
\begin{align}
N^{+}(n, q, t)&\triangleq \max_{\substack{\emph{\textbf{x}},\emph{\textbf{y}}\in \Sigma^{n}_{q} \\ \emph{\textbf{x}}\neq \emph{\textbf{y}}}}|\mathcal{I}_{t}(\emph{\textbf{x}})\cap \mathcal{I}_{t}(\emph{\textbf{y}})|\nonumber\\
&=\sum_{i=0}^{t-1}\binom{n+t}{i}(1-(-1)^{t-i})(q-1)^{i}.
\end{align}
This result was generalized in [9] for any two sequences with a given minimum Levenshtein distance. For integers $n, t\geq \ell\geq 0$, define
\[N^{+}(n, q, t, \ell)\triangleq \max_{\substack{\emph{\textbf{x}},\emph{\textbf{y}}\in \Sigma^{n}_{q} \\ d_{L}(\emph{\textbf{x}},\emph{\textbf{y}})\geq 2\ell}}|\mathcal{I}_{t}(\emph{\textbf{x}})\cap \mathcal{I}_{t}(\emph{\textbf{y}})|\]
The explicit formula is

$N^{+}(n, q, t, \ell)=$
\begin{equation}
\sum_{j=\ell}^{t}\sum_{i=0}^{t-j}\binom{2j}{j}\binom{t+j-i}{2j}\binom{n+t}{i}(-1)^{t+j-i}(q-1)^{i}.
\end{equation}
 We use these equations to characterize the intersection of two insertion balls.
\section{ The Single-Deletion Double-Insertion Ball Size}
 We derive the size of the error ball resulting from a single deletion and double insertions, i.e., the size of $\mathcal{B}_{0, 1, 2}(\boldsymbol{\omega})$ for all $\boldsymbol{\omega}\in \Sigma_{q}^{n}$.
\begin{Definition}
Let $n\geq2$ and $\boldsymbol{\omega}\in \Sigma^{n}_{q}$ be a sequence with $\rho(\boldsymbol{\omega})$ runs. Define $A(\boldsymbol{\omega})$ as the number of pairs of 1-subsequences like $\boldsymbol{\omega}^{(i)}$ and $\boldsymbol{\omega}^{(j)}$ such that $|\mathcal{I}_{2}(\boldsymbol{\omega}^{(i)})\cap \mathcal{I}_{2}(\boldsymbol{\omega}^{(j)})|=2|\mathcal{I}_{1}(\boldsymbol{\omega})|-2$; $B(\boldsymbol{\omega})$ as the number of pairs of 1-subsequences such that $|\mathcal{I}_{2}(\boldsymbol{\omega}^{(i)})\cap \mathcal{I}_{2}(\boldsymbol{\omega}^{(j)})|=|\mathcal{I}_{1}(\boldsymbol{\omega})|+2$; $C(\boldsymbol{\omega})$ as the number of pairs of 1-subsequences such that $|\mathcal{I}_{2}(\boldsymbol{\omega}^{(i)})\cap \mathcal{I}_{2}(\boldsymbol{\omega}^{(j)})|=|\mathcal{I}_{1}(\boldsymbol{\omega})|+1$ ; and $D(\boldsymbol{\omega})$ as the number of triples of 1-subsequences such that $|\mathcal{I}_{2}(\boldsymbol{\omega}^{(i)})\cap \mathcal{I}_{2}(\boldsymbol{\omega}^{(j)})\cap \mathcal{I}_{2}(\boldsymbol{\omega}^{(k)})|=|\mathcal{I}_{1}(\boldsymbol{\omega})|+1$.
\end{Definition}
\begin{Theorem}
  Let $\boldsymbol{\omega}\in \Sigma^{n}_{q}$ be a sequence with $\rho(\boldsymbol{\omega})$ runs and maximal alternating segments profile vector $[s_{1}, s_{2} \ldots, s_{\psi(\boldsymbol{\omega})}]$. Then the size of the single-deletion double-insertion ball is
  
 \[ |\mathcal{B}_{0, 1, 2}(\boldsymbol{\omega})|=\rho(\boldsymbol{\omega})\binom{n+1}{2}(q-1)^{2}\]
 \[-(A(\boldsymbol{\omega})-1)(1+(n+1)(q-1)) +2A(\boldsymbol{\omega})\]
 \[-2B(\boldsymbol{\omega})-C(\boldsymbol{\omega})+D(\boldsymbol{\omega}).\]
  \end{Theorem} 
\begin{Remark}
 For $\boldsymbol{\omega}\in \Sigma_{q}^{n}$ the expressions for $A(\boldsymbol{\omega})$ and $D(\boldsymbol{\omega})$ are given in Lemma 7 and Lemma 9  respectively. For $\boldsymbol{\omega}\in \Sigma_{2}^{n}$ the expressions for $B(\boldsymbol{\omega})$ and $C(\boldsymbol{\omega})$ are given in Lemma 8.
\end{Remark}
\begin{IEEEproof}
 We use the inclusion-exclusion principle. For each 1-subsequence of $\boldsymbol{\omega}$, we derive the insertion 2-sphere and then calculate the size of the union of these spheres. Based on Lemmas 2 and 3 (will be presented later in this section), the size of the intersection of two insertion 2-spheres of two 1-subsequence of $\boldsymbol{\omega}$ accepts four values: $2|\mathcal{I}_{1}(\boldsymbol{\omega})|-2, 2+|\mathcal{I}_{1}(\boldsymbol{\omega})|, 1+|\mathcal{I}_{1}(\boldsymbol{\omega})|,|\mathcal{I}_{1}(\boldsymbol{\omega})|$. By Lemma 5, the size of the intersection of three insertion 2-sphere of three 1-subsequence of $\boldsymbol{\omega}$ accepts two values $1+|\mathcal{I}_{1}(\boldsymbol{\omega})|,|\mathcal{I}_{1}(\boldsymbol{\omega})|$. By Lemma 6, the size of the intersection of four or more 1-subsequence is equal to $|\mathcal{I}_{1}(\boldsymbol{\omega})|$. Therefore,

$|\mathcal{B}_{0, 1, 2}(\boldsymbol{\omega})|=\rho(\boldsymbol{\omega})(1+(n+1)(q-1) 
+\binom{n+1}{2}(q-1)^{2})-(2|\mathcal{I}_{1}(\boldsymbol{\omega})|-2)A(\boldsymbol{\omega})
-(2+|\mathcal{I}_{1}(\boldsymbol{\omega})|)B(\boldsymbol{\omega})-(1+|\mathcal{I}_{1}(\boldsymbol{\omega})|)C(\boldsymbol{\omega})
-(\binom{\rho(\boldsymbol{\omega})}{2}-A(\boldsymbol{\omega})-B(\boldsymbol{\omega})-C(\boldsymbol{\omega}))|\mathcal{I}_{1}(\boldsymbol{\omega})|
+(1+|\mathcal{I}_{1}(\boldsymbol{\omega})|)D(\boldsymbol{\omega})+|\mathcal{I}_{1}(\boldsymbol{\omega})|(\binom{\rho(\boldsymbol{\omega})}{3}-D(\boldsymbol{\omega}))
+\sum_{i=4}^{\rho(\boldsymbol{\omega})}(-1)^{i-1}\binom{\rho(\boldsymbol{\omega})}{i}|\mathcal{I}_{1}(\boldsymbol{\omega})|=\rho(\boldsymbol{\omega})\binom{n+1}{2}(q-1)^{2}
-(A(\boldsymbol{\omega})-1)(1+(n+1)(q-1))+2A(\boldsymbol{\omega})-2B(\boldsymbol{\omega})
-C(\boldsymbol{\omega})+D(\boldsymbol{\omega}).$

The last equality is derived by use of binomial expansion.
\end{IEEEproof}
  First, we need to characterize the intersection of insertion 2-spheres, that is $|\mathcal{I}_{2}(\emph{\textbf{x}})\cap \mathcal{I}_{2}(\emph{\textbf{y}})|$ with the help of $|\mathcal{I}_{1}(\emph{\textbf{x}})\cap \mathcal{I}_{1}(\emph{\textbf{y}})|$. This has been obtained in [17] for binary sequences. 
\begin{Lemma} Let $n\geq 3$ and $\emph{\textbf{x}},\emph{\textbf{y}}\in \Sigma^{n}_{q}$. If $|\mathcal{I}_{1}(\emph{\textbf{x}})\cap \mathcal{I}_{1}(\emph{\textbf{y}})|=1$, then $1+(n+2)(q-1)\leq |\mathcal{I}_{2}(\emph{\textbf{x}})\cap \mathcal{I}_{2}(\emph{\textbf{y}})|\leq 3+(n+2)(q-1)$. In fact, when  \emph{\textbf{x}}$=$\textbf{\emph{u}}$\alpha\beta$\textbf{\emph{w}} and \emph{\textbf{y}}$=$\textbf{\emph{u}}$\beta\gamma$\textbf{\emph{w}}, then $|\mathcal{I}_{2}(\emph{\textbf{x}})\cap \mathcal{I}_{2}(\emph{\textbf{y}})|=3+(n+2)(q-1)$; and when \emph{x}$=$\textbf{\emph{u}}$\alpha\beta$\textbf{\emph{v}}$\gamma$\textbf{\emph{w}} and \emph{\textbf{y}}$=$\textbf{\emph{u}}$\beta$\textbf{\emph{v}}$\gamma\delta$\textbf{\emph{w}}, then we have
\begin{itemize}
  \item $|\mathcal{I}_{2}(\emph{\textbf{x}})\cap \mathcal{I}_{2}(\emph{\textbf{y}})|=1+(n+2)(q-1)$ if and only if $d_{H}(\alpha\beta$\textbf{\emph{v}},\textbf{\emph{v}}$\gamma\delta)\geq2$, and $\alpha\beta$\textbf{\emph{v}} and \textbf{\emph{v}}$\gamma\delta$ are neither Type-A-confusable, nor Type-B-confusable. 
  \item $|\mathcal{I}_{2}(\emph{\textbf{x}})\cap \mathcal{I}_{2}(\emph{\textbf{y}})|=2+(n+2)(q-1)$ if and only if $\alpha\beta$\textbf{\emph{v}} and \textbf{\emph{v}}$\gamma\delta$ are Type-B-confusable. 
  \item $|\mathcal{I}_{2}(\emph{\textbf{x}})\cap \mathcal{I}_{2}(\emph{\textbf{y}})|=3+(n+2)(q-1)$ if and only if $\alpha\beta$\textbf{\emph{v}} and \textbf{\emph{v}}$\gamma\delta$ are Type-A-confusable  or  $d_{H}(\alpha\beta$\textbf{\emph{v}},\textbf{\emph{v}}$\gamma\delta)=1$.
\end{itemize}
\end{Lemma}
\begin{IEEEproof}
 We conclude from Theorem 1 that \emph{\textbf{x}} and \emph{\textbf{y}} are Type-B-confusable. Let \emph{\textbf{x}}$=$\textbf{\emph{uvw}} and \emph{\textbf{y}}$=$\textbf{\emph{uv$^{'}$w}}. First, let $|$\textbf{\emph{v}}$|=2$ or equivalently \emph{\textbf{x}}$=$\textbf{\emph{u}}$\alpha\beta$\textbf{\emph{w}} and \emph{\textbf{y}}$=$\textbf{\emph{u}}$\beta\gamma$\textbf{\emph{w}} for some \textbf{\emph{u}},\textbf{\emph{w}}$\in \Sigma^{*}_{q}$. Also let $\mathcal{I}_{1}(\emph{\textbf{x}})\cap \mathcal{I}_{1}(\emph{\textbf{y}})=\{\emph{\textbf{z}}\}$. Then $\textbf{z}=\textbf{\emph{u}}\alpha\beta\gamma\textbf{\emph{w}}$. Let $S=\mathcal{I}_{2}(\emph{\textbf{x}})\cap \mathcal{I}_{2}(\emph{\textbf{y}})$. Clearly $\mathcal{I}_{1}(\emph{\textbf{z}})\subseteq S$. Thus, $S=\mathcal{I}_{1}(\emph{\textbf{z}})\cup J(\emph{\textbf{x}},\emph{\textbf{y}})$, where $J(\emph{\textbf{x}},\emph{\textbf{y}})=S\setminus\mathcal{I}_{1}(\emph{\textbf{z}})$. First, note that $\mathcal{I}_{2}(\emph{\textbf{x}})=\bigcup_{a+b+c=2}\mathcal{I}_{a}(\textbf{\emph{u}})\circ\mathcal{I}_{b}(\alpha\beta)\circ\mathcal{I}_{c}(\textbf{\emph{w}})$ and $\mathcal{I}_{2}(\emph{\textbf{y}})=\bigcup_{a+b+c=2}\mathcal{I}_{a}(\textbf{\emph{u}})\circ\mathcal{I}_{b}(\beta\gamma)\circ\mathcal{I}_{c}(\textbf{\emph{w}})$. Thus, $\mathcal{I}_{2}(\emph{\textbf{x}})\cap \mathcal{I}_{2}(\emph{\textbf{y}})=(\mathcal{I}_{1}(\textbf{\emph{u}})\circ\alpha\beta\gamma\textbf{\emph{w}})\cup (\textbf{\emph{u}}\alpha\beta\gamma\circ\mathcal{I}_{1}(\textbf{\emph{w}}))\cup (\textbf{\emph{u}}\circ(\mathcal{I}_{2}(\alpha\beta)\cap\mathcal{I}_{2}(\beta\gamma))\circ\textbf{\emph{w}})$. By deriving the two sets $\mathcal{I}_{2}(\alpha\beta)$, $\mathcal{I}_{2}(\beta\gamma)$ and their intersection we have $\mathcal{I}_{2}(\alpha\beta)\cap\mathcal{I}_{2}(\beta\gamma)=\mathcal{I}_{1}(\alpha\beta\gamma)\cup \{\beta\alpha\gamma\beta,\beta\gamma\alpha\beta\}$. Hence,  $\mathcal{I}_{2}(\emph{\textbf{x}})\cap \mathcal{I}_{2}(\emph{\textbf{y}})=\mathcal{I}_{1}(\textbf{\emph{u}}\alpha\beta\gamma\textbf{\emph{w}})\cup \{\textbf{\emph{u}}\beta\alpha\gamma\beta\textbf{\emph{w}},\textbf{\emph{u}}\beta\gamma\alpha\beta\textbf{\emph{w}}\}=\mathcal{I}_{1}(\emph{\textbf{z}})\cup J(\emph{\textbf{x}},\emph{\textbf{y}})$.
 
  Now, suppose that $|$\textbf{\emph{v}}$|\geq 3$. Then \emph{\textbf{x}} and \emph{\textbf{y}} have the following forms \emph{\textbf{x}}$=$\textbf{\emph{u}}$\alpha\beta$\textbf{\emph{v}}$\gamma$\textbf{\emph{w}} and \emph{\textbf{y}}$=$\textbf{\emph{u}}$\beta$\textbf{\emph{v}}$\gamma\delta$\textbf{\emph{w}}. Let $\mathcal{I}_{1}(\emph{\textbf{x}})\cap \mathcal{I}_{1}(\emph{\textbf{y}})=\{\emph{\textbf{z}}\}$. Then $\emph{\textbf{z}}=\textbf{\emph{u}}\alpha\beta\textbf{\emph{v}}\gamma\delta\textbf{\emph{w}}$ and $S=\mathcal{I}_{2}(\emph{\textbf{x}})\cap \mathcal{I}_{2}(\emph{\textbf{y}})$.  Clearly $\mathcal{I}_{1}(\emph{\textbf{z}})\subseteq S$. Suppose that $S=\mathcal{I}_{1}(\emph{\textbf{z}})\cup J(\emph{\textbf{x}},\emph{\textbf{y}})$ where $J(\emph{\textbf{x}},\emph{\textbf{y}})=S\setminus \mathcal{I}_{1}(\emph{\textbf{z}})$. Note that $\mathcal{I}_{2}(\emph{\textbf{x}})=\bigcup_{a+b+c=2}\mathcal{I}_{a}(\textbf{\emph{u}})\circ\mathcal{I}_{b}(\alpha\beta\textbf{\emph{v}}\gamma)\circ\mathcal{I}_{c}(\textbf{\emph{w}})$ and  $\mathcal{I}_{2}(\emph{\textbf{y}})=\bigcup_{a+b+c=2}\mathcal{I}_{a}(\textbf{\emph{u}})\circ\mathcal{I}_{b}(\beta\textbf{\emph{v}}\gamma\delta)\circ\mathcal{I}_{c}(\textbf{\emph{w}})$.
  Thus, $\mathcal{I}_{2}(\emph{\textbf{x}})\cap\mathcal{I}_{2}(\emph{\textbf{y}})=(\mathcal{I}_{1}(\textbf{\emph{u}})\circ\alpha\beta\textbf{\emph{v}}\gamma\delta\textbf{\emph{w}})\cup
  (\textbf{\emph{u}}\alpha\beta\textbf{\emph{v}}\gamma\delta\circ\mathcal{I}_{1}(\emph{\textbf{w}}))\cup (\textbf{\emph{u}}\circ(\mathcal{I}_{2}(\alpha\beta\textbf{\emph{v}}\gamma)\cap\mathcal{I}_{2}(\beta\textbf{\emph{v}}\gamma\delta))\circ\emph{\textbf{w}})$. Clearly, $\mathcal{I}_{1}(\alpha\beta\textbf{\emph{v}}\gamma\delta)\subseteq \mathcal{I}_{2}(\alpha\beta\textbf{\emph{v}}\gamma)\cap\mathcal{I}_{2}(\beta\textbf{\emph{v}}\gamma\delta)$. If we define $\mathcal{I}_{2}(\alpha\beta\textbf{\emph{v}}\gamma)\cap\mathcal{I}_{2}(\beta\textbf{\emph{v}}\gamma\delta)=\mathcal{I}_{1}(\alpha\beta\textbf{\emph{v}}\gamma\delta)\cup J(\alpha\beta\textbf{\emph{v}}\gamma,\beta\textbf{\emph{v}}\gamma\delta)$ such that $\mathcal{I}_{1}(\alpha\beta\textbf{\emph{v}}\gamma\delta)\cap J(\alpha\beta\textbf{\emph{v}}\gamma,\beta\textbf{\emph{v}}\gamma\delta)=\emptyset$ then $J(\emph{\textbf{x}},\emph{\textbf{y}})=\textbf{\emph{u}}\circ J(\alpha\beta\textbf{\emph{v}}\gamma,\beta\textbf{\emph{v}}\gamma\delta)\circ\textbf{\emph{w}}$.
  
  In order to characterize the subset $J(\emph{\textbf{x}},\emph{\textbf{y}})$, we note that the symbol after \textbf{\emph{u}} and the symbol before \textbf{\emph{w}} cannot be $\alpha$ and $\delta$ respectively. Otherwise, the resulting sequence have the forms $\textbf{\emph{u}}\alpha\circ(\mathcal{I}_{2}(\beta\textbf{\emph{v}}\gamma)\cap \mathcal{I}_{1}(\beta\textbf{\emph{v}}\gamma\delta))\circ\textbf{\emph{w}}=\textbf{\emph{u}}\alpha \circ\mathcal{I}_{1}(\beta\textbf{\emph{v}}\gamma\delta)\circ\textbf{\emph{w}}$ and  $\textbf{\emph{u}}\circ(\mathcal{I}_{1}(\alpha\beta\textbf{\emph{v}}\gamma)\cap \mathcal{I}_{2}(\beta\textbf{\emph{v}}\gamma))\circ\delta\textbf{\emph{w}}=\textbf{\emph{u}}\circ\mathcal{I}_{1}(\alpha\beta\textbf{\emph{v}}\gamma)\circ\delta\textbf{\emph{w}}$ respectively which make them an element in the set $\mathcal{I}_{1}(\emph{\textbf{z}})$. Suppose elements of $J(\emph{\textbf{x}},\emph{\textbf{y}})$ have the form  $\textbf{\emph{u}}(\tilde{\alpha}\circ\mathcal{I}_{1}(\alpha\beta\textbf{\emph{v}}\gamma)\cap\mathcal{I}_{1}(\beta\textbf{\emph{v}}\gamma\delta)\circ\tilde{\delta})$\textbf{\emph{w}}. In order to this sequence be an element of $\mathcal{I}_{2}(\emph{\textbf{x}})\cap\mathcal{I}_{2}(\emph{\textbf{y}})$, we should have $\tilde{\alpha}=\beta$ and  $\tilde{\delta}=\gamma$. Therefore, $J(\emph{\textbf{x}},\emph{\textbf{y}})$=\textbf{\emph{u}}$\beta\circ(\mathcal{I}_{1}(\alpha\beta\textbf{\emph{v}})\cap\mathcal{I}_{1}(\textbf{\emph{v}}\gamma\delta))\circ\gamma$\textbf{\emph{w}}. Finally, $|S|=|\mathcal{I}_{1}(\textbf{\emph{z}})|+|\mathcal{I}_{1}(\alpha\beta\textbf{\emph{v}})\cap\mathcal{I}_{1}(\textbf{\emph{v}}\gamma\delta)|$. The statement of Lemma is concluded from Theorem 1.
  \end{IEEEproof}   
  The next Lemma shows the connection between the size of the intersection of insertion 1-spheres with the size of the intersection of insertion 2-spheres. This Lemma is a generalization of [17, Lemma 12]. 
\begin{Lemma}
  Let  $n\geq 4$ and $\emph{\textbf{x}},\emph{\textbf{y}}\in \Sigma^{n}_{q}$. We have
 \begin{description}
    \item[(i)]  $|\mathcal{I}_{1}(\emph{\textbf{x}})\cap \mathcal{I}_{1}(\emph{\textbf{y}})|=0$ if and only if $|\mathcal{I}_{2}(\emph{\textbf{x}})\cap \mathcal{I}_{2}(\emph{\textbf{y}})|\leq 6$.
    \item[(ii)] $|\mathcal{I}_{1}(\emph{\textbf{x}})\cap \mathcal{I}_{1}(\emph{\textbf{y}})|=1$ if and only if $1+(n+2)(q-1)\leq |\mathcal{I}_{2}(\emph{\textbf{x}})\cap \mathcal{I}_{2}(\emph{\textbf{y}})|\leq 3+(n+2)(q-1)$.
    \item[(iii)] $|\mathcal{I}_{1}(\emph{\textbf{x}})\cap \mathcal{I}_{1}(\emph{\textbf{y}})|=2$ if and only if $|\mathcal{I}_{2}(\emph{\textbf{x}})\cap \mathcal{I}_{2}(\emph{\textbf{y}})|=2(n+2)(q-1)$. 
 \end{description}
\end{Lemma}
\begin{IEEEproof}
  First we prove the necessity. (i) Suppose that $|\mathcal{I}_{1}(\emph{\textbf{x}})\cap \mathcal{I}_{1}(\emph{\textbf{y}})|=0$ or equivalently $d_{L}(\emph{\textbf{x}},\emph{\textbf{y}})\geq 4$. Then we have $|\mathcal{I}_{2}(\emph{\textbf{x}})\cap \mathcal{I}_{2}(\emph{\textbf{y}})|\leq N^{+}(n, q, 2, 2)=\binom{4}{2}=6<1+(n+2)(q-1)$ by equation (6). (ii) If $|\mathcal{I}_{1}(\emph{\textbf{x}})\cap \mathcal{I}_{1}(\emph{\textbf{y}})|=1$, then by Lemma 2 we have $1+(n+2)(q-1)\leq |\mathcal{I}_{2}(\emph{\textbf{x}})\cap \mathcal{I}_{2}(\emph{\textbf{y}})|\leq 3+(n+2)(q-1)$.
  (iii) Suppose that $|\mathcal{I}_{1}(\emph{\textbf{x}})\cap \mathcal{I}_{1}(\emph{\textbf{y}})|=2$ or $\mathcal{I}_{1}($\emph{\textbf{x}})$\cap \mathcal{I}_{1}($\emph{\textbf{y}})$=\{$\emph{\textbf{z}},\emph{\textbf{w}}$\}$. Since \emph{\textbf{z}},\emph{\textbf{w}}$\in\Sigma_{q}^{n+1}$ we have $|\mathcal{I}_{1}(\emph{\textbf{z}})|=|\mathcal{I}_{1}(\emph{\textbf{w}})|=1+(n+2)(q-1)$ and $|\mathcal{I}_{2}($\emph{\textbf{x}}$)\cap \mathcal{I}_{2}($\emph{\textbf{y}}$)|\geq |\mathcal{I}_{1}($\emph{\textbf{z}})$\cup \mathcal{I}_{1}($\emph{\textbf{w}}$)|\geq 2(n+2)(q-1)$ (Note that \emph{\textbf{z}} and \emph{\textbf{w}} are Type-A-confusable). On the other hand, from (5) we have $|\mathcal{I}_{2}($\emph{\textbf{x}}$)\cap \mathcal{I}_{2}($\emph{\textbf{y}}$)|\leq 2(n+2)(q-1)$. We conclude that $|\mathcal{I}_{2}($\emph{\textbf{x}}$)\cap \mathcal{I}_{2}($\emph{\textbf{y}}$)|= 2(n+2)(q-1)$.
 
 Now we prove the sufficiency. If  $|\mathcal{I}_{2}($\emph{\textbf{x}}$)\cap \mathcal{I}_{2}($\emph{\textbf{y}}$)|= 2(n+2)(q-1)$ then \emph{\textbf{x}} and \emph{\textbf{y}} have the largest possible size of the intersection of two insertion 2-spheres. This happens when the two sequence have Hamming distance one or are Type-A-confusable. Therefore according to Theorem 1, $|\mathcal{I}_{1}($\emph{\textbf{x}}$)\cap \mathcal{I}_{1}($\emph{\textbf{y}}$)|= 2$. The claim (ii) is clear from Lemma 2. If  $|\mathcal{I}_{2}(\emph{\textbf{x}})\cap \mathcal{I}_{2}(\emph{\textbf{y}})|\leq 6$ then since $6<1+(n+2)(q-1)$ we must have $|\mathcal{I}_{1}(\emph{\textbf{x}})\cap \mathcal{I}_{1}(\emph{\textbf{y}})|=0$.
 \end{IEEEproof}
 The next Lemma completely characterizes the size of the intersection of two insertion 2-spheres of 1-subsequences which they are Type-B-confusable. For this Lemma we define an indicator notation: $\mathbb{1}[C]$ is equal to one, if the condition $C$ is true. Otherwise, it is zero.
\begin{Lemma}
  Let $n\geq4$ and $\boldsymbol{\omega} \in \Sigma^{n}_{q}$,
    then the size of the intersection of insertion 2-spheres of two 1-subsequences $\boldsymbol{\omega}^{(i)}$ and $\boldsymbol{\omega}^{(j)}$ (where $j-i\geq2$) that are Type-B-confusable is
  \begin{enumerate}
    \item If $j=i+2$ then
     \begin{math}
    |\mathcal{I}_{2}(\boldsymbol{\omega}^{(i)}) \cap \mathcal{I}_{2}(\boldsymbol{\omega}^{(j)})|=\left\{
    \begin{array}{ll}
      3+(n+1)(q-1), & \mbox{if $\ell_{i+1}=1,\sigma_{i}\neq\sigma_{j}$} .\\
      3+(n+1)(q-1), & \mbox{if $\ell_{i+1}=2,\sigma_{i}=\sigma_{j}$} .\\
      2+(n+1)(q-1), & \mbox{otherwise}.
    \end{array}
  \right.
\end{math}
    \item  If $j\geq i+3$ then
        \begin{math}
    |\mathcal{I}_{2}(\boldsymbol{\omega}^{(i)}) \cap \mathcal{I}_{2}(\boldsymbol{\omega}^{(j)})|=\left\{
    \begin{array}{ll}
      2+(n+1)(q-1), & \mbox{if $(C_{1}=1)$}\\
      2+(n+1)(q-1), & \mbox{if $(C_{2}=1) $}\\
      2+(n+1)(q-1), & \mbox{if $(C_{3}=1)$}\\
      3+(n+1)(q-1), & \mbox{if $(C_{4}=1)$}\\
      3+(n+1)(q-1), & \mbox{if $(C_{5}=1)$}\\
      1+(n+1)(q-1), & \mbox{otherwise}.\\
    \end{array}
  \right.
        \end{math}
  \end{enumerate}
where
$C_{1}=\mathbb{1}[\ell_{i+1}=\ldots=\ell_{t-1}=1,\ell_{t+1}=\ell_{t+3}=\ldots=\ell_{s-1}=1,\ell_{t}=\ell_{t+2}=\ldots=\ell_{s}=2,\ell_{s+1}=\ldots=\ell_{j-1}=1,i+1\leq t<t+1<s\leq j-1,\sigma_{k}=\sigma_{k+2},i\leq k\leq j-2]$.

$C_{2}=\mathbb{1}[\ell_{i+1}=\ldots=\ell_{t-1}=1,\ell_{t}\geq3,\ell_{t+1}=\ldots=\ell_{j-1}=1,i+1\leq t \leq j-1,\sigma_{k}=\sigma_{k+2},i\leq k\leq t-2, t\leq k\leq j-2]$.

$ C_{3}=\mathbb{1}[\ell_{i+1}=\ldots=\ell_{t-1}=1,\ell_{t}=2,\ell_{t+1}=\ldots=\ell_{j-1}=1,\sigma_{t-1}\neq\sigma_{t+1},i+1\leq t \leq j-1,\sigma_{k}=\sigma_{k+2},i\leq k\leq t-2,t\leq k\leq j-2]$.

$C_{4}=\mathbb{1}[\ell_{i+1}=\ldots=\ell_{t-1}=1,\ell_{t}=2,\ell_{t+1}=\ldots=\ell_{j-1}=1, i+1\leq t \leq j-1,\sigma_{k}=\sigma_{k+2},i\leq k\leq j-2]$.

$C_{5}=\mathbb{1}[\ell_{i+1}=\ldots=\ell_{j-1}=1,\sigma_{t}\neq\sigma_{t+2},i\leq t\leq j-2,\sigma_{k}=\sigma_{k+2},i\leq k\leq t-1,t+1\leq k\leq j-2]$.

\end{Lemma}
\begin{IEEEproof}
Since $\boldsymbol{\omega}^{(i)}$ and $\boldsymbol{\omega}^{(j)}$ are Type-B-confusable, by definition 5, we can assume that $\boldsymbol{\omega}^{(i)}=$\textbf{\emph{uvw}} and $\boldsymbol{\omega}^{(j)}=$\textbf{\emph{uv$^{'}$w}} for some sequences \textbf{\emph{u}}, \textbf{\emph{v}}, \textbf{\emph{w}} and \textbf{\emph{v$^{'}$}}, where $d_{H}$(\textbf{\emph{v}},\textbf{\emph{v$^{'}$}})$\geq2$. Then according to the definition of $\boldsymbol{\omega}^{(i)}$ and $\boldsymbol{\omega}^{(j)}$, we have  $\emph{\textbf{u}}=\sigma_{1}^{\ell_{1}}\ldots\sigma_{i}^{\ell_{i}-1}$, $\emph{\textbf{w}}=\sigma_{j}^{\ell_{j}-1}\ldots\sigma_{\rho(\boldsymbol{\omega})}^{\ell_{\rho(\boldsymbol{\omega})}}$ and $\emph{\textbf{v}}=\sigma_{i+1}^{\ell_{i+1}}\ldots\sigma_{j-1}^{\ell_{j-1}}\sigma_{j}$, \emph{\textbf{v}}$^{'}=\sigma_{i}\sigma_{i+1}^{\ell_{i+1}}\ldots\sigma_{j-1}^{\ell_{j-1}}$.

By Corollary 1 and the fact that $\boldsymbol{\omega}^{(i)}$ and $\boldsymbol{\omega}^{(j)}$ are Type-B-confusable, we conclude that at least one of the followings holds:
\begin{itemize}
  \item $\ell_{k}\geq2$ for some $i+1\leq k\leq j-1$;
  \item $\sigma_{k}\neq\sigma_{k+2}$ for some $i\leq k\leq j-2$.
\end{itemize}
 First we prove the first claim. Since $j=i+2$, we have $\emph{\textbf{v}}=\sigma_{i+1}^{\ell_{i+1}}\sigma_{i+2}$, \emph{\textbf{v}}$^{'}=\sigma_{i}\sigma_{i+1}^{\ell_{i+1}}$.
 
 If $\ell_{i+1}=1$, we must have $\sigma_{i}\neq\sigma_{j}$. Then the conclusion follows from Lemma 2. When  $\ell_{i+1}\geq2$, we have 
   $|$\textbf{\emph{v}}$|=|\emph{\textbf{v}}^{'}|\geq3$. Since $\boldsymbol{\omega}^{(i)}$ and $\boldsymbol{\omega}^{(j)}$ are Type-B-confusable, by Lemma 1, we can assume that \textbf{\emph{v}}$=\beta$\textbf{\emph{z}}$\gamma\delta$ and $\emph{\textbf{v}}^{'}=\alpha\beta\textbf{\emph{z}}\gamma$ for some $\alpha, \beta, \gamma, \delta\in \Sigma_{q}$ and some sequence \textbf{\emph{z}}, where $\alpha\neq\beta$ and $\gamma\neq\delta$. Then, we have $\alpha\beta$\textbf{\emph{z}}=$\sigma_{i}\sigma_{i+1}^{\ell_{i+1}-1}$ and \textbf{\emph{z}}$\gamma\delta=\sigma_{i+1}^{\ell_{i+1}-1}\sigma_{i+2}$. If $\ell_{i+1}=2$ and $\sigma_{i}=\sigma_{j}$, then $\alpha\beta$\textbf{\emph{z}} and \textbf{\emph{z}}$\gamma\delta$ are Type-A-confusable. If $\ell_{i+1}>2$ or $\sigma_{i}\neq\sigma_{j}$, then $\alpha\beta$\textbf{\emph{z}} and \textbf{\emph{z}}$\gamma\delta$ are Type-B-confusable. Now the conclusion follows from Lemma 2.

Next we prove the second claim. $j\geq i+3$, it follows that  $|$\textbf{\emph{v}}$|=|\emph{\textbf{v}}^{'}|\geq3$.  Since $\boldsymbol{\omega}^{(i)}$ and $\boldsymbol{\omega}^{(j)}$ are Type-B-confusable, by Lemma 1, we can assume that \textbf{\emph{v}}$=\beta$\textbf{\emph{z}}$\gamma\delta$ and $\emph{\textbf{v}}^{'}=\alpha\beta\textbf{\emph{z}}\gamma$ for some $\alpha, \beta, \gamma, \delta\in \Sigma_{q}$ and some sequence \textbf{\emph{z}}, where $\alpha\neq\beta$ and $\gamma\neq\delta$. Then, we have $\alpha\beta$\textbf{\emph{z}}$=\sigma_{i}\sigma_{i+1}^{\ell_{i+1}}\ldots\sigma_{j-2}^{\ell_{j-2}}\sigma_{j-1}^{\ell_{j-1}-1}$ and \textbf{\emph{z}}$\gamma\delta=\sigma_{i+1}^{\ell_{i+1}-1}\sigma_{i+2}^{\ell_{i+2}}\ldots\sigma_{j-1}^{\ell_{j-1}}\sigma_{j}$.
 
Now we investigate conditions when $\alpha\beta$\textbf{\emph{z}} and \textbf{\emph{z}}$\gamma\delta$ are Type-A-confusable or Type-B-confusable or have Hamming distance one. Let $\emph{\textbf{x}}=\alpha\beta\textbf{\emph{z}}=\emph{\textbf{u}}^{'}\emph{\textbf{z}}^{'}\emph{\textbf{w}}^{'}$ and $\emph{\textbf{y}}=\textbf{\emph{z}}\gamma\delta=\emph{\textbf{u}}^{''}\emph{\textbf{z}}^{''}\emph{\textbf{w}}^{''}$ such that $|\textbf{u}^{'}|=|\textbf{u}^{''}|$ and $|\textbf{w}^{'}|=|\textbf{w}^{''}|$ . Based on length of the runs between $i$ and $j$, there are four cases which we consider below.
\begin{enumerate}
  \item At least two runs with indices like $t$ and $s$ exist such that $i+1\leq t<s\leq j-1$ and $\ell_{t}, \ell_{s}\geq 2$. Suppose that $t$ be the smallest and $s$ the largest of such indices. Therefore \emph{\textbf{u}}$^{'}=\sigma_{i}\ldots\sigma_{t-2}$, \emph{\textbf{u}}$^{''}=\sigma_{i+2}\ldots\sigma_{t}$, \emph{\textbf{z}}$^{'}=\sigma_{t-1}\sigma_{t}^{\ell_{t}}\ldots\sigma_{s}^{\ell_{s}-1}$,  \emph{\textbf{z}}$^{''}=\sigma_{t}^{\ell_{t}-1}\ldots\sigma_{s}^{\ell_{s}}\sigma_{s+1}$, \emph{\textbf{w}}$^{'}=\sigma_{s}\ldots\sigma_{j-2}$ and \emph{\textbf{w}}$^{''}=\sigma_{s+2}\ldots\sigma_{j}$. Note that if $t=i+1$ then  \emph{\textbf{u}}$^{'}=\emptyset$ and if $s=j-1$ then \emph{\textbf{w}}$^{'}=\emptyset$ . Since  \emph{\textbf{z}}$^{'}_{1}\neq$ \emph{\textbf{z}}$^{''}_{1}$ and \emph{\textbf{z}}$^{'}_{|\emph{\textbf{z}}^{'}|}\neq$ \emph{\textbf{z}}$^{''}_{|\emph{\textbf{z}}^{''}|}$ we have $d_{H}(\emph{\textbf{x}}, \emph{\textbf{y}})\geq2$. Also \emph{\textbf{z}}$^{'}$ and \emph{\textbf{z}}$^{''}$ are not alternating sequences. So \emph{\textbf{x}} and \emph{\textbf{y}} can not be Type-A-confusable. Thus, they can only be Type-B-confusable. 
Suppose that $\textbf{\emph{u}}^{'}\neq\textbf{\emph{u}}^{''}$.  Let $\ell$ be the smallest index such that $\emph{\textbf{x}}_{\ell}\neq \emph{\textbf{y}}_{\ell+2}$. First suppose that $\ell\leq t-3$.
       We have $\emph{\textbf{x}}_{[t-i-2:t-i-1]}=\sigma_{t-3}\sigma_{t-2}\neq \textbf{\emph{y}}_{[t-i-1:t-i]}=\sigma_{t}\sigma_{t}$. For $\ell=t-2$, $\emph{\textbf{x}}_{t-i-1}=\sigma_{t-2}\neq \textbf{\emph{y}}_{t-i}=\sigma_{t}$.
       On the other hand \emph{\textbf{x}}$_{t-i}=\sigma_{t-1}\neq  \textbf{\emph{y}}_{t-i-1}=\sigma_{t}$. Therefore, based on definition 5, \emph{\textbf{x}} and \emph{\textbf{y}} can not be Type-B-confusable. Next, suppose that \textbf{\emph{w}}$^{'}\neq$\textbf{\emph{w}}$^{''}$. Let $\ell$ be the largest index such that $\emph{\textbf{x}}_{\ell}\neq \emph{\textbf{y}}_{\ell+2}$. First suppose that $\ell\geq |\emph{\textbf{x}}|+s-j+3$.
        We have \emph{\textbf{x}}$_{[|\emph{\textbf{x}}|+s-j+1:|\emph{\textbf{x}}|+s-j+2]}=\sigma_{s}\sigma_{s}\neq \textbf{\emph{y}}_{[|\emph{\textbf{x}}|+s-j+2:|\emph{\textbf{x}}|+s-j+3]}=\sigma_{s+2}\sigma_{s+3}$. If $\ell=|\emph{\textbf{x}}|+s-j+2$ then  $\emph{\textbf{x}}_{|\emph{\textbf{x}}|+s-j+1}=\sigma_{s}\neq \textbf{\emph{y}}_{|\emph{\textbf{x}}|+s-j+2}=\sigma_{s+2}$.
        On the other hand,  \emph{\textbf{x}}$_{|\emph{\textbf{x}}|+s-j+2}=\sigma_{s}\neq \emph{\textbf{y}}_{|\emph{\textbf{x}}|+s-j+1}=\sigma_{s+1}$. Therefore, based on definition 5, \emph{\textbf{x}} and \emph{\textbf{y}} can not be Type-B-confusable.
 Note that,  \emph{\textbf{z}}$^{'}_{1+\ell_{t}}=\sigma_{t}\neq$ \emph{\textbf{z}}$^{''}_{\ell_{t}}=\sigma_{t+1}$. We must check when \emph{\textbf{z}}$^{'}_{[1:|\emph{\textbf{z}}^{'}|-1]}=$ \emph{\textbf{z}}$^{''}_{[2:|\emph{\textbf{z}}^{''}|]}$. If the condition $\sigma_{t-1}\sigma_{t}^{\ell_{t}}\ldots\sigma_{s}^{\ell_{s}-2}=\sigma_{t}^{\ell_{t}-2}\ldots\sigma_{s}^{\ell_{s}}\sigma_{s+1}$ satisfies then $\ell_{t+1}=\ell_{t+3}=\ldots=\ell_{s-1}=1,\ell_{t}=\ell_{t+2}=\ldots=\ell_{s}=2$ and $\sigma_{k}=\sigma_{k+2}, t-1\leq k\leq s-1$. Also if \emph{\textbf{u}}$^{'}=$\emph{\textbf{u}}$^{''}$ then $\sigma_{k}=\sigma_{k+2}, i\leq k\leq t-2$ and if \emph{\textbf{w}}$^{'}=$\emph{\textbf{w}}$^{''}$ then $\sigma_{k}=\sigma_{k+2}, s\leq k\leq j-2$. In conclusion, if \emph{\textbf{x}} and \emph{\textbf{y}} are Type-B-confusable then condition $C_{1}$ is satisfied.
      Conversely, assume that the condition $C_{1}$ holds. Then $\mathcal{I}_{1}(\alpha\beta\textbf{\emph{z}})\cap\mathcal{I}_{1}(\textbf{\emph{z}}\gamma\delta)=\{\sigma_{i}\ldots\sigma_{t-2}\sigma_{t}\sigma_{t-1}\sigma^{2}_{t}\sigma_{t+1}\ldots\sigma_{s-1}\sigma_{s}^{2}\ldots\sigma_{j-2}\}$.
         which means that $\alpha\beta\textbf{\emph{z}}$ and $\textbf{\emph{z}}\gamma\delta$ are Type-B-confusable.
  \item Exactly one run with index $t$, $i+1\leq t\leq j-1$, exists such that $\ell_{t}\geq3$. In this case \emph{\textbf{u}}$^{'}=\sigma_{i}\ldots\sigma_{t-2}$, \emph{\textbf{u}}$^{''}=\sigma_{i+2}\ldots\sigma_{t}$,  \emph{\textbf{z}}$^{'}=\sigma_{t-1}\sigma_{t}^{\ell_{t}-1}$,  \emph{\textbf{z}}$^{''}=\sigma_{t}^{\ell_{t}-1}\sigma_{t+1}$, \emph{\textbf{w}}$^{'}=\sigma_{t}\ldots\sigma_{j-2}$ and \emph{\textbf{w}}$^{''}=\sigma_{t+2}\ldots\sigma_{j}$.  Since  \emph{\textbf{z}}$^{'}_{1}\neq$ \emph{\textbf{z}}$^{''}_{1}$ and \emph{\textbf{z}}$^{'}_{|\emph{\textbf{z}}^{'}|}\neq$ \emph{\textbf{z}}$^{''}_{|\emph{\textbf{z}}^{''}|}$ we have $d_{H}(\emph{\textbf{x}}, \emph{\textbf{y}})\geq2$. Also \emph{\textbf{z}}$^{'}$ and \emph{\textbf{z}}$^{''}$ are not alternating sequences. So \emph{\textbf{x}} and \emph{\textbf{y}} can not be Type-A-confusable. If $\emph{\textbf{u}}^{'}\neq \emph{\textbf{u}}^{''}$ or $\emph{\textbf{w}}^{'}\neq \emph{\textbf{w}}^{''}$ then similar to the case before we can prove that \emph{\textbf{x}} and \emph{\textbf{y}} are not Type-B-confusable.
       Since $\mathcal{I}_{1}($\emph{\textbf{z}}$^{'})\cap\mathcal{I}_{1}($\emph{\textbf{z}}$^{''})=\sigma_{t-1}\sigma_{t}^{\ell_{t}-1}\sigma_{t+1}$, then \emph{\textbf{x}} and \emph{\textbf{y}} are Type-B-confusable if \emph{\textbf{u}}$^{'}=\emph{\textbf{u}}^{''}$ and \emph{\textbf{w}}$^{'}=\emph{\textbf{w}}^{''}$ which is equivalent to condition $C_{2}$. Conversely, assume that the condition $C_{2}$ holds. Then $\mathcal{I}_{1}(\alpha\beta\textbf{\emph{z}})\cap\mathcal{I}_{1}(\textbf{\emph{z}}\gamma\delta)=\{\sigma_{i}\ldots\sigma_{t-2}\sigma_{t-1}\sigma_{t}^{\ell_{t}-1}\sigma_{t+1}\sigma_{t+2}\ldots\sigma_{j}$\}, which means that $\alpha\beta\textbf{\emph{z}}$ and $\textbf{\emph{z}}\gamma\delta$ are Type-B-confusable.
  \item Exactly one run with index $t$, $i+1\leq t\leq j-1$, exist such that $\ell_{t}=2$. In this case define the subsequences \emph{\textbf{u}}$^{'}$, \emph{\textbf{u}}$^{''}$, \emph{\textbf{w}}$^{'}$ and \emph{\textbf{w}}$^{''}$ same as the case before. Then  \emph{\textbf{z}}$^{'}=\sigma_{t-1}\sigma_{t}$ and 
      \emph{\textbf{z}}$^{''}=\sigma_{t}\sigma_{t+1}$. We must have $\emph{\textbf{u}}^{'}= \emph{\textbf{u}}^{''}$ and $\emph{\textbf{w}}^{'}= \emph{\textbf{w}}^{''}$ so that \emph{\textbf{x}} and \emph{\textbf{y}} can be Type-A-confusable or Type-B-confusable.
       If  \emph{\textbf{x}} and \emph{\textbf{y}} are Type-B-confusable then condition $C_{3}$ satisfies. If  \emph{\textbf{x}} and \emph{\textbf{y}} are Type-A-confusable then condition $C_{4}$ satisfies. Conversely, assume that the condition $C_{3}$ holds. Then  $\mathcal{I}_{1}(\alpha\beta\textbf{\emph{z}})\cap\mathcal{I}_{1}(\textbf{\emph{z}}\gamma\delta)=\{\sigma_{i}\ldots\sigma_{t-2}\sigma_{t-1}\sigma_{t}\sigma_{t+1}\sigma_{t+2}\ldots\sigma_{j}$\} and if the condition $C_{4}$ holds then  $\mathcal{I}_{1}(\alpha\beta\textbf{\emph{z}})\cap\mathcal{I}_{1}(\textbf{\emph{z}}\gamma\delta)=\{\sigma_{i}\ldots\sigma_{t-2}\sigma_{t-1}\sigma_{t}\sigma_{t+1}\sigma_{t+2}\ldots\sigma_{j}\\
      , \sigma_{i}\ldots\sigma_{t-2}\sigma_{t}\sigma_{t+1}\sigma_{t}\sigma_{t+2}\ldots\sigma_{j}$\}.
  \item If $i+1\leq t\leq j-1$ then  $\ell_{t}=1$. \emph{\textbf{x}}$=\sigma_{i}\ldots\sigma_{j-2}$ and \emph{\textbf{y}}$=\sigma_{i+2}\ldots\sigma_{j}$. Clearly, \emph{\textbf{x}} and \emph{\textbf{y}} can not be Type-A-confusable or Type-B-confusable. Then $d_{H}(\emph{\textbf{x}}, \emph{\textbf{y}})=1$ if and only if condition $C_{5}$ satisfies.
\end{enumerate}
\end{IEEEproof}
\begin{Lemma}
Let $\boldsymbol{\omega}=\sigma_{1}^{\ell_{1}}\sigma_{2}^{\ell_{2}}\ldots\sigma_{\rho(\boldsymbol{\omega})}^{\ell_{\rho(\boldsymbol{\omega})}}\in \Sigma^{n}_{q}$ where $\sigma_{i}\neq \sigma_{i+1}$.
   If $1\leq i<j<k\leq \rho(\boldsymbol{\omega})$ then the size of the intersection of insertion 2-spheres of three distinct 1-subsequences is
  
    $|\mathcal{I}_{2}(\boldsymbol{\omega}^{(i)}) \cap \mathcal{I}_{2}(\boldsymbol{\omega}^{(j)})\cap \mathcal{I}_{2}(\boldsymbol{\omega}^{(k)})|= 1+|\mathcal{I}_{1}(\boldsymbol{\omega})|$ if $|\mathcal{I}_{1}(\boldsymbol{\omega}^{(i)}) \cap \mathcal{I}_{1}(\boldsymbol{\omega}^{(j)})|=|\mathcal{I}_{1}(\boldsymbol{\omega}^{(j)})\cap \mathcal{I}_{1}(\boldsymbol{\omega}^{(k)})|=2$. Otherwise  $|\mathcal{I}_{2}(\boldsymbol{\omega}^{(i)}) \cap \mathcal{I}_{2}(\boldsymbol{\omega}^{(j)})\cap \mathcal{I}_{2}(\boldsymbol{\omega}^{(k)})|= |\mathcal{I}_{1}(\boldsymbol{\omega})|$.
\end{Lemma}
\emph{Proof:} The proof is given in Appendix B.
\begin{Lemma}
Let $\boldsymbol{\omega}=\sigma_{1}^{\ell_{1}}\sigma_{2}^{\ell_{2}}\ldots\sigma_{\rho(\boldsymbol{\omega})}^{\ell_{\rho(\boldsymbol{\omega})}}\in \Sigma^{n}_{q}$ be a sequence. If $1\leq i<j<k<l \leq \rho(\boldsymbol{\omega})$ then

$\mathcal{I}_{2}(\boldsymbol{\omega}^{(i)}) \cap \mathcal{I}_{2}(\boldsymbol{\omega}^{(j)})\cap \mathcal{I}_{2}(\boldsymbol{\omega}^{(k)})\cap \mathcal{I}_{2}(\boldsymbol{\omega}^{(l)})=\mathcal{I}_{1}(\boldsymbol{\omega})$.
\end{Lemma}
\begin{IEEEproof}
 Based on Lemma 5, the only way $|\mathcal{I}_{2}(\boldsymbol{\omega}^{(i)}) \cap \mathcal{I}_{2}(\boldsymbol{\omega}^{(j)})\cap \mathcal{I}_{2}(\boldsymbol{\omega}^{(k)})\cap \mathcal{I}_{2}(\boldsymbol{\omega}^{(l)})|>|\mathcal{I}_{1}(\boldsymbol{\omega})|$ is when the condition $|\mathcal{I}_{1}(\boldsymbol{\omega}^{(i)}) \cap \mathcal{I}_{1}(\boldsymbol{\omega}^{(j)})|=|\mathcal{I}_{1}(\boldsymbol{\omega}^{(j)})\cap \mathcal{I}_{1}(\boldsymbol{\omega}^{(k)})|=|\mathcal{I}_{1}(\boldsymbol{\omega}^{(k)})\cap \mathcal{I}_{1}(\boldsymbol{\omega}^{(l)})|=2$ is satisfied. Let $\mathcal{I}_{2}(\boldsymbol{\omega}^{(i)}) \cap \mathcal{I}_{2}(\boldsymbol{\omega}^{(j)})=\mathcal{I}_{1}(\boldsymbol{\omega})\cup \mathcal{I}_{1}(\emph{\textbf{z}})$, $\mathcal{I}_{2}(\boldsymbol{\omega}^{(j)}) \cap \mathcal{I}_{2}(\boldsymbol{\omega}^{(k)})=\mathcal{I}_{1}(\boldsymbol{\omega})\cup \mathcal{I}_{1}(\emph{\textbf{z}}^{'})$, and $\mathcal{I}_{2}(\boldsymbol{\omega}^{(k)}) \cap \mathcal{I}_{2}(\boldsymbol{\omega}^{(l)})=\mathcal{I}_{1}(\boldsymbol{\omega})\cup \mathcal{I}_{1}(\emph{\textbf{z}}^{''})$. Expressions for \emph{\textbf{z}} and \emph{\textbf{z}}$^{'}$ are described in proof of Lemma 5. Therefore $\mathcal{I}_{2}(\boldsymbol{\omega}^{(i)}) \cap \mathcal{I}_{2}(\boldsymbol{\omega}^{(j)})\cap \mathcal{I}_{2}(\boldsymbol{\omega}^{(k)})\cap \mathcal{I}_{2}(\boldsymbol{\omega}^{(l)})=\mathcal{I}_{1}(\boldsymbol{\omega})\cup (\mathcal{I}_{1}(\emph{\textbf{z}})\cap \mathcal{I}_{1}(\emph{\textbf{z}}^{'})\cap \mathcal{I}_{1}(\emph{\textbf{z}}^{''}))$.  If $l=k+1$ then $\emph{\textbf{z}}^{''}=\sigma_{1}^{\ell_{1}}\ldots\sigma_{k}^{\ell_{k}-1}\sigma_{k+1}\sigma_{k}\sigma_{k+1}^{\ell_{k+1}-1}\ldots\sigma_{\rho(\boldsymbol{\omega})}^{\ell_{\rho(\boldsymbol{\omega})}}$. If $\boldsymbol{\omega}^{k}$ and $\boldsymbol{\omega}^{l}$ are Type-A-confusable then 
$\emph{\textbf{z}}^{''}=\sigma_{1}^{\ell_{1}}\ldots\sigma_{k-1}^{\ell_{k-1}}\sigma_{k}^{\ell_{k}-1}\sigma_{k+1}\ldots\sigma_{l-1}\sigma_{l}^{\ell_{l}-1}\ldots\sigma_{\rho(\boldsymbol{\omega})}^{\ell_{\rho(\boldsymbol{\omega})}}$.
Based on the proof of Lemma 5, the general expression for $\mathcal{I}_{1}(\emph{\textbf{z}})\cap \mathcal{I}_{1}(\emph{\textbf{z}}^{'})$ is
 $\textbf{\emph{x}}=\emph{\textbf{u}}\sigma_{i+1}\sigma_{i}\sigma_{i+1}\ldots\sigma_{j-1}\sigma_{j}^{\ell_{j}-1}\sigma_{j+1}\ldots\sigma_{k}\sigma_{k-1}\emph{\textbf{w}}$, where \emph{\textbf{u}}$=\sigma_{1}^{\ell_{1}}\ldots\sigma_{i}^{\ell_{i}-1}$ and \emph{\textbf{w}}$=\sigma_{k}^{\ell_{k}-1}\ldots\sigma_{\rho(\boldsymbol{\omega})}^{\ell_{\rho(\boldsymbol{\omega})}}$. Now we show that $\mathcal{I}_{1}(\emph{\textbf{z}})\cap \mathcal{I}_{1}(\emph{\textbf{z}}^{'})\cap \mathcal{I}_{1}(\emph{\textbf{z}}^{''})=\{\emptyset\}$. Suppose on the contrary that
  \textbf{\emph{x}}$\in \mathcal{I}_{1}(\emph{\textbf{z}}^{''})$. By comparing the first symbol on the left of $\sigma_{l}^{\ell_{l}-1}$ in \textbf{\emph{x}} and $\emph{\textbf{z}}^{''}$, we see that this is impossible, since in $\emph{\textbf{z}}^{''}$, the symbol is $\sigma_{l-1}$ while it is $\sigma_{l}$ in \textbf{\emph{x}}.
\end{IEEEproof} 
\begin{Lemma}
Let $n\geq2$ and $\boldsymbol{\omega}=\sigma_{1}^{\ell_{1}}\sigma_{2}^{\ell_{2}}\ldots\sigma_{\rho(\boldsymbol{\omega})}^{\ell_{\rho(\boldsymbol{\omega})}}\in \Sigma^{n}_{q}$ be a sequence with maximal alternating segments profile vector $[s_{1}, s_{2} \ldots, s_{\psi(\boldsymbol{\omega})}]$. 
  Then $A(\boldsymbol{\omega})=\sum_{k=1}^{\psi(\boldsymbol{\omega})}\binom{s_{k}}{2}$.
\end{Lemma}
 \begin{IEEEproof}
Consider a particular maximal alternating segment with length $s_{i}$. Deleting each symbol from this segment creates a new subsequence from sequence $\boldsymbol{\omega}$. According to corollary 1,  any two 1-subsequences each derived by deleting a symbol inside the segment are either have Hamming distance one or are Type-A-confusable. Based on Theorem 1, the size of the intersection of their 1-insertion spheres is two. Therefore, based on part (iii) of Lemma 3, the size of the intersection of their 2-insertion sphere is $2|\mathcal{I}_{1}(\boldsymbol{\omega})|-2$. Two 1-subsequences derived by deleting a symbol from two different segments are Type-B-confusable.
\end{IEEEproof} 
\begin{Lemma}
Let $n\geq3$ and $\boldsymbol{\omega}\in \Sigma^{n}_{2}$ be a binary sequence with maximal alternating segments profile vector $[s_{1}, s_{2} \ldots, s_{\psi(\boldsymbol{\omega})}]$. 
\begin{itemize}
  \item The number of pairs of 1-subsequences $\boldsymbol{\omega}^{(i)}$ and  $\boldsymbol{\omega}^{(j)}$ such that $|\mathcal{I}_{2}(\boldsymbol{\omega}^{(i)})\cap \mathcal{I}_{2}(\boldsymbol{\omega}^{(j)})|=2+|\mathcal{I}_{1}(\boldsymbol{\omega})|$ is $B(\boldsymbol{\omega})=\sum_{k=1}^{\psi(\boldsymbol{\omega})-1}(s_{k}-1)(s_{k+1}-1)$.
  \item The number of pairs of 1-subsequences $\boldsymbol{\omega}^{(i)}$ and  $\boldsymbol{\omega}^{(j)}$ such that $|\mathcal{I}_{2}(\boldsymbol{\omega}^{(i)})\cap \mathcal{I}_{2}(\boldsymbol{\omega}^{(j)})|=1+|\mathcal{I}_{1}(\boldsymbol{\omega})|$ is 
      \(C(\boldsymbol{\omega})=\)
      \[\sum_{i=1}^{\psi(\boldsymbol{\omega})-2}\sum_{j=i+2}^{\psi(\boldsymbol{\omega})}(s_{i}-1)(\prod_{k=1}^{j-i-1}\mathbb{1}[s_{i+k}=3])(s_{j}-1)+\]
      \[\sum_{i=1}^{\psi(\boldsymbol{\omega})-2}\sum_{j=i+2}^{\psi(\boldsymbol{\omega})}(s_{i}-1)(\prod_{k=1}^{j-i-1}\mathbb{1}[s_{i+k}=1])(s_{j}-1).\]
\end{itemize}
\end{Lemma}
\begin{IEEEproof}
\begin{itemize}
\item Based on Lemma 4, for two 1-subsequences of the binary sequence $\boldsymbol{\omega}$ like $\boldsymbol{\omega}^{(i)}$ and  $\boldsymbol{\omega}^{(j)}$, the equality $|\mathcal{I}_{2}(\boldsymbol{\omega}^{(i)})\cap \mathcal{I}_{2}(\boldsymbol{\omega}^{(j)})|=2+|\mathcal{I}_{1}(\boldsymbol{\omega})|$ happens when $j=i+2, \ell_{i+1}=2$ or the condition $C_{4}$ is satisfied. Note that if the two 1-subsequences are derived by deleting a bit from two non-adjacent maximal alternating segments, then these two conditions can not be satisfied. Now, consider two adjacent maximal alternating segments of this sequence where the length of each of them is larger or equal than 2, denoted by $s_{k}$ and $s_{k+1}$. Let $\boldsymbol{\omega}^{(i)}$ be the 1-subsequence derived by deleting any bit from the first segment except the last bit of the first segment and $\boldsymbol{\omega}^{(j)}$ be the 1-subsequence derived by deleting any bit from the second segment except the first bit of the second segment. Now according to Lemma 4, we have $|\mathcal{I}_{2}(\boldsymbol{\omega}^{(i)})\cap \mathcal{I}_{2}(\boldsymbol{\omega}^{(j)})|=2+|\mathcal{I}_{1}(\boldsymbol{\omega})|$. Therefore the total number of such pairs is $\sum_{k=1}^{\psi(\boldsymbol{\omega})-1}(s_{k}-1)(s_{k+1}-1)$.
\item Consider two non-adjacent maximal alternating segments with lengths $s_{i}$ and $s_{j}$ where $j\geq i+2$. If the lengths of the maximal alternating segments between them is 3, then the condition $C_{1}$ is satisfied. If $\boldsymbol{\omega}^{(k)}$ be a 1-subsequence derived by deleting any bit from the segment with length $s_{i}$ except the last bit, and $\boldsymbol{\omega}^{(l)}$ be a 1-subsequence derived by deleting any bit from the segment with length $s_{j}$ except the first bit then $|\mathcal{I}_{2}(\boldsymbol{\omega}^{(k)})\cap \mathcal{I}_{2}(\boldsymbol{\omega}^{(l)})|=1+|\mathcal{I}_{1}(\boldsymbol{\omega})|$.
    
    If the lengths of the maximal alternating segments between two non-adjacent maximal alternating segments are one or equivalently a run with length greater than or equal to three exist between them, then condition $C_{2}$ is satisfied. The condition ($l=k+2,\ell_{k+1}\geq 3$) described in first part of Lemma 4, so that $|\mathcal{I}_{2}(\boldsymbol{\omega}^{(k)})\cap \mathcal{I}_{2}(\boldsymbol{\omega}^{(l)})|=1+|\mathcal{I}_{1}(\boldsymbol{\omega})|$, is a special case of condition $C_{2}$.
\end{itemize}
\end{IEEEproof}
\begin{Remark}
In general non-binary sequences, knowledge of maximal alternating segments profile vector is not enough for determining the values of $B(\boldsymbol{\omega})$ and $C(\boldsymbol{\omega})$. For example, let \emph{\textbf{x}}=1001 and \emph{\textbf{y}}=1002. Then the maximal alternating segments profile vector of both sequences is $[2, 2]$. We have $B(\emph{\textbf{x}})=1$ and $C(\emph{\textbf{x}})=0$ while $B(\emph{\textbf{y}})=0$ and $C(\emph{\textbf{y}})=1$.
\end{Remark}
Before computing the value of $D(\boldsymbol{\omega})$, we need a new definition.
\begin{Definition}
 The modified maximal alternating segments profile vector is derived when we omit the segments with length one from maximal  alternating segments profile vector. The number of such segments is denoted by $\psi^{'}(\boldsymbol{\omega})$.
\end{Definition}
\begin{Lemma}
  Let $n\geq 3$ and $\boldsymbol{\omega}=\sigma_{1}^{\ell_{1}}\sigma_{2}^{\ell_{2}}\ldots\sigma_{\rho(\boldsymbol{\omega})}^{\ell_{\rho(\boldsymbol{\omega})}}\in \Sigma^{n}_{q}$ be a sequence with modified maximal alternating segments profile vector $[s^{'}_{1}, s^{'}_{2} \ldots, s^{'}_{\psi^{'}(\boldsymbol{\omega})}]$. Then the number of triples of 1-subsequences $\boldsymbol{\omega}^{(i)}$, $\boldsymbol{\omega}^{(j)}$ and $\boldsymbol{\omega}^{(k)}$, $i<j<k$, such that  $|\mathcal{I}_{1}(\boldsymbol{\omega}^{(i)}) \cap \mathcal{I}_{1}(\boldsymbol{\omega}^{(j)})|=|\mathcal{I}_{1}(\boldsymbol{\omega}^{(j)}) \cap \mathcal{I}_{1}(\boldsymbol{\omega}^{(k)})|=2$ is
  \begin{equation*}
  D(\boldsymbol{\omega})=\sum_{l=1}^{\psi^{'}(\boldsymbol{\omega})}\binom{s^{'}_{l}}{3}+\sum_{l=1}^{\psi^{'}(\boldsymbol{\omega})-1}(s^{'}_{l}-1)(s^{'}_{l+1}-1).
  \end{equation*}
\end{Lemma} 
\begin{IEEEproof}
If the condition $|\mathcal{I}_{1}(\boldsymbol{\omega}^{(i)}) \cap \mathcal{I}_{1}(\boldsymbol{\omega}^{(j)})|=|\mathcal{I}_{1}(\boldsymbol{\omega}^{(j)}) \cap \mathcal{I}_{1}(\boldsymbol{\omega}^{(k)})|=2$ is satisfied then the pair of 1-subsequences $\boldsymbol{\omega}^{(i)},\boldsymbol{\omega}^{(j)}$ are derived by deleting  a symbol from the same segment and also the pair of 1-subsequences $\boldsymbol{\omega}^{(j)},\boldsymbol{\omega}^{(k)}$ are derived by deleting  a symbol from the same segment. Therefore,
we have only two cases. Either all three 1-subsequences derived by deleting a symbol from a single segment or $\boldsymbol{\omega}^{(i)}$ derived by deleting a symbol from segment $s^{'}_{l}$ except the last symbol, $\boldsymbol{\omega}^{(j)}$ could by derived by deleting the last symbol of $s^{'}_{l}$ or first symbol of $s^{'}_{l+1}$ and $\boldsymbol{\omega}^{(k)}$  derived by deleting a symbol from the segment $s^{'}_{l+1}$ except the first symbol. The number of triples of first case is $\sum_{l=1}^{\psi^{'}(\boldsymbol{\omega})}\binom{s^{'}_{l}}{3}$ and the number of the triples of the second case is $\sum_{l=1}^{\psi^{'}(\boldsymbol{\omega})-1}(s^{'}_{l}-1)(s^{'}_{l+1}-1)$. 
 \end{IEEEproof}
\begin{Example}
(i) Let $n=8$ and $\boldsymbol{\omega}=01011010$ be a sequence with maximal alternating segments profile vector [4, 4] and $\rho(\boldsymbol{\omega})=7$. We have $A(\boldsymbol{\omega})=2\binom{4}{2}=12$, $B(\boldsymbol{\omega})=(4-1)^{2}$, $C(\boldsymbol{\omega})=0$ and $D(\boldsymbol{\omega})=2\times\binom{4}{3}+(4-1)^{2}=17$. Therefore, $|\mathcal{B}_{0,1,2}(\boldsymbol{\omega})|=252(q-1)^2-99(q-1)+12$. In particular for $q=2$ we have $|\mathcal{B}_{0,1,2}(\boldsymbol{\omega})|=165$. 

(ii) Let $q=3$, $n=8$ and $\boldsymbol{\omega}=01110021$ be a sequence with maximal alternating segments profile vector [2, 1, 2, 2, 2] and $\rho(\boldsymbol{\omega})=5$. We have $A(\boldsymbol{\omega})=4, B(\boldsymbol{\omega})=1, C(\boldsymbol{\omega})=2, D(\boldsymbol{\omega})=3$. Therefore, $|\mathcal{B}_{0,1,2}(\boldsymbol{\omega})|=5\times36\times4-3\times19+8-2-2+3=670$.
\end{Example} 
\section{ The Single-Deletion Double-Substitution Ball Size}
Authors in [8] derived the single-deletion multiple-substitution ball size for binary sequences. For the single-deletion double-substitution case, we generalize their work for non-binary sequences. 
\begin{Theorem}
  Let $\boldsymbol{\omega}\in \Sigma^{n}_{q}$ be a sequence with $\rho(\boldsymbol{\omega})$ runs. Then the size of single-deletion double-substitution ball is 
  \begin{itemize}
    \item If $\rho(\boldsymbol{\omega})=1$ then $|\mathcal{B}_{2, 1, 0}(\boldsymbol{\omega})|=1+(n-1)(q-1)+\binom{n-1}{2}(q-1)^{2}$.
    \item If $\rho(\boldsymbol{\omega})=2$ then $|\mathcal{B}_{2, 1, 0}(\boldsymbol{\omega})|=1+(n-1)(q-1)+(n-2)^{2}(q-1)^{2}$.
    \item If $\rho(\boldsymbol{\omega})=3$ then $|\mathcal{B}_{2, 1, 0}(\boldsymbol{\omega})|=1+2(q-1)+\frac{1}{2}(3n-7)(n-2)(q-1)^{2}$.
    \item If $\rho(\boldsymbol{\omega})=4$ then $|\mathcal{B}_{2, 1, 0}(\boldsymbol{\omega})|=2-(n-4)(q-1)+(2n^{2}-9n+10)(q-1)^{2}$.
    \item If $\rho(\boldsymbol{\omega})\geq 5$ then $|\mathcal{B}_{2, 1, 0}(\boldsymbol{\omega})|=2+((3-\rho(\boldsymbol{\omega}))n+2\rho(\boldsymbol{\omega})-4)(q-1)+(\rho(\boldsymbol{\omega})\binom{n-1}{2}-(\rho(\boldsymbol{\omega})-1)(n-2))(q-1)^{2}$.
  \end{itemize}
\end{Theorem}
\begin{IEEEproof}
 The proof is based on the inclusion-exclusion principle. Clearly, if $\rho(\boldsymbol{\omega})=1$ then $|\mathcal{B}_{2, 1, 0}(\boldsymbol{\omega})|=1+(n-1)(q-1)+\binom{n-1}{2}(q-1)^{2}$. For $\rho(\boldsymbol{\omega})\geq 2$, the size of the intersection of 2 1-subsequences is given in Lemma 10 and the size of the 3 or more 1-subsequences is given in Lemma 11.
\begin{itemize}
  \item If $\rho(\boldsymbol{\omega})=2$ then $|\mathcal{B}_{2, 1, 0}(\boldsymbol{\omega})|=2|\mathcal{B}_{2, 0, 0}(\boldsymbol{\omega}^{(1)})|-|\mathcal{B}_{2, 0, 0}(\boldsymbol{\omega}^{(1)})\cap \mathcal{B}_{2, 0, 0}(\boldsymbol{\omega}^{(2)})|=2+2(n-1)(q-1)+2\binom{n-1}{2}(q-1)^{2}-q(1+(n-2)(q-1))=1+(n-1)(q-1)+(n-2)^{2}(q-1)^{2}$.
  \item If $\rho(\boldsymbol{\omega})=3$ then $|\mathcal{B}_{2, 1, 0}(\boldsymbol{\omega})|=3|\mathcal{B}_{2, 0, 0}(\boldsymbol{\omega}^{(1)})|-2|\mathcal{B}_{2, 0, 0}(\boldsymbol{\omega}^{(1)})\cap \mathcal{B}_{2, 0, 0}(\boldsymbol{\omega}^{2})|-|\mathcal{B}_{2, 0, 0}(\boldsymbol{\omega}^{(1)})\cap \mathcal{B}_{2, 0, 0}(\boldsymbol{\omega}^{(3)})|+|\mathcal{B}_{2, 0, 0}(\boldsymbol{\omega}^{(1)})\cap \mathcal{B}_{2, 0, 0}(\boldsymbol{\omega}^{(2)})\cap \mathcal{B}_{2, 0, 0}(\boldsymbol{\omega}^{(3)})|=3+3(n-1)(q-1)+3\binom{n-1}{2}(q-1)^{2}-2q(1+(n-2)(q-1))-q^{2}-2(n-3)(q-1)+q^{2}+(n-3)(q-1)=1+2(q-1)+\frac{1}{2}(3n-7)(n-2)(q-1)^{2}$.
  \item If $\rho(\boldsymbol{\omega})=4$ then $|\mathcal{B}_{2, 1, 0}(\boldsymbol{\omega})|=4|\mathcal{B}_{2, 0, 0}(\boldsymbol{\omega}^{(1)})|-3|\mathcal{B}_{2, 0, 0}(\boldsymbol{\omega}^{(1)})\cap \mathcal{B}_{2, 0, 0}(\boldsymbol{\omega}^{(2)})|-2|\mathcal{B}_{2, 0, 0}(\boldsymbol{\omega}^{(1)})\cap \mathcal{B}_{2, 0, 0}(\boldsymbol{\omega}^{(3)})|-|\mathcal{B}_{2, 0, 0}(\boldsymbol{\omega}^{(1)})\cap \mathcal{B}_{2, 0, 0}(\boldsymbol{\omega}^{(4)})|+2|\mathcal{B}_{2, 0, 0}(\boldsymbol{\omega}^{(1)})\cap \mathcal{B}_{2, 0, 0}(\boldsymbol{\omega}^{(2)})\cap \mathcal{B}_{2, 0, 0}(\boldsymbol{\omega}^{(3)})|+2|\mathcal{B}_{2, 0, 0}(\boldsymbol{\omega}^{(1)})\cap \mathcal{B}_{2, 0, 0}(\boldsymbol{\omega}^{(2)})\cap \mathcal{B}_{2, 0, 0}(\boldsymbol{\omega}^{(4)})|-|\mathcal{B}_{2, 0, 0}(\boldsymbol{\omega}^{(1)})\cap \mathcal{B}_{2, 0, 0}(\boldsymbol{\omega}^{(2)})\cap B_{2, 0, 0}(\boldsymbol{\omega}^{(3)})\cap \mathcal{B}_{2, 0, 0}(\boldsymbol{\omega}^{(4)})|=4+4(n-1)(q-1)+4\binom{n-1}{2}(q-1)^{2}-3q(1+(n-2)(q-1))-2(q^{2}+2(n-3)(q-1))-6(q-1)+2(q^{2}+(n-3)(q-1))+2(4q-3)-(3q-2)=2-(n-4)(q-1)+(2n^{2}-9n+10)(q-1)^{2}$.
  \item If $\rho(\boldsymbol{\omega})\geq 5$  then $|\mathcal{B}_{2, 1, 0}(\boldsymbol{\omega})|=\rho(\boldsymbol{\omega})|\mathcal{B}_{2, 0, 0}(\boldsymbol{\omega}^{(1)})|-(\rho(\boldsymbol{\omega})-1)|\mathcal{B}_{2, 0, 0}(\boldsymbol{\omega}^{(1)})\cap \mathcal{B}_{2, 0, 0}(\boldsymbol{\omega}^{(2)})|-(\rho(\boldsymbol{\omega})-2)|\mathcal{B}_{2, 0, 0}(\boldsymbol{\omega}^{(1)})\cap \mathcal{B}_{2, 0, 0}(\boldsymbol{\omega}^{(3)})|-(\rho(\boldsymbol{\omega})-3)|\mathcal{B}_{2, 0, 0}(\boldsymbol{\omega}^{(1)})\cap \mathcal{B}_{2, 0, 0}(\boldsymbol{\omega}^{(4)})|-(\rho(\boldsymbol{\omega})-4)|\mathcal{B}_{2, 0, 0}(\boldsymbol{\omega}^{(1)})\cap \mathcal{B}_{2, 0, 0}(\boldsymbol{\omega}^{(5)})|+(\rho(\boldsymbol{\omega})-2)|\mathcal{B}_{2, 0, 0}(\boldsymbol{\omega}^{(1)})\cap \mathcal{B}_{2, 0, 0}(\boldsymbol{\omega}^{(2)})\cap \mathcal{B}_{2, 0, 0}(\boldsymbol{\omega}^{(3)})|+2(\rho(\boldsymbol{\omega})-3)|\mathcal{B}_{2, 0, 0}(\boldsymbol{\omega}^{(1)})\cap \mathcal{B}_{2, 0, 0}(\boldsymbol{\omega}^{(2)})\cap \mathcal{B}_{2, 0, 0}(\boldsymbol{\omega}^{(4)})|+(\rho(\boldsymbol{\omega})-4)(2|\mathcal{B}_{2, 0, 0}(\boldsymbol{\omega}^{(1)})\cap \mathcal{B}_{2, 0, 0}(\boldsymbol{\omega}^{(2)})\cap \mathcal{B}_{2, 0, 0}(\boldsymbol{\omega}^{(5)})|+|\mathcal{B}_{2, 0, 0}(\boldsymbol{\omega}^{(1)})\cap \mathcal{B}_{2, 0, 0}(\boldsymbol{\omega}^{(3)})\cap \mathcal{B}_{2, 0, 0}(\boldsymbol{\omega}^{(5)})|)-2(\rho(\boldsymbol{\omega})-3)|\mathcal{B}_{2, 0, 0}(\boldsymbol{\omega}^{(1)})\cap \mathcal{B}_{2, 0, 0}(\boldsymbol{\omega}^{(2)})\cap \mathcal{B}_{2, 0, 0}(\boldsymbol{\omega}^{(3)})\cap \mathcal{B}_{2, 0, 0}(\boldsymbol{\omega}^{(4)}) |+(\rho(\boldsymbol{\omega})-4)(-2|\mathcal{B}_{2, 0, 0}(\boldsymbol{\omega}^{(1)})\cap \mathcal{B}_{2, 0, 0}(\boldsymbol{\omega}^{(3)})\cap \mathcal{B}_{2, 0, 0}(\boldsymbol{\omega}^{(4)})\cap \mathcal{B}_{2, 0, 0}(\boldsymbol{\omega}^{(5)}) |-|\mathcal{B}_{2, 0, 0}(\boldsymbol{\omega}^{(1)})\cap \mathcal{B}_{2, 0, 0}(\boldsymbol{\omega}^{(2)})\cap \mathcal{B}_{2, 0, 0}(\boldsymbol{\omega}^{(4)})\cap \mathcal{B}_{2, 0, 0}(\boldsymbol{\omega}^{(5)}) |+|\bigcap_{j=0}^{4}\mathcal{B}_{2, 0, 0}(\boldsymbol{\omega}^{(1+j)})|)=\rho(\boldsymbol{\omega})(1+(n-1)(q-1)+\binom{n-1}{2}(q-1)^{2})-(\rho(\boldsymbol{\omega})-1)q(1+(n-2)(q-1))-(\rho(\boldsymbol{\omega})-2)(q^{2}+2(n-3)(q-1))-6(\rho(\boldsymbol{\omega})-3)(q-1)-6(\rho(\boldsymbol{\omega})-4)
      +(\rho(\boldsymbol{\omega})-2)(q^{2}+(n-3)(q-1))+2(\rho(\boldsymbol{\omega})-3)(4q-3)+11(\rho(\boldsymbol{\omega})-4)-2(3q-2)(\rho(\boldsymbol{\omega})-3)-8(\rho(\boldsymbol{\omega})-4)+2(\rho(\boldsymbol{\omega})-4)=2+((3-\rho(\boldsymbol{\omega}))n+2\rho(\boldsymbol{\omega})-4)(q-1)+(\rho(\boldsymbol{\omega})\binom{n-1}{2}-(\rho(\boldsymbol{\omega})-1)(n-2))(q-1)^{2}$.
\end{itemize}
 \end{IEEEproof}
\begin{Lemma}
  Let $\textbf{\emph{x}},\textbf{\emph{y}}\in \Sigma^{n}_{q}$. Then the size of the intersection of their double-substitution balls is
  \begin{itemize}
    \item If $d_{H}(\textbf{\emph{x}},\textbf{\emph{y}})=1$ then $|\mathcal{B}_{2, 0, 0}(\textbf{\emph{x}})\cap \mathcal{B}_{2, 0, 0}(\textbf{\emph{y}})|=q(1+(n-1)(q-1))$.
    \item If $d_{H}(\textbf{\emph{x}},\textbf{\emph{y}})=2$ then $|\mathcal{B}_{2, 0, 0}(\textbf{\emph{x}})\cap \mathcal{B}_{2, 0, 0}(\textbf{\emph{y}})|=q^{2}+2(n-2)(q-1).$
    \item If $d_{H}(\textbf{\emph{x}},\textbf{\emph{y}})=3$ then $|\mathcal{B}_{2, 0, 0}(\textbf{\emph{x}})\cap \mathcal{B}_{2, 0, 0}(\textbf{\emph{y}})|=6(q-1).$
    \item If $d_{H}(\textbf{\emph{x}},\textbf{\emph{y}})=4$ then $|\mathcal{B}_{2, 0, 0}(\textbf{\emph{x}})\cap \mathcal{B}_{2, 0, 0}(\textbf{\emph{y}})|=6$.
    \item If $d_{H}(\textbf{\emph{x}},\textbf{\emph{y}})\geq 5$  then $|\mathcal{B}_{2, 0, 0}(\textbf{\emph{x}})\cap \mathcal{B}_{2, 0, 0}(\textbf{\emph{y}})|=0$.
  \end{itemize}
\end{Lemma}
\begin{IEEEproof} 
\begin{itemize}
\item
If $d_{H}(\textbf{\emph{x}},\textbf{\emph{y}})=1$ then \textbf{\emph{x}}$=$\textbf{\emph{u}}$\alpha$\textbf{\emph{v}} and   \textbf{\emph{y}}$=$\textbf{\emph{u}}$\beta$\textbf{\emph{v}} where $\alpha\neq\beta$. Then  $\mathcal{B}_{2, 0, 0}(\textbf{\emph{x}})\cap \mathcal{B}_{2, 0, 0}(\textbf{\emph{y}})=\{\tilde{\textbf{\emph{u}}}\gamma$\textbf{\emph{v}}$, $\textbf{\emph{u}}$\gamma\tilde{\textbf{\emph{v}}}\}$ where $\gamma\in \Sigma_{q}$, $d_{H}(\textbf{\emph{u}},\tilde{\textbf{\emph{u}}})\leq1$ and $d_{H}(\textbf{\emph{v}},\tilde{\textbf{\emph{v}}})\leq1$. 
\item
If $d_{H}(\textbf{\emph{x}},\textbf{\emph{y}})=2$ then \textbf{\emph{x}}$=$\textbf{\emph{u}}$\alpha_{1}$\textbf{\emph{v}}$\alpha_{2}$\textbf{\emph{w}} and \textbf{\emph{y}}$=$\textbf{\emph{u}}$\beta_{1}$\textbf{\emph{v}}$\beta_{2}$\textbf{\emph{w}} where $\alpha_{i}\neq\beta_{i}$. Let \textbf{\emph{z}}$\in \mathcal{B}_{2, 0, 0}(\textbf{\emph{x}})\cap \mathcal{B}_{2, 0, 0}(\textbf{\emph{y}})$. Consider the indices of two symbols where \textbf{\emph{x}} and \textbf{\emph{y}} are distinct. In sequence \textbf{\emph{z}}, either these two symbols are chosen freely from $\Sigma_{q}$  or one symbol chosen from \textbf{\emph{x}} and the other from \textbf{\emph{y}}. Also exactly one substitution must occurs in one of the subsequences \textbf{\emph{u}}, \textbf{\emph{v}} or \textbf{\emph{w}}. Then  $\mathcal{B}_{2, 0, 0}(\textbf{\emph{x}})\cap \mathcal{B}_{2, 0, 0}(\textbf{\emph{y}})=\{\tilde{\textbf{\emph{u}}}\alpha_{1}$\textbf{\emph{v}}$\beta_{2}$\textbf{\emph{w}}$, \textbf{\emph{u}}\alpha_{1}\tilde{\textbf{\emph{v}}}\beta_{2}$\textbf{\emph{w}}$
,\textbf{\emph{u}}\alpha_{1}\textbf{\emph{v}}\beta_{2}\tilde{\textbf{\emph{w}}},\tilde{\textbf{\emph{u}}}\alpha_{2}\textbf{\emph{v}}\beta_{1}\textbf{\emph{w}}\}\cup 
\{\textbf{\emph{u}}\alpha_{2}\tilde{\textbf{\emph{v}}}\beta_{1}\textbf{\emph{w}},\textbf{\emph{u}}\alpha_{2}\textbf{\emph{v}}\beta_{1}\tilde{\textbf{\emph{w}}}
,\textbf{\emph{u}}\gamma\textbf{\emph{v}}\delta\textbf{\emph{w}}\}$ where $\gamma,\delta\in \Sigma_{q}$, $d_{H}(\textbf{\emph{u}},\tilde{\textbf{\emph{u}}})=1$, $d_{H}(\textbf{\emph{v}},\tilde{\textbf{\emph{v}}})=1$ and $d_{H}(\textbf{\emph{w}},\tilde{\textbf{\emph{w}}})=1$ .
\item
If $d_{H}(\textbf{\emph{x}},\textbf{\emph{y}})=3$ then \textbf{\emph{x}}$=$\textbf{\emph{u}}$\alpha_{1}$\textbf{\emph{v}}$\alpha_{2}$\textbf{\emph{w}}$\alpha_{3}$\textbf{\emph{z}} and \textbf{\emph{y}}$=$\textbf{\emph{u}}$\beta_{1}$\textbf{\emph{v}}$\beta_{2}$\textbf{\emph{w}}$\beta_{3}$\textbf{\emph{z}} where $\alpha_{i}\neq\beta_{i}$. Let \textbf{\emph{a}}$\in \mathcal{B}_{2, 0, 0}(\textbf{\emph{x}})\cap \mathcal{B}_{2, 0, 0}(\textbf{\emph{y}})$. Consider the indices of three symbols where \textbf{\emph{x}} and \textbf{\emph{y}} are distinct. In sequence \textbf{\emph{a}}, one of these symbols is chosen freely. One chosen from \textbf{\emph{x}} and the other from \textbf{\emph{y}}. Then  $\mathcal{B}_{2, 0, 0}(\textbf{\emph{x}})\cap \mathcal{B}_{2, 0, 0}(\textbf{\emph{y}})=\{\textbf{\emph{u}}\gamma\textbf{\emph{v}}\alpha_{2}\textbf{\emph{w}}\beta_{3}\textbf{\emph{z}},
\textbf{\emph{u}}\gamma\textbf{\emph{v}}\beta_{2}\textbf{\emph{w}}\alpha_{3}\textbf{\emph{z}}
,\textbf{\emph{u}}\alpha_{1}\textbf{\emph{v}}\beta_{2}\textbf{\emph{w}}\gamma\textbf{\emph{z}}\}\cup
\{\textbf{\emph{u}}\beta_{1}\textbf{\emph{v}}\alpha_{2}\textbf{\emph{w}}\gamma\textbf{\emph{z}},
\textbf{\emph{u}}\alpha_{1}\textbf{\emph{v}}\gamma\textbf{\emph{w}}\beta_{3}\textbf{\emph{z}},
\textbf{\emph{u}}\beta_{1}\textbf{\emph{v}}\gamma\textbf{\emph{w}}\alpha_{3}\textbf{\emph{z}}\}$ where $\gamma\in\Sigma_{q}$. Note that these six sequences are counted twice and must be subtracted

 $\{\textbf{\emph{u}}\beta_{1}\textbf{\emph{v}}\alpha_{2}\textbf{\emph{w}}\beta_{3}\textbf{\emph{z}},
\textbf{\emph{u}}\beta_{1}\textbf{\emph{v}}\alpha_{2}\textbf{\emph{w}}\alpha_{3}\textbf{\emph{z}}
,\textbf{\emph{u}}\beta_{1}\textbf{\emph{v}}\beta_{2}\textbf{\emph{w}}\alpha_{3}\textbf{\emph{z}}\}\cup
\{\textbf{\emph{u}}\alpha_{1}\textbf{\emph{v}}\alpha_{2}\textbf{\emph{w}}\beta_{3}\textbf{\emph{z}}
,\textbf{\emph{u}}\alpha_{1}\textbf{\emph{v}}\beta_{2}\textbf{\emph{w}}\beta_{3}\textbf{\emph{z}}
,\textbf{\emph{u}}\alpha_{1}\textbf{\emph{v}}\beta_{2}\textbf{\emph{w}}\alpha_{3}\textbf{\emph{z}}\}$. Therefore the size of the intersection is $6(q-1)$.
\item
If $d_{H}(\textbf{\emph{x}},\textbf{\emph{y}})=4$ then \textbf{\emph{x}}$=$\textbf{\emph{t}}$\alpha_{1}$\textbf{\emph{u}}$\alpha_{2}$\textbf{\emph{v}}$\alpha_{3}$\textbf{\emph{w}}$\alpha_{4}$\textbf{\emph{z}} and 
\textbf{\emph{y}}$=$\textbf{\emph{t}}$\beta_{1}$\textbf{\emph{u}}$\beta_{2}$\textbf{\emph{v}}$\beta_{3}$\textbf{\emph{w}}$\beta_{4}$\textbf{\emph{z}}  where $\alpha_{i}\neq\beta_{i}$. Let \textbf{\emph{a}} $\in \mathcal{B}_{2, 0, 0}(\textbf{\emph{x}})\cap \mathcal{B}_{2, 0, 0}(\textbf{\emph{y}})$. Consider the indices of four symbols where \textbf{\emph{x}} and \textbf{\emph{y}} are distinct. In sequence \textbf{\emph{a}}, two of these symbols are chosen from \textbf{\emph{x}} and the other two from \textbf{\emph{y}}. Then  $\mathcal{B}_{2, 0, 0}(\textbf{\emph{x}})\cap \mathcal{B}_{2, 0, 0}(\textbf{\emph{y}})=\{\textbf{\emph{t}}\alpha_{1}\textbf{\emph{u}}\alpha_{2}\textbf{\emph{v}}\beta_{3}\textbf{\emph{w}}\beta_{4}\textbf{\emph{z}},
\textbf{\emph{t}}\alpha_{1}\textbf{\emph{u}}\beta_{2}\textbf{\emph{v}}\alpha_{3}\textbf{\emph{w}}\beta_{4}\textbf{\emph{z}},\\
\textbf{\emph{t}}\alpha_{1}\textbf{\emph{u}}\beta_{2}\textbf{\emph{v}}\beta_{3}\textbf{\emph{w}}\alpha_{4}\textbf{\emph{z}}
,\textbf{\emph{t}}\beta_{1}\textbf{\emph{u}}\alpha_{2}\textbf{\emph{v}}\alpha_{3}\textbf{\emph{w}}\beta_{4}\textbf{\emph{z}},\\
\textbf{\emph{t}}\beta_{1}\textbf{\emph{u}}\alpha_{2}\textbf{\emph{v}}\beta_{3}\textbf{\emph{w}}\alpha_{4}\textbf{\emph{z}}
,\textbf{\emph{t}}\beta_{1}\textbf{\emph{u}}\beta_{2}\textbf{\emph{v}}\alpha_{3}\textbf{\emph{w}}\alpha_{4}\textbf{\emph{z}}\}$.
\item
If $d_{H}(\textbf{\emph{x}},\textbf{\emph{y}})\geq5$ then $\mathcal{B}_{2, 0, 0}(\textbf{\emph{x}})\cap \mathcal{B}_{2, 0, 0}(\textbf{\emph{y}})=\emptyset$.
\end{itemize}
 \end{IEEEproof}
\begin{Lemma}
 Let $\boldsymbol{\omega}\in \Sigma^{n}_{q}$.
   Then the size of the intersection of double-substitution balls of three, four or five 1-subsequences is
 \begin{description}
  \item[(i)] $|\mathcal{B}_{2, 0, 0}(\boldsymbol{\omega}^{(i)})\cap \mathcal{B}_{2, 0, 0}(\boldsymbol{\omega}^{(i+1)})\cap \mathcal{B}_{2, 0, 0}(\boldsymbol{\omega}^{(i+2)})|=q^{2}+(n-3)(q-1)$.
  \item[(ii)] $|\mathcal{B}_{2, 0, 0}(\boldsymbol{\omega}^{(i)})\cap \mathcal{B}_{2, 0, 0}(\boldsymbol{\omega}^{(i+1)})\cap \mathcal{B}_{2, 0, 0}(\boldsymbol{\omega}^{(i+3)})|=|\mathcal{B}_{2, 0, 0}(\boldsymbol{\omega}^{(i)})\cap \mathcal{B}_{2, 0, 0}(\boldsymbol{\omega}^{(i+2)})\cap \mathcal{B}_{2, 0, 0}(\boldsymbol{\omega}^{(i+3)})|=4q-3$.
  \item[(iii)] $|\mathcal{B}_{2, 0, 0}(\boldsymbol{\omega}^{(i)})\cap \mathcal{B}_{2, 0, 0}(\boldsymbol{\omega}^{(i+1)})\cap \mathcal{B}_{2, 0, 0}(\boldsymbol{\omega}^{(i+4)})|=|\mathcal{B}_{2, 0, 0}(\boldsymbol{\omega}^{(i)})\cap \mathcal{B}_{2, 0, 0}(\boldsymbol{\omega}^{(i+3)})\cap \mathcal{B}_{2, 0, 0}(\boldsymbol{\omega}^{(i+4)})|=3$.
  \item[(iv)] $|\mathcal{B}_{2, 0, 0}(\boldsymbol{\omega}^{(i)})\cap \mathcal{B}_{2, 0, 0}(\boldsymbol{\omega}^{(i+2)})\cap \mathcal{B}_{2, 0, 0}(\boldsymbol{\omega}^{(i+4)})|=5$.
  \item[(v)] $|\mathcal{B}_{2, 0, 0}(\boldsymbol{\omega}^{(i)})\cap \mathcal{B}_{2, 0, 0}(\boldsymbol{\omega}^{(i+1)})\cap \mathcal{B}_{2, 0, 0}(\boldsymbol{\omega}^{(i+2)})\cap \mathcal{B}_{2, 0, 0}(\boldsymbol{\omega}^{(i+3)})|=3q-2$.
  \item[(vi)] $|\mathcal{B}_{2, 0, 0}(\boldsymbol{\omega}^{(i)})\cap \mathcal{B}_{2, 0, 0}(\boldsymbol{\omega}^{(i+2)})\cap \mathcal{B}_{2, 0, 0}(\boldsymbol{\omega}^{(i+3)})\cap \mathcal{B}_{2, 0, 0}(\boldsymbol{\omega}^{(i+4)})|=3$.
  \item[(vii)] $|\mathcal{B}_{2, 0, 0}(\boldsymbol{\omega}^{(i)})\cap \mathcal{B}_{2, 0, 0}(\boldsymbol{\omega}^{(i+1)})\cap \mathcal{B}_{2, 0, 0}(\boldsymbol{\omega}^{(i+3)})\cap \mathcal{B}_{2, 0, 0}(\boldsymbol{\omega}^{(i+4)})|=2$.
  \item[(viii)] $|\bigcap_{j=0}^{4}\mathcal{B}_{2, 0, 0}(\boldsymbol{\omega}^{(i+j)})|=2$.
 \end{description}
\end{Lemma}
\begin{IEEEproof}
The proof is given in appendix C.
\end{IEEEproof}
\section{ The Single-Insertion Single-Substitution Ball Size}
Recently the authors in [8] derived the error ball size and the maximum intersection of two error balls for single-insertion single-substitution channel for the binary alphabet. In this section, we generalize their work to the non-binary alphabet. The \emph{runs-profile vector} for a sequence $\boldsymbol{\omega}\in \Sigma_{q}^{n}$ with $\rho(\boldsymbol{\omega})$ runs is the vector $\ell(\boldsymbol{\omega})=(\ell_{1},\ell_{2},\ldots,\ell_{\rho(\boldsymbol{\omega})})$, which determines the length of each of the runs of $\boldsymbol{\omega}$.
\begin{Definition}
  Let $\boldsymbol{\omega}=\sigma_{1}^{\ell_{1}}\ldots\sigma_{\rho(\boldsymbol{\omega})}^{\ell_{\rho(\boldsymbol{\omega})}}\in\Sigma_{q}^{n}$ be a sequence. Define $\boldsymbol{\omega}^{((i-1)(q-1)+j)}$, $i\in [1, n]$, $j\in\{1, \ldots,q-1\}$ as the sequence created by inserting $\tilde{\omega}_{i}$ after $\omega_{i}$ in $\boldsymbol{\omega}$ where $\tilde{\omega}_{i}\neq\omega_{i}$. Specifically, define $\boldsymbol{\omega}^{((\sum_{k=1}^{h-1}\ell_{k})(q-1))}=\sigma_{1}^{\ell_{1}}\ldots \sigma_{h}^{\ell_{h}+1}\ldots\sigma_{\rho(\boldsymbol{\omega})}^{\ell_{\rho(\boldsymbol{\omega})}}$.
   Also, let $\boldsymbol{\omega}^{(0)}$ be the sequence generated by inserting $\omega_{1}$ and $\boldsymbol{\omega}^{(-1)}$, $\ldots$, $\boldsymbol{\omega}^{(-(q-1))}$  be the sequences generated by inserting $\tilde{\omega}_{1}\in \Sigma_{q}\setminus \{\omega_{1}\}$ at the beginning of $\boldsymbol{\omega}$.
\end{Definition}
\begin{Lemma}
Suppose that the symbol with index $i$ falls in the $h$-th run and the symbol with index $l$, $(l>h)$, falls in the $h+t$-th run $(t\geq0)$. The Hamming distance between two elements of $\mathcal{I}_{1}(\boldsymbol{\omega})$ is
\begin{enumerate}
  \item  $d_{H}(\boldsymbol{\omega}^{(0)},\boldsymbol{\omega}^{(-i)})=d_{H}(\boldsymbol{\omega}^{(-i)},\boldsymbol{\omega}^{(-k)})=d_{H}(\boldsymbol{\omega}^{(i-1)(q-1)+j},\boldsymbol{\omega}^{(i-1)(q-1)+j^{\prime}})=1$.
  \item $d_{H}(\boldsymbol{\omega}^{(0)},\boldsymbol{\omega}^{((i-1)(q-1)+j)})=h$.
  \item $d_{H}(\boldsymbol{\omega}^{(-k)},\boldsymbol{\omega}^{((i-1)(q-1)+j)})=h+1$.
  \item  Let $i=\sum_{k=1}^{h}\ell_{k}$. If $j=q-1$ then $d_{H}(\boldsymbol{\omega}^{(i(q-1))},\boldsymbol{\omega}^{((l-1)(q-1)+j^{\prime})})=t$ and if $j\neq q-1$ then $d_{H}(\boldsymbol{\omega}^{((i-1)(q-1)+j)},\boldsymbol{\omega}^{((l-1)(q-1)+j^{\prime})})=t+1$.
  \item If $\sum_{k=1}^{h-1}\ell_{k}+1\leq i<\sum_{j=1}^{h}\ell_{k}$ then  $d_{H}(\boldsymbol{\omega}^{((i-1)(q-1)+j)},\boldsymbol{\omega}^{((l-1)(q-1)+j^{\prime})})=t+2$.
\end{enumerate}
\end{Lemma}
\begin{IEEEproof}
\begin{enumerate}
  \item Clearly the two sequences differ only in one symbol.
  \item Let $\boldsymbol{\omega}^{((i-1)(q-1)+j)}=\sigma_{1}^{\ell_{1}}\ldots\sigma_{h}^{a}\alpha\sigma_{h}^{\ell_{h}-a}\sigma_{\rho(\boldsymbol{\omega})}^{\ell_{\rho(\boldsymbol{\omega})}}$ where $\alpha\neq \sigma_{h}$. By comparing the symbols of $\boldsymbol{\omega}^{((i-1)(q-1)+j)}$ and $\boldsymbol{\omega}^{(0)}$ we realize that the two sequences are different in indices $\{\ell_{1}+1,\ell_{1}+\ell_{2}+1,\ldots,\ell_{1}+\ldots+\ell_{h-1}+1,\ell_{1}+\ldots+\ell_{h-1}+a+1\}$.
  \item By comparing the symbols of $\boldsymbol{\omega}^{((i-1)(q-1)+j)}$ and $\boldsymbol{\omega}^{(-k)}$, we realize that the two sequences are different in indices $\{1, \ell_{1}+1,\ell_{1}+\ell_{2}+1,\ldots,\ell_{1}+\ldots+\ell_{h-1}+1,\ell_{1}+\ldots+\ell_{h-1}+a+1\}$.
  \item Let $j=q-1$. Based on definition 8, $\boldsymbol{\omega}^{(i(q-1))}=\sigma_{1}^{\ell_{1}}\ldots\sigma_{h}^{\ell_{h}}\sigma_{h+1}^{\ell_{h+1}+1}\ldots\sigma_{\rho(\boldsymbol{\omega})}^{\ell_{\rho(\boldsymbol{\omega})}}$. Let $\boldsymbol{\omega}^{((l-1)(q-1)+j^{\prime})})=\sigma_{1}^{\ell_{1}}\ldots\sigma_{h+t}^{b}\beta\sigma_{h+t}^{\ell_{h+t}-b}\ldots\sigma_{\rho(\boldsymbol{\omega})}^{\ell_{\rho(\boldsymbol{\omega})}}$ where $\beta\neq\sigma_{h+t}$ and $l=\sum_{k=1}^{h+t-1}\ell_{k}+b$. By comparing the symbols of the two sequences we realize that they are different in indices $\{\sum_{k=1}^{h+1}\ell_{k}+1,\ldots,\sum_{k=1}^{h+t-1}\ell_{k}+1,\sum_{k=1}^{h+t-1}\ell_{k}+b+1\}$. If $j\neq q-1$ then the indices of symbols in which two sequences are different are $\{\sum_{k=1}^{h}\ell_{k}+1,\sum_{k=1}^{h+1}\ell_{k}+1,\ldots,\sum_{k=1}^{h+t-1}\ell_{k}+1,\sum_{k=1}^{h+t-1}\ell_{k}+b+1\}$.
  \item Let $\boldsymbol{\omega}^{((i-1)(q-1)+j)}=\sigma_{1}^{\ell_{1}}\ldots\sigma_{h}^{a}\alpha\sigma_{h}^{\ell_{h}-a}\ldots\sigma_{\rho(\boldsymbol{\omega})}^{\ell_{\rho(\boldsymbol{\omega})}}$ and \\
      $\boldsymbol{\omega}^{((l-1)(q-1)+j^{\prime})}=\sigma_{1}^{\ell_{1}}\ldots\sigma_{h+t}^{b}\beta\sigma_{h+t}^{\ell_{h+t}-b}\ldots\sigma_{\rho(\boldsymbol{\omega})}^{\ell_{\rho(\boldsymbol{\omega})}}$ where $1\leq a\leq \ell_{h}$ and $\alpha\neq \sigma_{h}$. By comparing the symbols of the two sequences we realize that they are different in indices $\{\sum_{k=1}^{h-1}\ell_{k}+a+1,\sum_{k=1}^{h}\ell_{k}+1,\sum_{k=1}^{h+1}\ell_{k}+1,\ldots,\sum_{k=1}^{h+t-1}\ell_{k}+1,\sum_{k=1}^{h+t-1}\ell_{k}+b+1\}$.
\end{enumerate}
\end{IEEEproof}
\begin{Theorem}
Let $\boldsymbol{\omega} \in \Sigma^{n}_{q}$ be a sequence with runs-profile vector $\ell(\boldsymbol{\omega})=(\ell_{1},\ell_{2},\ldots,\ell_{\rho(\boldsymbol{\omega})})$. Then, the size of its single-insertion single-substitution ball is given by
\small
\begin{equation*}
|\mathcal{B}_{1, 0, 1}(\boldsymbol{\omega})|=1+(n+1)(q-1)
+\left(n^{2}-\sum_{j=1}^{\rho(\boldsymbol{\omega})}\binom{\ell_{j}}{2}\right)(q-1)^{2}.
\end{equation*}
\normalsize
\end{Theorem}
\begin{IEEEproof}
 The single-insertion single-substitution ball  is
 \begin{equation*}
 \mathcal{B}_{1, 0, 1}(\boldsymbol{\omega})=\bigcup_{\emph{\textbf{z}}\in \mathcal{I}_{1}(\boldsymbol{\omega})} \mathcal{B}_{1, 0, 0}(\emph{\textbf{z}})=\bigcup_{l=-(q-1)}^{n(q-1)}\mathcal{B}_{1, 0, 0}(\boldsymbol{\omega}^{(l)})
  \end{equation*}
  \begin{equation*}
 =\bigcup_{l=-(q-1)}^{n(q-1)}\left(\mathcal{B}_{1, 0, 0}(\boldsymbol{\omega}^{(l)})\setminus \bigcup_{k=-(q-1)}^{l-1}\mathcal{B}_{1, 0, 0}(\boldsymbol{\omega}^{(k)})\right).
 \end{equation*}
 Hence, $|\mathcal{B}_{1, 0, 1}(\boldsymbol{\omega})|$ is equal to the following,
  \small
 \begin{equation*}
    |\mathcal{B}_{1, 0, 1}(\boldsymbol{\omega})|= 
\sum_{l=-(q-1)}^{n(q-1)}\left|\mathcal{B}_{1, 0, 0}(\boldsymbol{\omega}^{(l)})\setminus \bigcup_{k=-(q-1)}^{l-1}\mathcal{B}_{1, 0,0}(\boldsymbol{\omega}^{(k)})\right|
\end{equation*}
 \begin{equation*}
 =\sum_{l=-(q-1)}^{n(q-1)}|\mathcal{B}_{1, 0, 0}(\boldsymbol{\omega}^{(l)})|-
 \left|\mathcal{B}_{1, 0, 0}(\boldsymbol{\omega}^{(l)})\cap \bigcup_{k=-(q-1)}^{l-1}\mathcal{B}_{1, 0, 0}(\boldsymbol{\omega}^{(k)})\right|.
 \end{equation*}
 \normalsize
For any sequence $\emph{\textbf{z}}\in \Sigma_{q}^{n+1}$, $|\mathcal{B}_{1, 0, 0}(\emph{\textbf{z}})|=1+(n+1)(q-1)$ and based on Lemma 13, for $-q+2\leq l \leq 0$ we have $|\mathcal{B}_{1, 0, 0}(\boldsymbol{\omega}^{(l)})\cap \bigcup_{k=-(q-1)}^{l-1}\mathcal{B}_{1, 0, 0}(\boldsymbol{\omega}^{(k)})|=q$. Also, for $0<l\leq n(q-1)$, let $l=(i-1)(q-1)+j^{\prime}$
 and the $i$-th symbol is the $j$-th one in the $h$-run. Then, $|\mathcal{B}_{1, 0, 0}(\boldsymbol{\omega}^{(l)})\cap \bigcup_{k=-(q-1)}^{l-1}\mathcal{B}_{1, 0, 0}(\boldsymbol{\omega}^{(k)})|=j(q-1)+q$.
 Thus,
\[|\mathcal{B}_{1, 0, 1}(\boldsymbol{\omega})|=(1+(n+1)(q-1))^{2}-q(q-1)-\]
\[\sum_{l=1}^{n(q-1)}\left|\mathcal{B}_{1, 0, 0}(\boldsymbol{\omega}^{(l)})\cap \bigcup_{k=-(q-1)}^{l-1}\mathcal{B}_{1, 0, 0}(\boldsymbol{\omega}^{(k)})\right|\]
\[=1+(2n+1)(q-1)+(n^{2}+2n)(q-1)^{2}-\]
\[(q-1)\sum_{i=1}^{\rho(\boldsymbol{\omega})}\sum_{j=1}^{\ell_{i}}(j(q-1)+q)\]
\[=1+(n+1)(q-1)+(n^{2}-\sum_{i=1}^{\rho(\boldsymbol{\omega})}\binom{\ell_{i}}{2})(q-1)^{2}.\]
\end{IEEEproof}
\begin{Proposition}
The size of the intersection of two single-substitution balls of two sequences, $\emph{\textbf{z}},\tilde{\emph{\textbf{z}}}\in \Sigma_{q}^{n}$ is
\begin{itemize}
  \item If $d_{H}(\emph{\textbf{z}},\tilde{\emph{\textbf{z}}})=1$ then  $|\mathcal{B}_{1, 0, 0}(\emph{\textbf{z}})\cap \mathcal{B}_{1, 0, 0}(\tilde{\emph{\textbf{z}}})|=q$,
  \item If $d_{H}(\emph{\textbf{z}},\tilde{\emph{\textbf{z}}})=2$ then  $|\mathcal{B}_{1, 0, 0}(\emph{\textbf{z}})\cap \mathcal{B}_{1, 0, 0}(\tilde{\emph{\textbf{z}}})|=2$,
  \item If $d_{H}(\emph{\textbf{z}},\tilde{\emph{\textbf{z}}})>2$ then $\mathcal{B}_{1, 0, 0}(\emph{\textbf{z}})\cap \mathcal{B}_{1, 0, 0}(\tilde{\emph{\textbf{z}}})=\emptyset$.
\end{itemize}
\end{Proposition}
\begin{Lemma}
The size of the following intersections are
\begin{itemize}
  \item If $-(q-1)+1\leq i\leq 0$ then $|\mathcal{B}_{1, 0, 0}(\boldsymbol{\omega}^{(i)})\cap \bigcup_{j=-(q-1)}^{i-1}\mathcal{B}_{1, 0, 0}(\boldsymbol{\omega}^{(j)})|=q$.
  \item Let $1\leq i\leq n$ and the $i$-th symbol is the $j$-th one in the $h$-th run. 
  For $k=(i-1)(q-1)+j^{\prime}$ where $j^{\prime}\in \{1,\ldots, q-1\}$ we have $|\mathcal{B}_{1, 0, 0}(\boldsymbol{\omega}^{(k)})\cap \bigcup_{j^{\prime\prime}=-(q-1)}^{k-1}\mathcal{B}_{1, 0, 0}(\boldsymbol{\omega}^{(j^{\prime\prime})})|=j(q-1)+q$.

\end{itemize}
\end{Lemma}
\begin{IEEEproof}
\begin{itemize}
  \item Based on proposition 1, for any two distinct sequences like $\boldsymbol{\omega}^{(i)}$ and $\boldsymbol{\omega}^{(j)}$ such that $-(q-1)\leq i,j\leq 0$ we have $\mathcal{B}_{1, 0, 0}(\boldsymbol{\omega}^{(i)})\cap \mathcal{B}_{1, 0, 0}(\boldsymbol{\omega}^{(j)})=\{\boldsymbol{\omega}^{(0)},\boldsymbol{\omega}^{(-1)},\ldots,\boldsymbol{\omega}^{(-(q-1))}\}$. Thus, $|\mathcal{B}_{1, 0, 0}(\boldsymbol{\omega}^{(i)})\cap \bigcup_{j=-(q-1)}^{i-1}\mathcal{B}_{1, 0, 0}(\boldsymbol{\omega}^{(j)})|=q$.
  \item First suppose that,\\ $\boldsymbol{\omega}^{(k)}=\sigma_{1}^{\ell_{1}}\ldots\sigma_{h}^{j}\alpha\sigma_{h}^{\ell_{h}-j}\ldots\sigma_{\rho(\boldsymbol{\omega})}^{\ell_{\rho(\boldsymbol{\omega})}}$.
  Based on Lemma 1 and Proposition 1, the following claims are true,
  
  1) Let $i\geq 1$. We have $\mathcal{B}_{1, 0, 0}(\boldsymbol{\omega}^{(0)})\cap \mathcal{B}_{1, 0, 0}(\boldsymbol{\omega}^{((i-1)(q-1)+j^{\prime}})\neq \emptyset$ if and only if the $i$-bit falls in the first or second run.
  
  2) Let $i,k\geq 1$. We have $\mathcal{B}_{1, 0, 0}(\boldsymbol{\omega}^{(-k)})\cap \mathcal{B}_{1, 0, 0}(\boldsymbol{\omega}^{((i-1)(q-1)+j^{\prime}})\neq \emptyset$ if and only if the $i$-bit falls in the first run.
  
  3) Let $1\leq l\leq i$ and the $l$-th symbol falls in the $h-t$-run. We have $\mathcal{B}_{1, 0, 0}(\boldsymbol{\omega}^{((i-1)(q-1)+j^{\prime})})\cap\mathcal{B}_{1, 0, 0}(\boldsymbol{\omega}^{((l-1)(q-1)+j^{\prime\prime})})\neq \emptyset$ if and only if $0\leq t\leq 2$. If $t=2$ then $\boldsymbol{\omega}^{((l-1)(q-1)+j^{\prime\prime})}=\sigma_{1}^{\ell_{1}}\ldots\sigma_{h-1}^{\ell_{h-1}+1}\sigma_{h}^{\ell_{h}}\ldots\sigma_{\rho(\boldsymbol{\omega})}^{\ell_{\rho(\boldsymbol{\omega})}}$. If $t=1$ then  $\boldsymbol{\omega}^{((l-1)(q-1)+j^{\prime\prime})}=\sigma_{1}^{\ell_{1}}\ldots\sigma_{h-1}^{\ell_{h-1}}\beta\sigma_{h}^{\ell_{h}}\ldots\sigma_{\rho(\boldsymbol{\omega})}^{\ell_{\rho(\boldsymbol{\omega})}}$ where $\beta\neq\sigma_{h-1}$. 
  
  In conclusion, $\mathcal{B}_{1, 0, 0}(\boldsymbol{\omega}^{((i-1)(q-1)+j^{\prime})})\cap\mathcal{B}_{1, 0, 0}(\boldsymbol{\omega}^{((l-1)(q-1)+j^{\prime\prime})})\neq\emptyset$ if and only if $\boldsymbol{\omega}^{((l-1)(q-1)+j^{\prime\prime})}$ has one of the two forms $\sigma_{1}^{\ell_{1}}\ldots\sigma_{h}^{j}\beta\sigma_{h}^{\ell_{h}-j}\ldots\sigma_{\rho(\boldsymbol{\omega})}^{\ell_{\rho(\boldsymbol{\omega})}}$ where $\beta\neq\alpha$ or $\sigma_{1}^{\ell_{1}}\ldots\sigma_{h}^{k}\beta\sigma_{h}^{\ell_{h}-k}\ldots\sigma_{\rho(\boldsymbol{\omega})}^{\ell_{\rho(\boldsymbol{\omega})}}$ where $0\leq k<j$. If $k>0$ then $\beta\neq \sigma_{h}$. First note that $\mathcal{B}_{1, 0, 0}(\sigma_{1}^{\ell_{1}}\ldots\sigma_{h}^{j}\alpha\sigma_{h}^{\ell_{h}-j}\ldots\sigma_{\rho(\boldsymbol{\omega})}^{\ell_{\rho(\boldsymbol{\omega})}})\cap\mathcal{B}_{1, 0, 0}(\sigma_{1}^{\ell_{1}}\ldots\sigma_{h}^{j}\beta\sigma_{h}^{\ell_{h}-j}\ldots\sigma_{\rho(\boldsymbol{\omega})}^{\ell_{\rho(\boldsymbol{\omega})}})=\mathcal{B}_{1, 0, 0}(\sigma_{1}^{\ell_{1}}\ldots\sigma_{h}^{j}\alpha\sigma_{h}^{\ell_{h}-j}\ldots\sigma_{\rho(\boldsymbol{\omega})}^{\ell_{\rho(\boldsymbol{\omega})}})\cap\mathcal{B}_{1, 0, 0}(\sigma_{1}^{\ell_{1}}\ldots\sigma_{h}^{\ell_{h}+1}\ldots\sigma_{\rho(\boldsymbol{\omega})}^{\ell_{\rho(\boldsymbol{\omega})}})$. The size of these intersections is $q$. On the other hand, $\mathcal{B}_{1, 0, 0}(\sigma_{1}^{\ell_{1}}\ldots\sigma_{h}^{j}\alpha\sigma_{h}^{\ell_{h}-j}\ldots\sigma_{\rho(\boldsymbol{\omega})}^{\ell_{\rho(\boldsymbol{\omega})}})\cap\mathcal{B}_{1, 0, 0}(\sigma_{1}^{\ell_{1}}\ldots\sigma_{h}^{k}\beta\sigma_{h}^{\ell_{h}-k}\ldots\sigma_{\rho(\boldsymbol{\omega})}^{\ell_{\rho(\boldsymbol{\omega})}})=\{\sigma_{1}^{\ell_{1}}\ldots\sigma_{h}^{\ell_{h}+1}\ldots\sigma_{\rho(\boldsymbol{\omega})}^{\ell_{\rho(\boldsymbol{\omega})}},\\
  \sigma_{1}^{\ell_{1}}\ldots\sigma_{h}^{k}\beta\sigma_{h}^{j-k-1}\alpha\sigma_{h}^{\ell_{h}-j}\ldots\sigma_{\rho(\boldsymbol{\omega})}^{\ell_{\rho(\boldsymbol{\omega})}}\}$. Since $0\leq k\leq j-1$ and $\beta\neq \sigma_{h}$, there are $j(q-1)$ intersections and each intersection give us a new sequence. In conclusion, $|\mathcal{B}_{1, 0, 0}(\boldsymbol{\omega}^{(k)})\cap \bigcup_{j^{\prime\prime}=-(q-1)}^{k-1}\mathcal{B}_{1, 0, 0}(\boldsymbol{\omega}^{(j^{\prime\prime})})|=j(q-1)+q$. 
\end{itemize}
\end{IEEEproof}

\section{ Concluding Remarks }
In this paper, we investigated the size of error balls for channels with three edits. We first derived the size of the single-deletion double-insertion ball and then extended the results to single-deletion double-substitution and single-insertion single-substitution. Our results for single-insertion single-substitution and single-deletion double-substitution balls confirms the result given in [8] for binary sequences. The size of the error balls are also derived by computer search and confirm the results presented in this paper.

 Although we made some progress in studying the size of error balls for channels with multiple types of errors, several problems remain open. In particular, finding the maximum and minimum size of the error balls for various combinations of errors is of great interest. It seems that the minimum size for all balls occurs when the sequence is constant, but finding the maximum size of the error balls is more challenging. We leave the derivation of the size of the error ball for other channels with triple edits, such as single-substitution double-deletion, single-substitution double-insertion, single-insertion double-deletion, and single-substitution single-insertion single-deletion, as our future work.

\appendices
\section{Proof of Theorem 1}
Let \emph{\textbf{x}}$=$\emph{\textbf{uvw}} and \emph{\textbf{y}}$=$\emph{\textbf{uv$^{'}$w}} be Type-B-confusable sequences. If $|\emph{\textbf{v}}|=2$ then assume that  \emph{\textbf{x}}$=$\emph{\textbf{u}}$\alpha\beta$\emph{\textbf{w}} and \emph{\textbf{y}}$=$\emph{\textbf{u}}$\beta\gamma$\emph{\textbf{w}} where $\alpha\neq\beta\neq\gamma$. In this case, $\mathcal{I}_{1}(\emph{\textbf{x}})\cap \mathcal{I}_{1}(\emph{\textbf{y}})=\{\emph{\textbf{u}}\alpha\beta\gamma\emph{\textbf{w}}\}$ and $\mathcal{D}_{1}($\emph{\textbf{x}}$)\cap \mathcal{D}_{1}($\emph{\textbf{y}})$=\{$\emph{\textbf{u}}$\beta$\emph{\textbf{w}}$\}$. If $|\emph{\textbf{v}}|\geq 3$ then let \emph{\textbf{x}}$=$\emph{\textbf{u}}$\alpha\beta$\emph{\textbf{z}}$\gamma$\emph{\textbf{w}} and \emph{\textbf{y}}$=$\emph{\textbf{u}}$\beta$\emph{\textbf{z}}$\gamma\delta$\emph{\textbf{w}} where $\alpha\neq\beta$ and $\gamma\neq\delta$. In this case, $\mathcal{I}_{1}($\emph{\textbf{x}}$)\cap \mathcal{I}_{1}($\emph{\textbf{y}}$)=\{$\emph{\textbf{u}}$\alpha\beta$\emph{\textbf{z}}$\gamma\delta$\emph{\textbf{w}}$\}$ and $\mathcal{D}_{1}($\emph{\textbf{x}}$)\cap \mathcal{D}_{1}($\emph{\textbf{y}}$)=\{$\emph{\textbf{u}}$\beta$\emph{\textbf{z}}$\gamma$\emph{\textbf{w}}$\}$.

Conversely, suppose that $|\mathcal{I}_{1}(\emph{\textbf{x}})\cap \mathcal{I}_{1}(\emph{\textbf{y}})|=|\mathcal{D}_{1}(\emph{\textbf{x}})\cap \mathcal{D}_{1}(\emph{\textbf{y}})|=1$ where \emph{\textbf{x}}, \emph{\textbf{y}}$\in\Sigma_{q}^{n}$. Therefore \emph{\textbf{x}} and \emph{\textbf{y}} are not Type-A-confusable and $d_{H}(\emph{\textbf{x}},\emph{\textbf{y}})\geq 2$. We can assume that \emph{\textbf{x}}$=$\emph{\textbf{u}}$\alpha$\emph{\textbf{d}}$\gamma$\emph{\textbf{w}}, \emph{\textbf{y}}$=$\emph{\textbf{u}}$\beta$\emph{\textbf{e}}$\delta$\emph{\textbf{w}} where $\alpha, \beta, \gamma, \delta\in \Sigma_{q}$ and $\alpha\neq\beta, \gamma\neq\delta$. 

Suppose that \emph{\textbf{d}}$=$\emph{\textbf{e}}$=\emptyset$. If $\alpha=\gamma$ or $\beta=\delta$ then $|\mathcal{I}_{1}(\emph{\textbf{x}})\cap \mathcal{I}_{1}(\emph{\textbf{y}})|=|\mathcal{D}_{1}(\emph{\textbf{x}})\cap \mathcal{D}_{1}(\emph{\textbf{y}})|=0$ which is a contradiction. If $\alpha=\delta$ and $\beta=\gamma$ then \emph{\textbf{x}} and \emph{\textbf{y}} are Type-A-confusable which is a contradiction. If $\alpha\neq\delta$ and $\beta\neq\gamma$ then $|\mathcal{I}_{1}(\emph{\textbf{x}})\cap \mathcal{I}_{1}(\emph{\textbf{y}})|=|\mathcal{D}_{1}(\emph{\textbf{x}})\cap \mathcal{D}_{1}(\emph{\textbf{y}})|=0$ which is a contradiction. If $\alpha\neq\delta$ and $\beta=\gamma$ then \emph{\textbf{x}} and \emph{\textbf{y}} are Type-B-confusable. If $\beta\neq\gamma$ and $\alpha=\delta$ then \emph{\textbf{x}} and \emph{\textbf{y}} are Type-B-confusable.

Now suppose that  \emph{\textbf{d}} and  \emph{\textbf{e}} are nonempty. Let \emph{\textbf{z}}$=z_{1}z_{2}\ldots z_{n+1}$ and  \emph{\textbf{s}}$=s_{1}s_{2}\ldots s_{n-1}$ such that $\{$\emph{\textbf{z}}$\}=\mathcal{I}_{1}(\emph{\textbf{x}})\cap \mathcal{I}_{1}(\emph{\textbf{y}})$ and $\{$\emph{\textbf{s}}$\}=\mathcal{D}_{1}(\emph{\textbf{x}})\cap \mathcal{D}_{1}(\emph{\textbf{y}})$. Matching positions of symbols in \emph{\textbf{x}} and \emph{\textbf{y}} imply that the insertion or deletion cannot occur in subsequences \emph{\textbf{u}} and \emph{\textbf{w}}. 
 In sequence \emph{\textbf{z}}, if the symbol after the end of \emph{\textbf{u}} is $\alpha$ and the symbol before the beginning of \emph{\textbf{w}} is $\delta$ then \emph{\textbf{z}}$=$\emph{\textbf{u}}$\alpha\beta$\emph{\textbf{e}}$\delta$\emph{\textbf{w}}$=$\emph{\textbf{u}}$\alpha$\emph{\textbf{d}}$\gamma\delta$\emph{\textbf{w}}.
If sequence \emph{\textbf{s}} is derived by deleting the symbol $\alpha$ from \emph{\textbf{x}} and $\delta$ from \emph{\textbf{y}} then \emph{\textbf{s}}$=$\emph{\textbf{ud}}$\gamma$\emph{\textbf{w}}$=$\emph{\textbf{u}}$\beta$\emph{\textbf{ew}}.
  In both cases, \emph{\textbf{d}}$\gamma=\beta$\emph{\textbf{e}}, and there exist a sequence like \emph{\textbf{v}} such that  \emph{\textbf{d}}=$\beta$\emph{\textbf{v}} and \emph{\textbf{e}}$=$\emph{\textbf{v}}$\gamma$. Therefore, \emph{\textbf{x}}$=$\emph{\textbf{u}}$\alpha\beta$\emph{\textbf{v}}$\gamma$\emph{\textbf{w}} and \emph{\textbf{y}}$=$\emph{\textbf{u}}$\beta$\emph{\textbf{v}}$\gamma\delta$\emph{\textbf{w}}. 
 
 In sequence \emph{\textbf{z}}, if the symbol after the end of \emph{\textbf{u}} is $\beta$ and the symbol before the beginning of \emph{\textbf{w}} is $\gamma$ then \emph{\textbf{z}}$=$\emph{\textbf{u}}$\beta\alpha$\emph{\textbf{d}}$\gamma$\emph{\textbf{w}}$=$\emph{\textbf{u}}$\beta$\emph{\textbf{e}}$\delta\gamma$\emph{\textbf{w}}.
 If sequence \emph{\textbf{s}} is derived by deleting the symbol $\gamma$ from \emph{\textbf{x}} and $\beta$ from \emph{\textbf{y}} then \emph{\textbf{s}}$=$\emph{\textbf{u}}$\alpha$\emph{\textbf{dw}}$=$\emph{\textbf{ue}}$\delta$\emph{\textbf{w}}.
  In both cases, \emph{\textbf{e}}$\delta=\alpha$\emph{\textbf{d}}, and there exist a sequence like \emph{\textbf{v}} such that  \emph{\textbf{d}}=\emph{\textbf{v}}$\delta$ and \emph{\textbf{e}}$=\alpha$\emph{\textbf{v}}. Therefore, \emph{\textbf{x}}$=$\emph{\textbf{u}}$\alpha$\emph{\textbf{v}}$\delta\gamma$\emph{\textbf{w}} and \emph{\textbf{y}}$=$\emph{\textbf{u}}$\beta\alpha$\emph{\textbf{v}}$\delta$\emph{\textbf{w}}. In conclusion,  \emph{\textbf{x}} and  \emph{\textbf{y}} are always Type-B-confusable. 
\section{Proof of Lemma 5}
In order to prove the first part of the lemma, we suppose that $\mathcal{I}_{1}(\boldsymbol{\omega}^{(i)})\cap \mathcal{I}_{1}(\boldsymbol{\omega}^{(j)})=\{\boldsymbol{\omega}, \emph{\textbf{z}}\}$ and $\mathcal{I}_{1}(\boldsymbol{\omega}^{(j)})\cap \mathcal{I}_{1}(\boldsymbol{\omega}^{(k)})=\{\boldsymbol{\omega}, \emph{\textbf{z}}^{'}\}$. Then according to Lemma 3 we have $\mathcal{I}_{2}(\boldsymbol{\omega}^{(i)})\cap \mathcal{I}_{2}(\boldsymbol{\omega}^{(j)})\cap \mathcal{I}_{2}(\boldsymbol{\omega}^{(k)})=(\mathcal{I}_{1}(\boldsymbol{\omega})\cup\mathcal{I}_{1}(\emph{\textbf{z}}))\cap (\mathcal{I}_{1}(\boldsymbol{\omega})\cup\mathcal{I}_{1}(\emph{\textbf{z}}^{'}))=\mathcal{I}_{1}(\boldsymbol{\omega})\cup(\mathcal{I}_{1}(\emph{\textbf{z}})\cap \mathcal{I}_{1}(\textbf{\emph{z}}^{'}))$. For brevity we define \emph{\textbf{u}}$=\sigma_{1}^{\ell_{1}}\ldots\sigma_{i}^{\ell_{i}-1}$ and \emph{\textbf{w}}$=\sigma_{k}^{\ell_{k}-1}\ldots\sigma_{\rho(\boldsymbol{\omega})}^{\ell_{\rho(\boldsymbol{\omega})}}$. We have four cases

   (i) $d_{H}(\boldsymbol{\omega}^{(i)},\boldsymbol{\omega}^{(j)})=d_{H}(\boldsymbol{\omega}^{(j)},\boldsymbol{\omega}^{(k)})=1$. In this case $\emph{\textbf{z}}= \emph{\textbf{u}}\sigma_{i+1}\sigma_{i}\sigma_{i+1}^{\ell_{i+1}-1}\ldots\sigma_{\rho(\boldsymbol{\omega})}^{\ell_{\rho(\boldsymbol{\omega})}}$ and $\emph{\textbf{z}}^{'}=\sigma_{1}^{\ell_{1}}\ldots\sigma_{i}^{\ell_{i}}\sigma_{i+1}^{\ell_{i+1}-1}\sigma_{i+2}\sigma_{i+1}\sigma_{i+2}^{\ell_{i+2}-1}\ldots\sigma_{\rho(\boldsymbol{\omega})}^{\ell_{\rho(\boldsymbol{\omega})}}$. Therefore we have $\mathcal{I}_{1}(\emph{\textbf{z}})\cap \mathcal{I}_{1}(\emph{\textbf{z}}^{'})=\{\emph{\textbf{u}}\sigma_{i+1}\sigma_{i}\sigma_{i+1}^{\ell_{i+1}-1}\sigma_{i+2}\sigma_{i+1}\emph{\textbf{w}}\}.$
   
  Note that $\emph{\textbf{u}}\sigma_{i+1}\sigma_{i}\sigma_{i+1}^{\ell_{i+1}-1}\sigma_{i+2}\sigma_{i+1}\emph{\textbf{w}}\notin \mathcal{I}_{1}(\boldsymbol{\omega})$ since $\boldsymbol{\omega}\notin \mathcal{D}_{1}(\emph{\textbf{u}}\sigma_{i+1}\sigma_{i}\sigma_{i+1}^{\ell_{i+1}-1}\sigma_{i+2}\sigma_{i+1}\emph{\textbf{w}})$.
   
  (ii) $d_{H}(\boldsymbol{\omega}^{(i)},\boldsymbol{\omega}^{(j)})=1$, $\boldsymbol{\omega}^{(j)}$ and $\boldsymbol{\omega}^{(k)}$ are Type-A-confusable. Thus, 
  
  $\boldsymbol{\omega}=\sigma_{1}^{\ell_{1}}\ldots\sigma_{i}^{\ell_{i}}\sigma_{i+1}^{\ell_{i+1}}\sigma_{i+2}\ldots\sigma_{k-1}\sigma_{k}^{\ell_{k}}\ldots\sigma_{\rho(\boldsymbol{\omega})}^{\ell_{\rho(\boldsymbol{\omega})}}$. We have 
  
  $\emph{\textbf{z}}=\sigma_{1}^{\ell_{1}}\ldots\sigma_{i}^{\ell_{i}-1}\sigma_{i+1}\sigma_{i}\sigma_{i+1}^{\ell_{i+1}-1}\sigma_{i+2}\ldots\sigma_{\rho(\omega)}^{\ell_{\rho(\omega)}}$ and 
      
  $\emph{\textbf{z}}^{'}=\sigma_{1}^{\ell_{1}}\ldots\sigma_{i}^{\ell_{i}}\sigma_{i+1}^{\ell_{i+1}-1}\sigma_{i+2}\ldots\sigma_{k}\sigma_{k-1}\emph{\textbf{w}}$. 
  
   If $\ell_{j}=2$ and $\sigma_{l}=\sigma_{l+2}, i\leq l\leq k-2$ then the two sequences \emph{\textbf{z}} and $\emph{\textbf{z}}^{'}$ are Type-A-confusable and  $\mathcal{I}_{1}(\emph{\textbf{z}})\cap \mathcal{I}_{1}(\emph{\textbf{z}}^{'})=\{\emph{\textbf{u}}\sigma_{i+1}\sigma_{i}\sigma_{i+1}\sigma_{i+2}\ldots\sigma_{k}\sigma_{k-1}\emph{\textbf{w}}, 
    \emph{\textbf{u}}\sigma_{i}\sigma_{i+1}\sigma_{i}\ldots\sigma_{k}\emph{\textbf{w}}$\} but $ \emph{\textbf{u}}\sigma_{i}\sigma_{i+1}\sigma_{i}\ldots\sigma_{k}\emph{\textbf{w}}\in \mathcal{I}_{1}(\boldsymbol{\omega})$.
        Otherwise \textbf{z} and $\emph{\textbf{z}}^{'}$ are Type-B-confusable and $\mathcal{I}_{1}(\emph{\textbf{z}})\cap \mathcal{I}_{1}(\emph{\textbf{z}}^{'})=\{\emph{\textbf{u}}\sigma_{i+1}\sigma_{i}\sigma_{i+1}^{\ell_{i+1}-1}\sigma_{i+2}\ldots\sigma_{k}\sigma_{k-1}\emph{\textbf{w}}\}.$
     Note that $\emph{\textbf{u}}\sigma_{i+1}\sigma_{i}\sigma_{i+1}^{\ell_{i+1}-1}\sigma_{i+2}\ldots\sigma_{k}\sigma_{k-1}\emph{\textbf{w}}\notin \mathcal{I}_{1}(\boldsymbol{\omega})$ since $\boldsymbol{\omega}\notin \mathcal{D}_{1}(\emph{\textbf{u}}\sigma_{i+1}\sigma_{i}\sigma_{i+1}^{\ell_{i+1}-1}\sigma_{i+2}\ldots\sigma_{k}\sigma_{k-1}\emph{\textbf{w}}).$
      
   (iii) $\boldsymbol{\omega}^{(i)}$ and $\boldsymbol{\omega}^{(j)}$ are Type-A-confusable and $d_{H}(\boldsymbol{\omega}^{(j)},\boldsymbol{\omega}^{(k)})=1$. Then
    
    $\emph{\textbf{z}}=\emph{\textbf{u}}\sigma_{i+1}\ldots\sigma_{j}\sigma_{j-1}\sigma_{j}^{\ell_{j}-1}\ldots\sigma_{\rho(\omega)}^{\ell_{\rho(\omega)}}$ and
    
    $\emph{\textbf{z}}^{'}=\sigma_{1}^{\ell_{1}}\ldots\sigma_{j}^{\ell_{j}-1}\sigma_{j+1}\sigma_{j}\sigma_{j+1}^{\ell_{j+1}-1}\ldots\sigma_{\rho(\omega)}^{\ell_{\rho(\omega)}}$.
    
  If $\ell_{j}=2$ and  $\sigma_{l}=\sigma_{l+2}, i\leq l\leq k-2$ then the two sequences \emph{\textbf{z}} and $\emph{\textbf{z}}^{'}$ are Type-A-confusable and  $\mathcal{I}_{1}(\emph{\textbf{z}})\cap \mathcal{I}_{1}(\emph{\textbf{z}}^{'})=\{\emph{\textbf{u}}\sigma_{i+1}\sigma_{i}\ldots\sigma_{j}\sigma_{j+1}\sigma_{j}\emph{\textbf{w}},
    \emph{\textbf{u}}\sigma_{i}\sigma_{i+1}\ldots\sigma_{j}\sigma_{j-1}\sigma_{j}\sigma_{j+1}\emph{\textbf{w}}\}$, but $\emph{\textbf{u}}\sigma_{i}\sigma_{i+1}\ldots\sigma_{j}\sigma_{j-1}\sigma_{j}\sigma_{j+1}\emph{\textbf{w}}\in \mathcal{I}_{1}(\boldsymbol{\omega})$. Otherwise they are Type-B-confusable and $\mathcal{I}_{1}(\emph{\textbf{z}})\cap \mathcal{I}_{1}(\emph{\textbf{z}}^{'})=\{\emph{\textbf{u}}\sigma_{i+1}\sigma_{i}\ldots\sigma_{j-1}\sigma_{j}^{\ell_{j}-1}\sigma_{j+1}\sigma_{j}\emph{\textbf{w}}\}$.
   Note that $\emph{\textbf{u}}\sigma_{i+1}\sigma_{i}\ldots\sigma_{j-1}\sigma_{j}^{\ell_{j}-1}\sigma_{j+1}\sigma_{j}\emph{\textbf{w}}\notin \mathcal{I}_{1}(\boldsymbol{\omega})$ since $\boldsymbol{\omega}\notin \mathcal{D}_{1}(\emph{\textbf{u}}\sigma_{i+1}\sigma_{i}\ldots\sigma_{j-1}\sigma_{j}^{\ell_{j}-1}\sigma_{j+1}\sigma_{j}\emph{\textbf{w}})$.
    
  (iv) The two pairs of $\boldsymbol{\omega}^{(i)}$, $\boldsymbol{\omega}^{(j)}$ and  $\boldsymbol{\omega}^{(j)}$, $\boldsymbol{\omega}^{(k)}$ are Type-A-confusable. Therefore 
  
  $\boldsymbol{\omega}=\emph{\textbf{u}}\sigma_{i}\sigma_{i+1}\ldots\sigma_{j-1}\sigma_{j}^{\ell_{j}}\sigma_{j+1}\ldots\sigma_{k-1}\sigma_{k}\emph{\textbf{w}}$. We have
  
   $\emph{\textbf{z}}= \emph{\textbf{u}}\sigma_{i+1}\sigma_{i}\sigma_{i+1}\ldots\sigma_{j-1}\sigma_{j}^{\ell_{j}-1}\sigma_{j+1}\ldots\sigma_{k-1}\sigma_{k}\emph{\textbf{w}}$ and 
   
   $\emph{\textbf{z}}^{'}= \emph{\textbf{u}}\sigma_{i}\sigma_{i+1}\ldots\sigma_{j-1}\sigma_{j}^{\ell_{j}-1}\sigma_{j+1}\ldots\sigma_{k}\sigma_{k-1}\emph{\textbf{w}}$. 
   
 If  $\ell_{j}=2$ and $\sigma_{l}=\sigma_{l+2}, i\leq l\leq k-2$ then the two sequences \emph{\textbf{z}} and $\emph{\textbf{z}}^{'}$ are Type-A-confusable and  $\mathcal{I}_{1}(\emph{\textbf{z}})\cap \mathcal{I}_{1}(\emph{\textbf{z}}^{'})=\{\emph{\textbf{u}}\sigma_{i+1}\sigma_{i}\ldots\sigma_{j}\ldots\sigma_{k}\sigma_{k-1}\emph{\textbf{w}},
   \emph{\textbf{u}}\sigma_{i}\ldots\sigma_{k}\sigma_{k-1}\sigma_{k}\emph{\textbf{w}}\}$ but $ \emph{\textbf{u}}\sigma_{i}\ldots\sigma_{k}\sigma_{k-1}\sigma_{k}\emph{\textbf{w}}\in \mathcal{I}_{1}(\boldsymbol{\omega})$.
  
   Otherwise \emph{\textbf{z}} and $\emph{\textbf{z}}^{'}$ are Type-B-confusable and 
   $\mathcal{I}_{1}(\emph{\textbf{z}})\cap \mathcal{I}_{1}(\emph{\textbf{z}}^{'})= \{ \emph{\textbf{u}}\sigma_{i+1}\sigma_{i}\sigma_{i+1}\ldots\sigma_{j-1}\sigma_{j}^{\ell_{j}-1}\sigma_{j+1}\ldots\sigma_{k}\sigma_{k-1}\emph{\textbf{w}}\}$. Note that $
\emph{\textbf{u}}\sigma_{i+1}\sigma_{i}\sigma_{i+1}\ldots\sigma_{j-1}\sigma_{j}^{\ell_{j}-1}\sigma_{j+1}\ldots\sigma_{k}\sigma_{k-1}\emph{\textbf{w}}\notin \mathcal{I}_{1}(\boldsymbol{\omega})$ since $\boldsymbol{\omega}\notin \mathcal{D}_{1}(\emph{\textbf{u}}\sigma_{i+1}\sigma_{i}\sigma_{i+1}\ldots\sigma_{j-1}\sigma_{j}^{\ell_{j}-1}\sigma_{j+1}\ldots\sigma_{k}\sigma_{k-1}\emph{\textbf{w}})$.

In order to complete the proof, it is sufficient to show that if at least one of the conditions $|\mathcal{I}_{1}(\boldsymbol{\omega}^{(i)})\cap \mathcal{I}_{1}(\boldsymbol{\omega}^{(j)})|=2$ or $|\mathcal{I}_{1}(\boldsymbol{\omega}^{(j)})\cap \mathcal{I}_{1}(\boldsymbol{\omega}^{(k)})|=2$ does not satisfy, then $\mathcal{I}_{2}(\boldsymbol{\omega}^{(i)})\cap \mathcal{I}_{2}(\boldsymbol{\omega}^{(j)})\cap \mathcal{I}_{2}(\boldsymbol{\omega}^{(k)})=\mathcal{I}_{1}(\boldsymbol{\omega})$.
We have three cases which we consider below:
\begin{itemize}
  \item Let  $\mathcal{I}_{1}(\boldsymbol{\omega}^{(i)})\cap \mathcal{I}_{1}(\boldsymbol{\omega}^{(j)})=\{\boldsymbol{\omega}\}$ and $\mathcal{I}_{1}(\boldsymbol{\omega}^{(j)})\cap \mathcal{I}_{1}(\boldsymbol{\omega}^{(k)})=\{\boldsymbol{\omega}, \emph{\textbf{z}}\}$. Therefore $\mathcal{I}_{2}(\boldsymbol{\omega}^{(j)})\cap \mathcal{I}_{2}(\boldsymbol{\omega}^{(k)})= \mathcal{I}_{1}(\boldsymbol{\omega})\cup \mathcal{I}_{1}(\emph{\textbf{z}})$. If $d_{H}(\boldsymbol{\omega}^{(j)},\boldsymbol{\omega}^{(k)})=1$ then $\emph{\textbf{z}}=\sigma_{1}^{\ell_{1}}\ldots\sigma_{j}^{\ell_{j}-1}\sigma_{j+1}\sigma_{j}\sigma_{j+1}^{\ell_{j+1}-1}\ldots\sigma_{\rho(\boldsymbol{\omega})}^{\ell_{\rho(\boldsymbol{\omega})}}$. 
      If $\boldsymbol{\omega}^{(j)}$ and $\boldsymbol{\omega}^{(k)}$ be Type-A-confusable then 
     $\emph{\textbf{z}}=\sigma_{1}^{\ell_{1}}\ldots\sigma_{j}^{\ell_{j}-1}\sigma_{j+1}\sigma_{j}\sigma_{j+1}\ldots\sigma_{k-1}\sigma_{k}^{\ell_{k}-1}\ldots\sigma_{\rho(\boldsymbol{\omega})}^{\ell_{\rho(\boldsymbol{\omega})}}$.
     If $\mathcal{I}_{2}(\boldsymbol{\omega}^{(i)})\cap \mathcal{I}_{2}(\boldsymbol{\omega}^{(j)})=\mathcal{I}_{1}(\boldsymbol{\omega})\cup J(\boldsymbol{\omega}^{(i)},\boldsymbol{\omega}^{(j)})$ then  $\mathcal{I}_{2}(\boldsymbol{\omega}^{(i)})\cap \mathcal{I}_{2}(\boldsymbol{\omega}^{(j)})\cap \mathcal{I}_{2}(\boldsymbol{\omega}^{(k)})=\mathcal{I}_{1}(\boldsymbol{\omega})\cup ( \mathcal{I}_{1}(\emph{\textbf{z}})\cap J(\boldsymbol{\omega}^{(i)},\boldsymbol{\omega}^{(j)}))$.
     
   For this case define \emph{\textbf{u}}$=\sigma_{1}^{\ell_{1}}\ldots\sigma_{i}^{\ell_{i}-1}$ and \emph{\textbf{w}}$=\sigma_{j}^{\ell_{j}-1}\ldots\sigma_{\rho(\boldsymbol{\omega})}^{\ell_{\rho(\boldsymbol{\omega})}}$. If $\boldsymbol{\omega}^{(i)}=\textbf{\emph{u}}\beta\gamma\textbf{\emph{w}}$ and $\boldsymbol{\omega}^{(j)}=\textbf{\emph{u}}\alpha\beta\textbf{\emph{w}}$ then $ J(\boldsymbol{\omega}^{(i)},\boldsymbol{\omega}^{(j)})=\{\textbf{\emph{u}}\beta\alpha\gamma\beta\textbf{\emph{w}}, \textbf{\emph{u}}\beta\gamma\alpha\beta\textbf{\emph{w}}\}$. If $\boldsymbol{\omega}^{(i)}=\textbf{\emph{u}}\beta\textbf{\emph{v}}\gamma\delta\textbf{\emph{w}}$ and $\boldsymbol{\omega}^{(j)}=\textbf{\emph{u}}\alpha\beta\textbf{\emph{v}}\gamma\textbf{\emph{w}}$ then $ J(\boldsymbol{\omega}^{(i)},\boldsymbol{\omega}^{(j)})=\textbf{\emph{u}}\beta(\mathcal{I}_{1}(\alpha\beta\textbf{\emph{v}})\cap\mathcal{I}_{1}(\textbf{\emph{v}}\gamma\delta))\gamma\textbf{\emph{w}}$. 
     Based on Lemma 4, there are several subcases which we consider below
    
       1) $j=i+2, \ell_{i+1}=1, \sigma_{i}\neq\sigma_{i+2}$. In this subcase $ J(\boldsymbol{\omega}^{(i)},\boldsymbol{\omega}^{(j)})=\{\emph{\textbf{u}}\sigma_{i+1}\sigma_{i}\sigma_{i+2}\sigma_{i+1}\emph{\textbf{w}}, \emph{\textbf{u}}\sigma_{i+1}\sigma_{i+2}\sigma_{i+1}\emph{\textbf{w}}\}$.
        
      2)  $j=i+2, \ell_{i+1}=2, \sigma_{i}=\sigma_{i+2}$. In this subcase $ J(\boldsymbol{\omega}^{(i)},\boldsymbol{\omega}^{(j)})=\{\emph{\textbf{u}}\sigma_{i+1}\sigma_{i}\sigma_{i+1}\sigma_{i+2}\sigma_{i+1}\emph{\textbf{w}},\emph{\textbf{u}}\sigma_{i+1}^{2}\sigma_{i}\sigma_{i+1}^{2}\emph{\textbf{w}}\}$.
       
       3) $j=i+2,(\ell_{i+1}=2,\sigma_{i}\neq\sigma_{i+2})\lor (\ell_{i+1}\geq 3)$. In this subcase $ J(\boldsymbol{\omega}^{(i)},\boldsymbol{\omega}^{(j)})=\{\emph{\textbf{u}}\sigma_{i+1}\sigma_{i}\sigma_{i+1}^{\ell_{i+1}-1}\sigma_{i+2}\sigma_{i+1}\emph{\textbf{w}}\}$.
       
       4) $j\geq i+3, \ell_{i+1}=\ldots=\ell_{t-1}=1,\ell_{t+1}=\ell_{t+3}=\ldots=\ell_{s-1}=1,\ell_{t}=\ell_{t+2}=\ldots=\ell_{s}=2,\ell_{s+1}=\ldots=\ell_{j-1}=1,i+1\leq t<t+1<s\leq j-1,\sigma_{l}=\sigma_{l+2},i\leq l\leq j-2$. In this subcase, based on the notations defined in the proof of Lemma 4, we have
           $J(\boldsymbol{\omega}^{(i)},\boldsymbol{\omega}^{(j)})=\{\emph{\textbf{u}}\sigma_{i+1}\emph{\textbf{u}}^{'}\sigma_{t}\sigma_{t-1}\sigma_{t}^{2}\ldots\sigma_{s-2}^{2}\sigma_{s-1}\sigma_{s}\emph{\textbf{w}}^{'}\sigma_{j-1}\emph{\textbf{w}}\}$.
           
     5) $j\geq i+3,\ell_{i+1}=\ldots=\ell_{t-1}=1,\ell_{t}\geq3,\ell_{t+1}=\ldots=\ell_{j-1}=1,i+1\leq t \leq j-1,\sigma_{l}=\sigma_{l+2},i\leq l\leq t-2, t\leq l\leq j-2$. In this subcase, based on the notations defined in the proof of Lemma 4, we have
       $J(\boldsymbol{\omega}^{(i)},\boldsymbol{\omega}^{(j)})=\{\emph{\textbf{u}}\sigma_{i+1}\emph{\textbf{u}}^{'}\sigma_{t-1}\sigma_{t}^{\ell_{t}-1}\sigma_{t+1}\emph{\textbf{w}}^{'}\sigma_{j-1}\emph{\textbf{w}}\}$.
       
     6) $j\geq i+3, \ell_{i+1}=\ldots=\ell_{t-1}=1,\ell_{t}=2,\ell_{t+1}=\ldots=\ell_{j-1}=1,\sigma_{t-1}\neq\sigma_{t+1},i+1\leq t \leq j-1,\sigma_{k}=\sigma_{k+2},i\leq k\leq t-2,t\leq k\leq j-2$. In this subcase,
      $J(\boldsymbol{\omega}^{(i)},\boldsymbol{\omega}^{(j)})=\{\emph{\textbf{u}}\sigma_{i+1}\sigma_{i}\ldots\sigma_{j}\sigma_{j-1}\emph{\textbf{w}}\}$.
      
    7)  $j\geq i+3,\ell_{i+1}=\ldots=\ell_{t-1}=1,\ell_{t}=2,\ell_{t+1}=\ldots=\ell_{j-1}=1, i+1\leq t \leq j-1,\sigma_{l}=\sigma_{l+2},i\leq l\leq j-2$. In this subcase,
            $J(\boldsymbol{\omega}^{(i)},\boldsymbol{\omega}^{(j)})=\{\emph{\textbf{u}}\sigma_{i+1}\sigma_{i}\ldots\sigma_{j}\sigma_{j-1}\emph{\textbf{w}}, \emph{\textbf{u}}\sigma_{i+1}\emph{\textbf{u}}^{'}\sigma_{t}\sigma_{t+1}\sigma_{t}\emph{\textbf{w}}^{'}\sigma_{j-1}\emph{\textbf{w}}\}$.
           
    8) $j\geq i+3,\ell_{i+1}=\ldots=\ell_{j-1}=1,\sigma_{t}\neq\sigma_{t+2},i\leq t\leq j-2,\sigma_{l}=\sigma_{l+2},i\leq l\leq t-1,t+1\leq l\leq j-2$. In this subcase,  $J(\boldsymbol{\omega}^{(i)},\boldsymbol{\omega}^{(j)})=\{\emph{\textbf{u}}\sigma_{i+1}\sigma_{i}\ldots\sigma_{t}\sigma_{t+2}\sigma_{t+1}\ldots\sigma_{j-1}\emph{\textbf{w}}$
      
     $,\emph{\textbf{u}}\sigma_{i+1}\ldots\sigma_{t+2}\sigma_{t}\sigma_{t+3}\ldots\sigma_{j}\sigma_{j-1}\emph{\textbf{w}}\}$.
      
     Now, we must show that always $J(\boldsymbol{\omega}^{(i)},\boldsymbol{\omega}^{(j)})\cap \mathcal{I}_{1}(\emph{\textbf{z}})=\emptyset$. Suppose on the contrary that $J(\boldsymbol{\omega}^{(i)},\boldsymbol{\omega}^{(j)})\cap \mathcal{I}_{1}(\emph{\textbf{z}})\neq\emptyset$. By comparing the first symbol on the left of $\sigma_{k}^{\ell_{k}-1}$ in \emph{\textbf{z}} and elements of  $J(\boldsymbol{\omega}^{(i)},\boldsymbol{\omega}^{(j)})$ we see that this is impossible, since in \emph{\textbf{z}}, the symbol is $\sigma_{k-1}$ while it is $\sigma_{k}$ in elements of $J(\boldsymbol{\omega}^{(i)},\boldsymbol{\omega}^{(j)})$. 
  \item Let  $\mathcal{I}_{1}(\boldsymbol{\omega}^{(i)})\cap \mathcal{I}_{1}(\boldsymbol{\omega}^{(j)})=\{\boldsymbol{\omega}, \emph{\textbf{z}}\}$ and $\mathcal{I}_{1}(\boldsymbol{\omega}^{(j)})\cap \mathcal{I}_{1}(\boldsymbol{\omega}^{(k)})=\{\boldsymbol{\omega}\}$. Therefore $\mathcal{I}_{2}(\boldsymbol{\omega}^{(i)})\cap \mathcal{I}_{2}(\boldsymbol{\omega}^{(j)})= \mathcal{I}_{1}(\boldsymbol{\omega})\cup \mathcal{I}_{1}(\emph{\textbf{z}})$. If $d_{H}(\boldsymbol{\omega}^{(i)},\boldsymbol{\omega}^{(j)})=1$ then $\emph{\textbf{z}}=\sigma_{1}^{\ell_{1}}\ldots\sigma_{i}^{\ell_{i}-1}\sigma_{i+1}\sigma_{i}\sigma_{i+1}^{\ell_{i+1}-1}\ldots\sigma_{\rho(\boldsymbol{\omega})}^{\ell_{\rho(\boldsymbol{\omega})}}$. If $\boldsymbol{\omega}^{(i)}$ and $\boldsymbol{\omega}^{(j)}$ be Type-A-confusable then $\emph{\textbf{z}}=\sigma_{1}^{\ell_{1}}\ldots\sigma_{i}^{\ell_{i}-1}\sigma_{i+1}\sigma_{i}\sigma_{i+1}\ldots\sigma_{j-1}\sigma_{j}^{\ell_{j}-1}\ldots\sigma_{\rho(\boldsymbol{\omega})}^{\ell_{\rho(\boldsymbol{\omega})}}$.
     If $\mathcal{I}_{2}(\boldsymbol{\omega}^{(j)})\cap \mathcal{I}_{2}(\boldsymbol{\omega}^{(k)})=\mathcal{I}_{1}(\boldsymbol{\omega})\cup J(\boldsymbol{\omega}^{(j)},\boldsymbol{\omega}^{(k)})$ then  $\mathcal{I}_{2}(\boldsymbol{\omega}^{(i)})\cap \mathcal{I}_{2}(\boldsymbol{\omega}^{(j)})\cap \mathcal{I}_{2}(\boldsymbol{\omega}^{(k)})=\mathcal{I}_{1}(\boldsymbol{\omega})\cup ( \mathcal{I}_{1}(\emph{\textbf{z}})\cap J(\boldsymbol{\omega}^{(j)},\boldsymbol{\omega}^{(k)}))$. If $\boldsymbol{\omega}^{(j)}=\textbf{\emph{u}}\alpha\beta\textbf{\emph{w}}$ and $\boldsymbol{\omega}^{(k)}=\textbf{\emph{u}}\beta\gamma\textbf{\emph{w}}$ then $ J(\boldsymbol{\omega}^{(j)},\boldsymbol{\omega}^{(k)})=\{\textbf{\emph{u}}\beta\alpha\gamma\beta\textbf{\emph{w}}, \textbf{\emph{u}}\beta\gamma\alpha\beta\textbf{\emph{w}}\}$. If $\boldsymbol{\omega}^{(j)}=\textbf{\emph{u}}\beta\textbf{\emph{v}}\gamma\delta\textbf{\emph{w}}$ and $\boldsymbol{\omega}^{(k)}=\textbf{\emph{u}}\alpha\beta\textbf{\emph{v}}\gamma\textbf{\emph{w}}$ then $ J(\boldsymbol{\omega}^{(j)},\boldsymbol{\omega}^{(k)})=\textbf{\emph{u}}\beta(\mathcal{I}_{1}(\alpha\beta\textbf{\emph{v}})\cap\mathcal{I}_{1}(\textbf{\emph{v}}\gamma\delta))\gamma\textbf{\emph{w}}$. There are several subcases which we consider in below. For this case define \textbf{\emph{u}}$=\sigma_{1}^{\ell_{1}}\ldots\sigma_{j}^{\ell_{j}-1}$, \textbf{\emph{u}}$^{'}=\sigma_{j}\ldots\sigma_{t-2}$, \textbf{\emph{w}}$^{'}=\sigma_{t+2}\ldots\sigma_{k}$ \emph{\textbf{w}}$=\sigma_{k}^{\ell_{k}-1}\ldots\sigma_{\rho(\boldsymbol{\omega})}^{\ell_{\rho(\boldsymbol{\omega})}}$.
     
       1) $k=j+2, \ell_{j+1}=1, \sigma_{j}\neq \sigma_{k}$. In this subcase $J(\boldsymbol{\omega}^{(j)},\boldsymbol{\omega}^{(k)})=\{\textbf{\emph{u}}\sigma_{j+1}\sigma_{k}\sigma_{j}\sigma_{j+1}\textbf{\emph{w}}, \textbf{\emph{u}}\sigma_{j+1}\sigma_{j}\sigma_{k}\sigma_{j+1}\textbf{\emph{w}}\}$.
       2) $k=j+2, \ell_{j+1}=2, \sigma_{j}=\sigma_{k}$. In this subcase  $J(\boldsymbol{\omega}^{(j)},\boldsymbol{\omega}^{(k)})=\{\textbf{\emph{u}}\sigma_{j+1}\sigma_{j}\sigma_{j+1}\sigma_{j}\sigma_{j+1}\textbf{\emph{w}}, \textbf{\emph{u}}\sigma_{j+1}^{2}\sigma_{j}\sigma_{j+1}^{2}\textbf{\emph{w}}\}$.
       
       3) $k=j+2$, $(\ell_{j+1}=2,\sigma_{j}\neq\sigma_{k})\lor (\ell_{j+1}\geq 3)$. In this subcase
        
       $J(\boldsymbol{\omega}^{(j)},\boldsymbol{\omega}^{(k)})=\textbf{\emph{u}}\sigma_{j+1}\sigma_{j}\sigma_{j+1}^{\ell_{j+1}-1}\sigma_{k}\sigma_{j+1}\textbf{\emph{w}}$.
 
        4) $k\geq j+3, \ell_{j+1}=\ldots=\ell_{t-1}=1,\ell_{t+1}=\ell_{t+3}=\ldots=\ell_{s-1}=1,\ell_{t}=\ell_{t+2}=\ldots=\ell_{s}=2,\ell_{s+1}=\ldots=\ell_{k-1}=1,j+1\leq t<t+1<s\leq k-1,\sigma_{l}=\sigma_{l+2},i\leq l\leq j-2$. In this subcase, we have
           $J(\boldsymbol{\omega}^{(j)},\boldsymbol{\omega}^{(k)})=\{\emph{\textbf{u}}\sigma_{j+1}\emph{\textbf{u}}^{'}\sigma_{t}\sigma_{t-1}\sigma_{t}^{2}\ldots\sigma_{s-2}^{2}\sigma_{s-1}\sigma_{s}\emph{\textbf{w}}^{'}\sigma_{k-1}\emph{\textbf{w}}\}$.
        
        5) $k\geq j+3,\ell_{j+1}=\ldots=\ell_{t-1}=1,\ell_{t}\geq3,\ell_{t+1}=\ldots=\ell_{k-1}=1,j+1\leq t \leq k-1,\sigma_{l}=\sigma_{l+2},i\leq l\leq t-2, t\leq l\leq k-2$. In this subcase,
       $J(\boldsymbol{\omega}^{(j)},\boldsymbol{\omega}^{(k)})=\{\emph{\textbf{u}}\sigma_{j+1}\emph{\textbf{u}}^{'}\sigma_{t-1}\sigma_{t}^{\ell_{t}-1}\sigma_{t+1}\emph{\textbf{w}}^{'}\sigma_{k-1}\emph{\textbf{w}}\}$.
        
                6) $k\geq j+3$, $\ell_{j+1}=\ldots=\ell_{t-1}=1,\ell_{t}=2,\ell_{t+1}=\ldots=\ell_{k-1}=1$, $\sigma_{t-1}\neq\sigma_{t+1},\sigma_{l}=\sigma_{l+2},j\leq l\leq t-2, t\leq l\leq k-2 $. In this subcase
           
         $J(\boldsymbol{\omega}^{(j)},\boldsymbol{\omega}^{(k)})=\textbf{\emph{u}}\sigma_{j+1}\sigma_{j}\ldots\sigma_{k}\sigma_{k-1}\textbf{\emph{w}}$.
      
       7) $k\geq j+3$, $\ell_{j+1}=\ldots=\ell_{t-1}=1,\ell_{t}=2,\ell_{t+1}=\ldots=\ell_{k-1}=1$, $\sigma_{l}=\sigma_{l+2},j\leq l\leq k-2$. In this subcase
        $J(\boldsymbol{\omega}^{(j)},\boldsymbol{\omega}^{(k)})=\{\textbf{\emph{u}}\sigma_{j+1}\sigma_{j}\ldots\sigma_{k}\sigma_{k-1}\textbf{\emph{w}}$
       $,\textbf{\emph{u}}\sigma_{j+1}\sigma_{j+2}\ldots\sigma_{t}^{2}\sigma_{t+1}\sigma_{t}^{2}\sigma_{t+1}\ldots\sigma_{k-1}\textbf{\emph{w}}\}$
       
        8)  $k\geq j+3$, $\ell_{j+1}=\ldots=\ell_{k-1}=1$, $\sigma_{t}\neq\sigma_{t+2},\sigma_{l}=\sigma_{l+2},j\leq l\leq t-1,t+1\leq l\leq k-2$. In this subcase
          $J(\boldsymbol{\omega}^{(j)},\boldsymbol{\omega}^{(k)})=\{\textbf{\emph{u}}\sigma_{j+1}\sigma_{j}\ldots\sigma_{t}\sigma_{t+2}\sigma_{t+1}\ldots\sigma_{k-2}\sigma_{k-1}\textbf{\emph{w}}$        
       $,\textbf{\emph{u}}\sigma_{j+1}\sigma_{j+2}\ldots\sigma_{t+2}\sigma_{t}\sigma_{t+3}\ldots\sigma_{k}\sigma_{k-1}\textbf{\emph{w}}\}$.  
         
     Now, we must show that always $J(\boldsymbol{\omega}^{(j)},\boldsymbol{\omega}^{(k)})\cap \mathcal{I}_{1}(\emph{\textbf{z}})=\emptyset$. Suppose on the contrary that $J(\boldsymbol{\omega}^{(j)},\boldsymbol{\omega}^{(k)})\cap \mathcal{I}_{1}(\emph{\textbf{z}})\neq\emptyset$. By comparing the first symbol on the left of $\sigma_{k}^{\ell_{k}-1}$ in \emph{\textbf{z}} and elements of  $J(\boldsymbol{\omega}^{(i)},\boldsymbol{\omega}^{(j)})$ we see that this is impossible, since in \emph{\textbf{z}}, the symbol is $\sigma_{k}$ while it is $\sigma_{k-1}$ in elements of $J(\boldsymbol{\omega}^{(i)},\boldsymbol{\omega}^{(j)})$.
  \item Let  $\mathcal{I}_{1}(\boldsymbol{\omega}^{i})\cap \mathcal{I}_{1}(\boldsymbol{\omega}^{j})=\{\boldsymbol{\omega}\}$ and $\mathcal{I}_{1}(\boldsymbol{\omega}^{j})\cap \mathcal{I}_{1}(\boldsymbol{\omega}^{k})=\{\boldsymbol{\omega}\}$. Therefore  $\mathcal{I}_{2}(\boldsymbol{\omega}^{i})\cap \mathcal{I}_{2}(\boldsymbol{\omega}^{j})=\mathcal{I}_{1}(\boldsymbol{\omega})\cup J(\boldsymbol{\omega}^{(i)},\boldsymbol{\omega}^{(j)})$ and  $\mathcal{I}_{2}(\boldsymbol{\omega}^{j})\cap \mathcal{I}_{2}(\boldsymbol{\omega}^{k})=\mathcal{I}_{1}(\boldsymbol{\omega})\cup J(\boldsymbol{\omega}^{(j)},\boldsymbol{\omega}^{(k)})$. 
     The expressions for $J(\boldsymbol{\omega}^{(i)},\boldsymbol{\omega}^{(j)})$ are described in the first case and the expressions for $J(\boldsymbol{\omega}^{(j)},\boldsymbol{\omega}^{(k)})$ are described in the second case. We must show that always $ J(\boldsymbol{\omega}^{(i)},\boldsymbol{\omega}^{(j)})\cap J(\boldsymbol{\omega}^{(j)},\boldsymbol{\omega}^{(k)})=\emptyset$. While in the elements of $J(\boldsymbol{\omega}^{(i)},\boldsymbol{\omega}^{(j)})$ the symbol on the left of $\sigma_{k}^{\ell_{k}-1}$ is $\sigma_{k}$ it is $\sigma_{k-1}$ in $J(\boldsymbol{\omega}^{(j)},\boldsymbol{\omega}^{(k)})$.
\end{itemize}
\section{Proof of Lemma 11}
First note that we do not need to consider other intersections, since other sets of 1-subsequences definitely contain a pair of 1-subsequences of Hamming distance at least five. We prove the claims above by describing the set of sequences belong to the intersection of substitution balls.
\begin{description}
  \item[(i)] Let $\boldsymbol{\omega}^{(i)}=$\textbf{\emph{u}}$\alpha$\textbf{\emph{v}}$\gamma$\textbf{\emph{w}}, $\boldsymbol{\omega}^{(i+1)}=\textbf{\emph{u}}\beta\textbf{\emph{v}}\gamma\textbf{\emph{w}}$ and $\boldsymbol{\omega}^{(i+2)}=\textbf{\emph{u}}\beta\textbf{\emph{v}}\delta\textbf{\emph{w}}$, where $\alpha=\delta=\sigma_{i+1}$, $\beta=\sigma_{i}$ and $\gamma=\sigma_{i+2}$. Let \textbf{\emph{z}} $\in \mathcal{B}_{2, 0, 0}(\boldsymbol{\omega}^{(i)})\cap \mathcal{B}_{2, 0, 0}(\boldsymbol{\omega}^{(i+1)})\cap \mathcal{B}_{2, 0, 0}(\boldsymbol{\omega}^{(i+2)})$. Consider the indices of two symbols where $\boldsymbol{\omega}^{(i)}$ and $\boldsymbol{\omega}^{(i+2)}$ are distinct. Then in sequence \textbf{\emph{z}}, either these two symbols are chosen freely from $\Sigma_{q}$ or first symbol is chosen from  $\boldsymbol{\omega}^{(i+2)}$ while the second symbol is chosen from  $\boldsymbol{\omega}^{(i)}$. Also exactly one substitution must occurs in one of the subsequences \textbf{\emph{u}}, \textbf{\emph{v}} or \textbf{\emph{w}} . Then $\mathcal{B}_{2, 0, 0}(\boldsymbol{\omega}^{(i)})\cap \mathcal{B}_{2, 0, 0}(\boldsymbol{\omega}^{(i+1)})\cap \mathcal{B}_{2, 0, 0}(\boldsymbol{\omega}^{(i+2)})=\{\textbf{\emph{u}}\theta\textbf{\emph{v}}\phi\textbf{\emph{w}},\tilde{\textbf{\emph{u}}}\beta\textbf{\emph{v}}\gamma\textbf{\emph{w}}
      ,\textbf{\emph{u}}\beta\tilde{\textbf{\emph{v}}}\gamma\textbf{\emph{w}},\textbf{\emph{u}}\beta\textbf{\emph{v}}\gamma\tilde{\textbf{\emph{w}}}\}$ where $\theta, \phi \in \Sigma_{q}$ and  $d_{H}(\textbf{\emph{u}},\tilde{\textbf{\emph{u}}})=1$, $d_{H}(\textbf{\emph{v}},\tilde{\textbf{\emph{v}}})=1$ and $d_{H}(\textbf{\emph{w}},\tilde{\textbf{\emph{w}}})=1$.
  \item[(ii)] Let $\boldsymbol{\omega}^{(i)}=$\textbf{\emph{u}}$\alpha_{1}$\textbf{\emph{v}}$\alpha_{2}$\textbf{\emph{w}}$\alpha_{3}$\textbf{\emph{z}},\\ $\boldsymbol{\omega}^{(i+1)}=$\textbf{\emph{u}}$\beta_{1}$\textbf{\emph{v}}$\alpha_{2}$\textbf{\emph{w}}$\alpha_{3}$\textbf{\emph{z}},\\
      $\boldsymbol{\omega}^{(i+2)}=$\textbf{\emph{u}}$\beta_{1}$\textbf{\emph{v}}$\beta_{2}$\textbf{\emph{w}}$\alpha_{3}$\textbf{\emph{z}},\\
      $\boldsymbol{\omega}^{(i+3)}=$\textbf{\emph{u}}$\beta_{1}$\textbf{\emph{v}}$\beta_{2}$\textbf{\emph{w}}$\beta_{3}$\textbf{\emph{z}} where $\alpha_{i}\neq\beta_{i}$. Based on claim 3 of Lemma 10, we calculate $\mathcal{B}_{2, 0, 0}(\boldsymbol{\omega}^{(i)})\cap \mathcal{B}_{2, 0, 0}(\boldsymbol{\omega}^{(i+3)})$. After checking which sequences are also elements of $\mathcal{B}_{2, 0, 0}(\boldsymbol{\omega}^{(i+1)})$ we have $\mathcal{B}_{2, 0, 0}(\boldsymbol{\omega}^{(i)})\cap \mathcal{B}_{2, 0, 0}(\boldsymbol{\omega}^{(i+1)})\cap \mathcal{B}_{2, 0, 0}(\boldsymbol{\omega}^{(i+3)})=$\{\textbf{\emph{u}}$\beta_{1}$\textbf{\emph{v}}$\alpha_{2}$\textbf{\emph{w}}$\gamma$\textbf{\emph{z}},
      \textbf{\emph{u}}$\beta_{1}$\textbf{\emph{v}}$\delta$\textbf{\emph{w}}$\alpha_{3}$\textbf{\emph{z}},
      \textbf{\emph{u}}$\theta$\textbf{\emph{v}}$\alpha_{2}$\textbf{\emph{w}}$\beta_{3}$\textbf{\emph{z}},
       \textbf{\emph{u}}$\theta$\textbf{\emph{v}}$\beta_{2}$\textbf{\emph{w}}$\alpha_{3}$\textbf{\emph{z}}\} where $\gamma\in\Sigma_{q}$, $\delta\in \Sigma_{q}\backslash \{\alpha_{2}\}$, $\theta\in \Sigma_{q} \backslash \{\beta_{1}\}$. It is similar to prove $|\mathcal{B}_{2, 0, 0}(\boldsymbol{\omega}^{(i)})\cap \mathcal{B}_{2, 0, 0}(\boldsymbol{\omega}^{(i+2)})\cap \mathcal{B}_{2, 0, 0}(\boldsymbol{\omega}^{(i+3)})|=4q-3$.
  \item[(iii)] Let $\boldsymbol{\omega}^{(i)}=\textbf{\emph{t}}\alpha_{1}\textbf{\emph{u}}\alpha_{2}\textbf{\emph{v}}\alpha_{3}\textbf{\emph{w}}\alpha_{4}\textbf{\emph{z}}$,    
  $\boldsymbol{\omega}^{(i+1)}=\textbf{\emph{t}}\beta_{1}\textbf{\emph{u}}\alpha_{2}\textbf{\emph{v}}\alpha_{3}\textbf{\emph{w}}\alpha_{4}\textbf{\emph{z}}$,
  $\boldsymbol{\omega}^{(i+2)}=\textbf{\emph{t}}\beta_{1}\textbf{\emph{u}}\beta_{2}\textbf{\emph{v}}\alpha_{3}\textbf{\emph{w}}\alpha_{4}\textbf{\emph{z}}$,
  $\boldsymbol{\omega}^{(i+3)}=\textbf{\emph{t}}\beta_{1}\textbf{\emph{u}}\beta_{2}\textbf{\emph{v}}\beta_{3}\textbf{\emph{w}}\alpha_{4}\textbf{\emph{z}}$,
  $\boldsymbol{\omega}^{(i+4)}=\textbf{\emph{t}}\beta_{1}\textbf{\emph{u}}\beta_{2}\textbf{\emph{v}}\beta_{3}\textbf{\emph{w}}\beta_{4}\textbf{\emph{z}}$. Based on claim 4 of Lemma 10, we calculate $\mathcal{B}_{2, 0, 0}(\boldsymbol{\omega}^{(i)})\cap \mathcal{B}_{2, 0, 0}(\boldsymbol{\omega}^{(i+4)})$. After checking which sequences are also elements of $\mathcal{B}_{2, 0, 0}(\boldsymbol{\omega}^{(i+1)})$ we have
  
  $\mathcal{B}_{2, 0, 0}(\boldsymbol{\omega}^{(i)})\cap \mathcal{B}_{2, 0, 0}(\boldsymbol{\omega}^{(i+1)})\cap \mathcal{B}_{2, 0, 0}(\boldsymbol{\omega}^{(i+4)})=\{$\textbf{\emph{t}}$\beta_{1}$\textbf{\emph{u}}$\alpha_{2}$\textbf{\emph{v}}$\alpha_{3}$\textbf{\emph{w}}$\beta_{4}$\textbf{\emph{z}},
  \textbf{\emph{t}}$\beta_{1}$\textbf{\emph{u}}$\alpha_{2}$\textbf{\emph{v}}$\beta_{3}$\textbf{\emph{w}}$\alpha_{4}$\textbf{\emph{z}},
  \textbf{\emph{t}}$\beta_{1}$\textbf{\emph{u}}$\beta_{2}$\textbf{\emph{v}}$\alpha_{3}$\textbf{\emph{w}}$\alpha_{4}$\textbf{\emph{z}} $\}$.
  \item[(iv)] Similarly $\mathcal{B}_{2, 0, 0}(\boldsymbol{\omega}^{(i)})\cap \mathcal{B}_{2, 0, 0}(\boldsymbol{\omega}^{(i+2)})\cap \mathcal{B}_{2, 0, 0}(\boldsymbol{\omega}^{(i+4)})=\{$\textbf{\emph{t}}$\beta_{1}$\textbf{\emph{u}}$\alpha_{2}$\textbf{\emph{v}}$\alpha_{3}$\textbf{\emph{w}}$\beta_{4}$\textbf{\emph{z}},
  \textbf{\emph{t}}$\beta_{1}$\textbf{\emph{u}}$\alpha_{2}$\textbf{\emph{v}}$\beta_{3}$\textbf{\emph{w}}$\alpha_{4}$\textbf{\emph{z}},
  \textbf{\emph{t}}$\beta_{1}$\textbf{\emph{u}}$\beta_{2}$\textbf{\emph{v}}$\alpha_{3}$\textbf{\emph{w}}$\alpha_{4}$\textbf{\emph{z}},
   \textbf{\emph{t}}$\alpha_{1}$\textbf{\emph{u}}$\beta_{2}$\textbf{\emph{v}}$\alpha_{3}$\textbf{\emph{w}}$\beta_{4}$\textbf{\emph{z}},
   \textbf{\emph{t}}$\alpha_{1}$\textbf{\emph{u}}$\beta_{2}$\textbf{\emph{v}}$\beta_{3}$\textbf{\emph{w}}$\alpha_{4}$\textbf{\emph{z}}$\}$.
  \item[(v)] We use the notation of the claim (ii), Based on claim 3 of Lemma 10, we calculate $\mathcal{B}_{2, 0, 0}(\boldsymbol{\omega}^{(i)})\cap \mathcal{B}_{2, 0, 0}(\boldsymbol{\omega}^{(i+3)})$. After checking which sequences are also elements of $\mathcal{B}_{2, 0, 0}(\boldsymbol{\omega}^{(i+1)})$ and $\mathcal{B}_{2, 0, 0}(\boldsymbol{\omega}^{(i+2)})$ , we have $\mathcal{B}_{2, 0, 0}(\boldsymbol{\omega}^{(i)})\cap \mathcal{B}_{2, 0, 0}(\boldsymbol{\omega}^{(i+1)})\cap \mathcal{B}_{2, 0, 0}(\boldsymbol{\omega}^{(i+2)})\cap \mathcal{B}_{2, 0, 0}(\boldsymbol{\omega}^{(i+3)})=\{$\textbf{\emph{u}}$\beta_{1}$\textbf{\emph{v}}$\alpha_{2}$\textbf{\emph{w}}$\gamma$\textbf{\emph{z}},
     \textbf{\emph{u}}$\beta_{1}$\textbf{\emph{v}}$\delta$\textbf{\emph{w}}$\alpha_{3}$\textbf{\emph{z}},
     \textbf{\emph{u}}$\theta$\textbf{\emph{v}}$\beta_{2}$\textbf{\emph{w}}$\alpha_{3}$\textbf{\emph{z}}$\}$, where $\gamma\in\Sigma_{q}$, $\delta\in \Sigma_{q}\backslash\{\alpha_{2}\}$, $\theta\in \Sigma_{q}\backslash\{\beta_{1}\}$.
  \item[(vi)] We use the notation of the claim (iii), Based on claim 4 of Lemma 10, we calculate $\mathcal{B}_{2, 0, 0}(\boldsymbol{\omega}^{(i)})\cap \mathcal{B}_{2, 0, 0}(\boldsymbol{\omega}^{(i+4)})$. After checking which sequences are also elements of $\mathcal{B}_{2, 0, 0}(\boldsymbol{\omega}^{(i+2)})$ and $\mathcal{B}_{2, 0, 0}(\boldsymbol{\omega}^{(i+3)})$ we have $\mathcal{B}_{2, 0, 0}(\boldsymbol{\omega}^{(i)})\cap \mathcal{B}_{2, 0, 0}(\boldsymbol{\omega}^{(i+2)})\cap \mathcal{B}_{2, 0, 0}(\boldsymbol{\omega}^{(i+3)})\cap \mathcal{B}_{2, 0, 0}(\boldsymbol{\omega}^{(i+4)})=\{$\textbf{\emph{t}}$\alpha_{1}$\textbf{\emph{u}}$\beta_{2}$\textbf{\emph{v}}$\beta_{3}$\textbf{\emph{w}}$\alpha_{4}$\textbf{\emph{z}},
      \textbf{\emph{t}}$\beta_{1}$\textbf{\emph{u}}$\alpha_{2}$\textbf{\emph{v}}$\beta_{3}$\textbf{\emph{w}}$\alpha_{4}$\textbf{\emph{z}},
      \textbf{\emph{t}}$\beta_{1}$\textbf{\emph{u}}$\beta_{2}$\textbf{\emph{v}}$\alpha_{3}$\textbf{\emph{w}}$\alpha_{4}$\textbf{\emph{z}}$\}$.
  \item[(vii)] And $\mathcal{B}_{2, 0, 0}(\boldsymbol{\omega}^{(i)})\cap \mathcal{B}_{2, 0, 0}(\boldsymbol{\omega}^{(i+1)})\cap \mathcal{B}_{2, 0, 0}(\boldsymbol{\omega}^{(i+3)})\cap \mathcal{B}_{2, 0, 0}(\boldsymbol{\omega}^{(i+4)})=\{ \textbf{\emph{t}}\beta_{1}\textbf{\emph{u}}\alpha_{2}\textbf{\emph{v}}\beta_{3}\textbf{\emph{w}}\alpha_{4}\textbf{\emph{z}},
      \textbf{\emph{t}}\beta_{1}\textbf{\emph{u}}\beta_{2}\textbf{\emph{v}}\alpha_{3}\textbf{\emph{w}}\alpha_{4}\textbf{\emph{z}}\}$.
  \item[(viii)] And finally $\bigcap_{j=0}^{4}\mathcal{B}_{2, 0, 0}(\boldsymbol{\omega}^{(i+j)})=\{$ \textbf{\emph{t}}$\beta_{1}$\textbf{\emph{u}}$\alpha_{2}$\textbf{\emph{v}}$\beta_{3}$\textbf{\emph{w}}$\alpha_{4}$\textbf{\emph{z}},
      \textbf{\emph{t}}$\beta_{1}$\textbf{\emph{u}}$\beta_{2}$\textbf{\emph{v}}$\alpha_{3}$\textbf{\emph{w}}$\alpha_{4}$\textbf{\emph{z}}$\}$
  
\end{description}

\ifCLASSOPTIONcaptionsoff
  \newpage
\fi

\end{document}